\documentclass[twocolumn]{aastex63}
\usepackage{amsmath}
\usepackage{amssymb}
\usepackage{chngcntr}
\counterwithout{figure}{section}

\received{}
\revised{}
\accepted{}
\submitjournal{ApJ}

\shorttitle{Optical intranight variability PSDs}
\shortauthors{A. Goyal}

\begin{document}

\title{Optical variability power spectrum analysis of blazar sources on intranight timescales}

\correspondingauthor{Arti Goyal}
\email{arti.goyal@uj.edu.pl}

\author[0000-0002-2224-6664]{Arti Goyal}
\affiliation{Astronomical Observatory of the Jagiellonian University, Orla 171, 30-244 Krakow, Poland}

\begin{abstract}
We report the first results of a systematic investigation to characterize blazar variability power spectral densities (PSDs) at optical frequencies using densely sampled (5--15 minutes integration time), high photometric accuracy ($\lesssim$0.2--0.5\%) R-band intranight light curves, covering timescales ranging from several hours to $\sim$15\,minutes. Our sample consists of 14 optically bright blazars, including nine BL Lacertae objects (BL Lacs) and five flat-spectrum radio quasars (FSRQs) which have shown statistically significant variability during 29 monitoring sessions. We model the intranight PSDs as simple power--laws and derive the best-fit slope along with uncertainty using the `power spectral response' method. Our main results are the following: (1) on 19 out of 29 monitoring sessions, the intranight PSDs show an acceptable fit to simple power-laws at the rejection confidence $\leq$ 90\%; (2) for these 19 instances, the PSD slopes show a large range from 1.4 to 4.0, consistent with statistical characters of red (slope$\sim$2) and black (slope$\geq$3) noise stochastic processes; (3) the average PSD slopes for the BL Lacs and FSRQs are indistinguishable from one another; (4) the normalization of intranight PSDs for individual blazar sources which were monitored on more than one occasion turns out to be consistent with one another with a few exceptions. The average PSD slope, 2.9$\pm$0.3 (1$\sigma$ uncertainty) is steeper than the red-noise type character of variability found on longer timescales (many decades to days), indicative of a cutoff in the variability spectrum on timescales around a few days at the synchrotron frequencies of the emission spectrum.
        
\end{abstract}


\section{Introduction}\label{sec:intro}
     
Intense emission and rapid flux variability provide important clues to understand the underlying physical process operating in the relativistic, magnetized jets of blazar sources.  This subset of active galactic nuclei (AGN) is comprised of BL Lacertae objects (BL Lacs) and the high optical polarization flat-spectrum radio quasars (FSRQs), with jets launched from a supermassive black hole (SMBH)--accretion disk systems\citep[for a recent review see,][]{Hovatta19}.  The flux variability is observed at all frequencies of the electromagnetic spectrum on both long-term (decades to $\simeq$ day; \citealt{Ulrich97, Aller99, Ghosh00, Gopal-Krishna11, Goyal12, Gupta16, Gaur19, Hess17, Abdo10a}) and intranight timescales ($\leq$day; \citealt{Aharonian07, Albert07, Goyal13b, Bachev15, Ackermann16, Nalewajko17, Zhu18, Shukla18}). The two-component broadband spectral energy distribution is nonthermal radiation arising within the relativistic jet \citep[][]{Ghisellini08}. Within the leptonic scenario, the particle (electron and positron) pairs, accelerated to GeV/TeV energies, produce synchrotron radiation in the presence of magnetic field at lower frequencies (radio--to--optical/X-rays) and inverse Comptonization of the seed photons (same or thermal from the accretion disk) by the synchrotron emitting particles produce emission at higher frequencies (X-rays--to--TeV $\gamma-$rays). Alternatively, direct synchrotron radiation by the protons accelerated to PeV/EeV energies or the emission from secondaries can give rise to high energy radiation within the hadronic scenario \citep[e.g.,][]{Blandford19}. Well--defined flare emission has often been attributed to particle acceleration mechanisms related to shocks in the jet \citep[][]{Spada01, Marscher08} while turbulence can mimic observed fluctuations on long-term as well as small timescales \citep[][]{Marscher14}. Annihilation of magnetic field lines at the reconnection sites within the jet plasma can also impart energy to the particles \citep[][]{Giannios13, Sironi15}; this scenario is supported by the recent detection of minute-like variability at GeV energies for the blazar 3C\,279 \citep[][]{Shukla20}. On the other hand, flux changes on the intranight timescales have also been associated with variable Doppler boosting factors related to changes in the viewing angle of the emitting plasma \citep[e.g.,][]{Gopal-Krishna12}, although in such a scenario frequency--independent variability is expected \citep[see, in this context,][]{Pasierb20}.   

\begin{deluxetable*}{cccccccc}
\tablenum{1}
\tablecaption{Sample properties.\label{tab:sample}}
\tablewidth{0pt}
\tabletypesize{\small}
\tablehead{
\colhead{IAU name} & \colhead{RA(J2000)} & \colhead{Dec(J2000)} & SED & $z$ &  V--mag & $M_{BH}$ & Reference for $M_{BH}$  \\
\colhead{} &  \colhead{(h m s)} & \colhead{(d $\prime$ ${\prime\prime}$)} & \colhead{}  &  \colhead{} &  \colhead{} &  \colhead{(M$_\odot$)} & \colhead{} 
}
\decimalcolnumbers
\startdata
0109$+$224 & 01 12 05.824 & $+$22 44 38.78 &       BL Lac$^c$     & 0.265$^f$ &  15.66 & --                       &   \\ 
0235$+$164 & 02 38 38.930 & $+$16 36 59.27 &       FSRQ$^c$       & 0.940$^f$ &  15.50 & 2.0$\times$10${^8}$& \citet{Raiteri07}   \\ 
0420$-$014 & 04 23 15.800 & $-$01 20 33.06 &       FSRQ$^c$       & 0.915$^f$ &  17.00 & 7.9$\times$10${^8}$& \citet{Liang03}   \\ 
0716$+$714 & 07 21 53.448 & $+$71 20 36.36 &       BL Lac$^c$     & 0.300$^f$ &  15.50 & 1.3$\times$10${^8}$& \citet{Liang03}   \\ 
0806$+$315 & 08 09 13.440 & $+$31 22 22.90 &       BL Lac$^d$     & 0.220$^g$ &  15.70 & --                 &   \\ 
0806$+$524 & 08 09 49.186 & $+$52 18 58.25 &       BL Lac$^e$     & 0.138$^h$ &  15.59 & 7.9$\times$10${^8}$& \citet{Wu02}   \\ 
0851$+$202$^a$ & 08 54 48.874 & $+$20 06 30.64 &   BL Lac$^c$     & 0.306$^f$ &  15.43 & 1.5$\times$10${^8}$& \citet{Liang03}   \\ 
1011$+$496 & 10 15 04.139 & $+$49 26 00.70 &       BL Lac$^c$     & 0.200$^f$ &  16.15 & 2.1$\times$10${^8}$& \citet{Wu02}   \\ 
1156$+$295 & 11 59 31.833 & $+$29 14 43.82 &       FSRQ$^c$       & 0.729$^f$ &  14.41 & 7.9$\times$10${^8}$& \citet{Liang03}   \\ 
1216$-$010 & 12 18 34.929 & $-$01 19 54.34 &       BL Lac$^e$     & 0.415$^i$ &  15.64 & --                 &   \\ 
1219$+$285 & 12 21 31.690 & $+$28 13 58.50 &       BL Lac$^c$     & 0.102$^f$ &  16.11 & 2.5$\times$10${^7}$& \citet{Liang03}  \\ 
1253$-$055$^b$ & 12 56 11.166 & $-$05 47 21.52 &   FSRQ$^c$       & 0.538$^f$ &  17.75 & 7.9$\times$10${^8}$& \citet{Sbarrato12}  \\ 
1510$-$089 & 15 12 50.532 & $-$09 05 59.82 &       FSRQ$^d$       & 0.360$^j$ &  16.54 & 4.0$\times$10${^8}$& \citet{Sbarrato12}  \\ 
1553$+$113 & 15 55 43.044 & $+$11 11 24.36 &       BL Lac$^c$     & 0.360$^f$ &  15.00 & -- \\ 
\enddata
\tablecomments{
(1) the name of the blazar following the IAU convention. $^a$ also known as OJ\,287; $^b$ also known as 3C\,279; 
(2) right ascension; 
(3) declination;
(4) SED classification. $^c$\citet[][]{Healey08}; $^d$\citet[][]{Veron06}; $^e$\citet[][]{Plotkin08}; 
(5) spectroscopic redshift. $^f$\citet[][]{Healey08};  $^g$\citet{Falco98}; $^h$\citet{Bade98}; $^i$\citet{Dunlop89};  $^j$\citet{Thompson90}.
(6) typical optical V--band magnitude \citep{Veron10};
(7) mass of the SMBH; 
(8) reference for the mass of the SMBH.
}
\end{deluxetable*}

The noise-like appearance of blazar light curves have prompted efforts to investigate variability power spectral densities (PSDs) which is a distribution of variability amplitudes over different Fourier frequencies (=timescale$^{-1}$). The blazar PSDs are mostly represented by power-law shapes defined as P($\nu_k$) $\propto$ $\nu_k^{-\beta}$ where $\beta$(=1--3) is the slope and $\nu_k$ is the temporal frequency which indicate that variability is a {\it correlated} colored--noise type stochastic processes \citep[see,][and references therein]{Goyal20}. Specifically, $\beta$$\simeq$1, $\simeq$2 and $\gtrsim$3 are known as long-memory/pink--noise, damped--random walk/red--noise, and black--noise type stochastic processes while $\beta$$\simeq$0 corresponds to {\it uncorrelated}, white--noise type stochastic process \citep[][]{Press78, Schroeder91}. For a colored noise--type stationary stochastic process, one expects the slope of PSDs to change to 0 on longer timescales to preserve the finite variance of the process, leading to a relaxation timescale beyond which the variations should be generated due to uncorrelated processes. Moreover, it also means that different random realizations of the process will have different statistical moments (e.g., mean, sigma) due to statistical fluctuation of the process itself and not due to the change of nature of the process which indicates that the process is weakly non-stationary \citep[][]{Vaughan03}. Fluctuations resulting from such stochastic processes obey certain probability distributions, so the light curves tend to produce predictable PSDs. PSD slope and normalization, as well as the breaks, are of particular interest as they carry information about the parameters of the stochastic process and the `characteristic timescales' in the system which can be related to physical parameters shaping the variability, such as the size of the emission zone or the particle cooling timescales \citep[][]{Sobolewska14, Finke14, Chen16}. The noise-like appearance of light curves has been modeled where the aggregate flux arises from many cells behind a shock \citep[][see also, \citet{Marscher14} who models the flux and polarization light curves but does not provide PSDs]{Calafut15, Pollack16} against the emission from well-defined flares which could be attributed to single emission zones \citep[][]{Hughes85, Abdo10b}. The models of \citet[][]{Calafut15} and \cite[][]{Pollack16} compute the light curves and the PSDs, which are shaped by the combination bulk Lorentz factor fluctuations and the turbulence within the jets. In their hydrodynamic simulations of 2D jets, the changes in bulk Lorentz factor produce PSD slopes in the range 2.1 to 2.9 while the turbulence produces PSD slopes in the range 1.7 to 2.3, respectively \citep[][]{Pollack16}. The model of \citet[][]{O'Riordan17}, on the other hand, hypothesizes that the turbulence in the magnetically arrested disk (MAD) shapes the variability of synchrotron and IC emission components from the jet, with a cutoff of variability power at the timescales governed by the light crossing time of the event horizon of the SMBH. 

Unlike the long--term variability timescales where the slopes of PSDs of multiwavelength variability have been estimated for large samples of blazar sources \citep[in particular, $\beta$$\sim$2 for radio and optical and  $\beta$$\sim$1 for $\gamma-$rays,][]{Max-Moerbeck14a, Park17, Nilsson18, Meyer19, Goyal20}, such studies remain scarce on intranight timescales. This is due to the fact that it requires continuous pointing of an observing facility to a single target for many hours which is usually not feasible due to scheduling constraints, weather, limited photon sensitivity (at high energies), etc. The intranight PSD slopes at GeV and TeV energies exhibited $\beta\sim$1 and 2, respectively, for the blazars 3C\,279 and PKS\,2155$-$304, respectively, using the {\it Fermi-}LAT and the High Energy Stereoscopic System data, but only when the blazars were in a flaring state \citep{Aharonian07, Ackermann16}. \citet[][]{Zhang19} obtained the X-ray PSD slopes of 1.5, 3.1 and 1.4 using the 40--180\,ks long {\it Suzaku} observations. In another study, \citet{Bhattacharyya20} obtained  intranight X-ray PSD slopes equal to 2.7, 2.6, 1.9 and, 2.7 for the Mrk\,421 and 2.2, 2.8, and 2.9, respectively, for the PKS\,2155$-$304 using the 30--90\,ks long XMM--{\it Newton} observations. \citet[][]{Goyal17} obtained the optical intranight PSD slopes in the range 1.5--4.1 for five monitoring sessions of the BL Lac PKS\,0735+178. \citet{Wehrle13} and \citet{Wehrle19} obtained the PSD slopes of blazar sources using the {\it Kepler}--satellite in the range 1.2--3.8 on timescales ranging from half a day to few months. Recently, \citet{Raiteri20} obtained the PSD slope $\sim$2.0 using the Transiting Exoplanet Survey Satellite ({\it TESS})--2\,min integration time light curve for the blazar 0716+714 at variability timescales between a month and few minutes. 

In this respect, a few optical observatories with 1--2\,m class optical telescopes and fitted with CCDs have been devoted to blazar/AGN monitoring programs since 1990 \citep[see, for a review,][]{Gopal-Krishna18}. The goal of the present study is to characterize the intranight variability of a large sample of blazars using the ARIES monitoring program which was carried out between 1998 to 2010, the results of which are presented in \citet[][]{Goyal13b}. The paper is organized as follows. Sample selection is given in Section~\ref{sec:sample} while Section~\ref{sec:analysis} provides the details on the analysis method, in particular, the derivation of power spectral densities and the estimation of best-fit PSD shapes using extensive numerical simulations of light curves. Section~\ref{sec:results} provides the main results while a discussion and conclusions are given in Section~\ref{sec:discussions}.  

\begin{figure*}[ht!]
\hbox{
\includegraphics[width=0.30\textwidth]{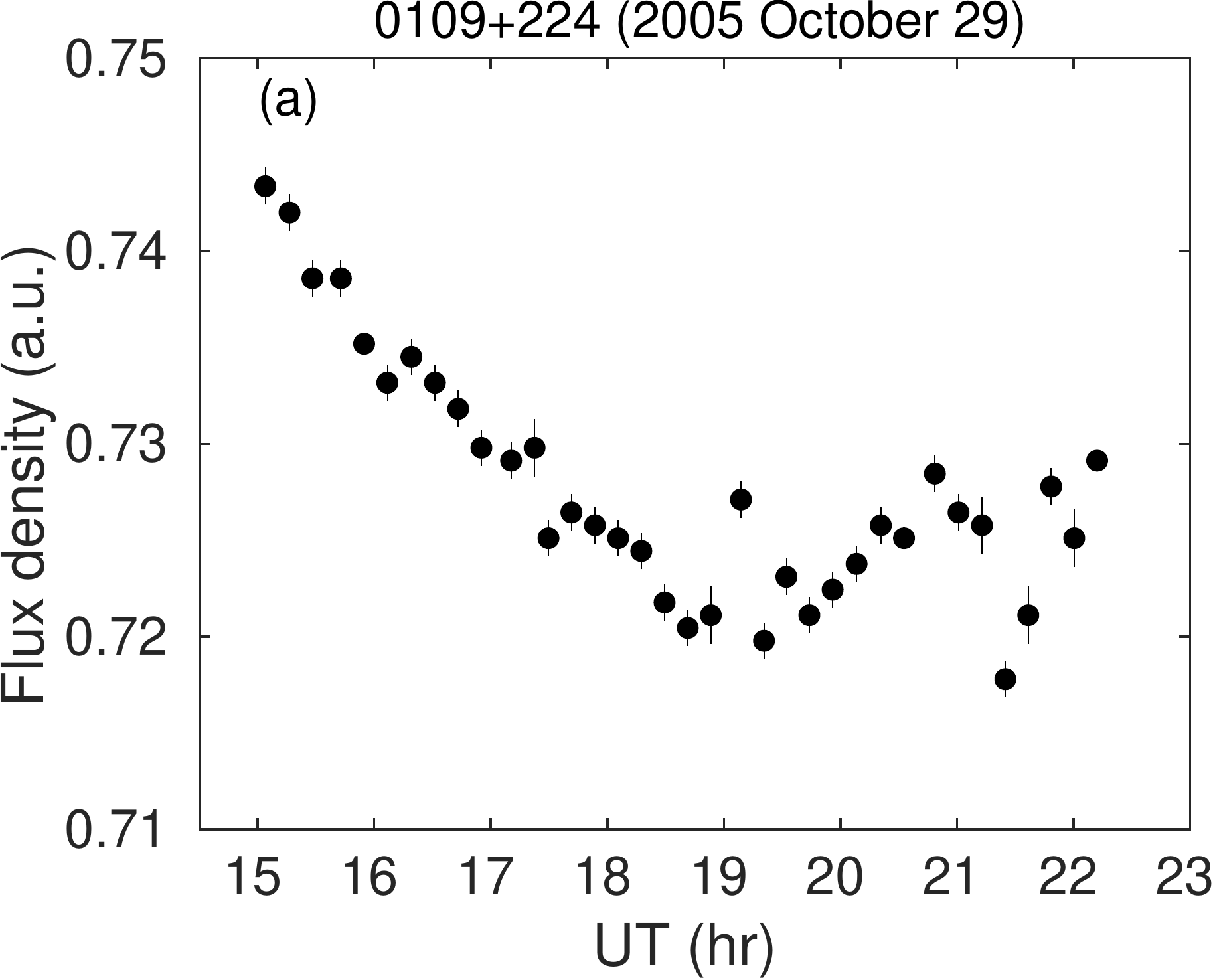}
\includegraphics[width=0.30\textwidth]{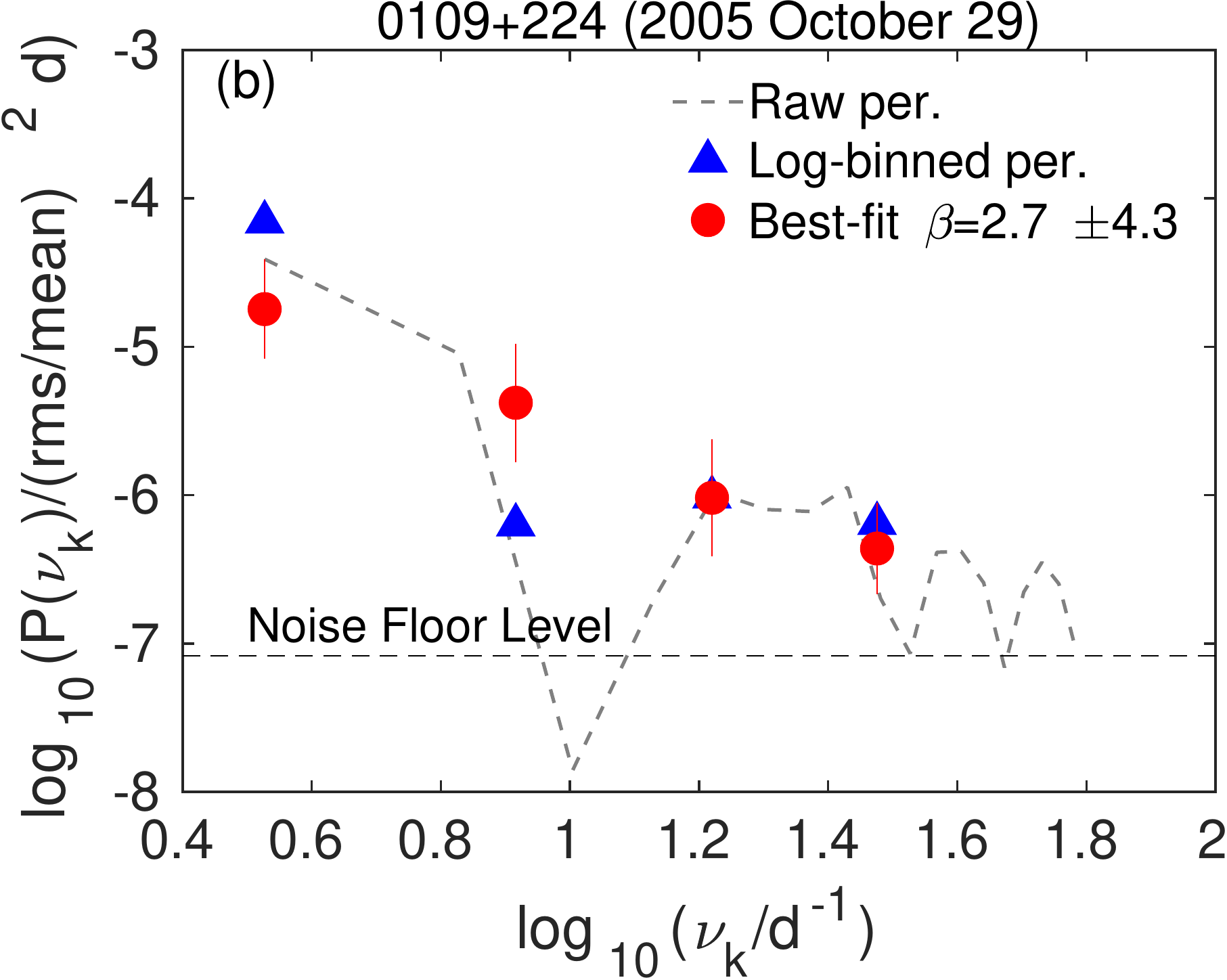}
\includegraphics[width=0.30\textwidth]{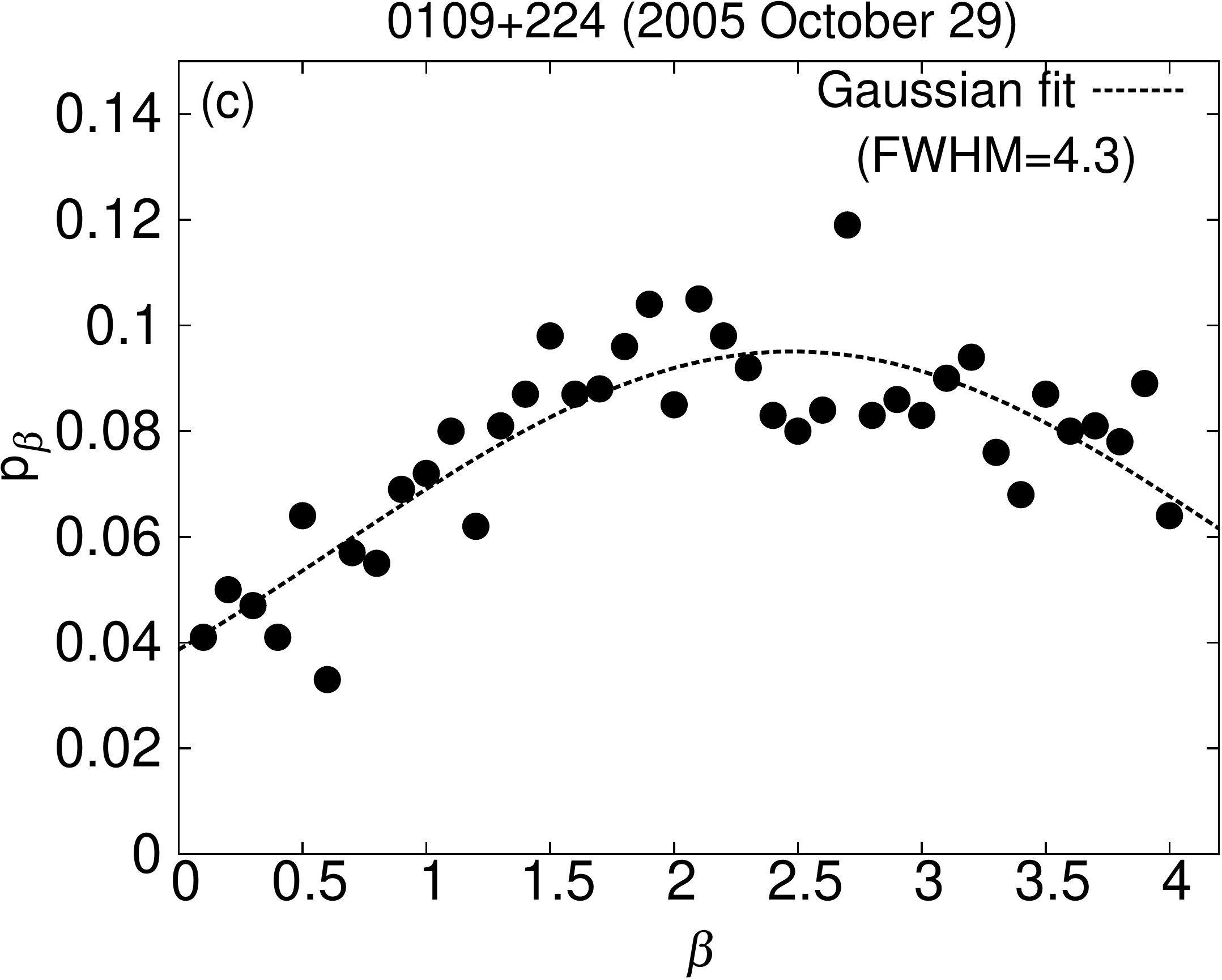}
}
\hbox{
\includegraphics[width=0.30\textwidth]{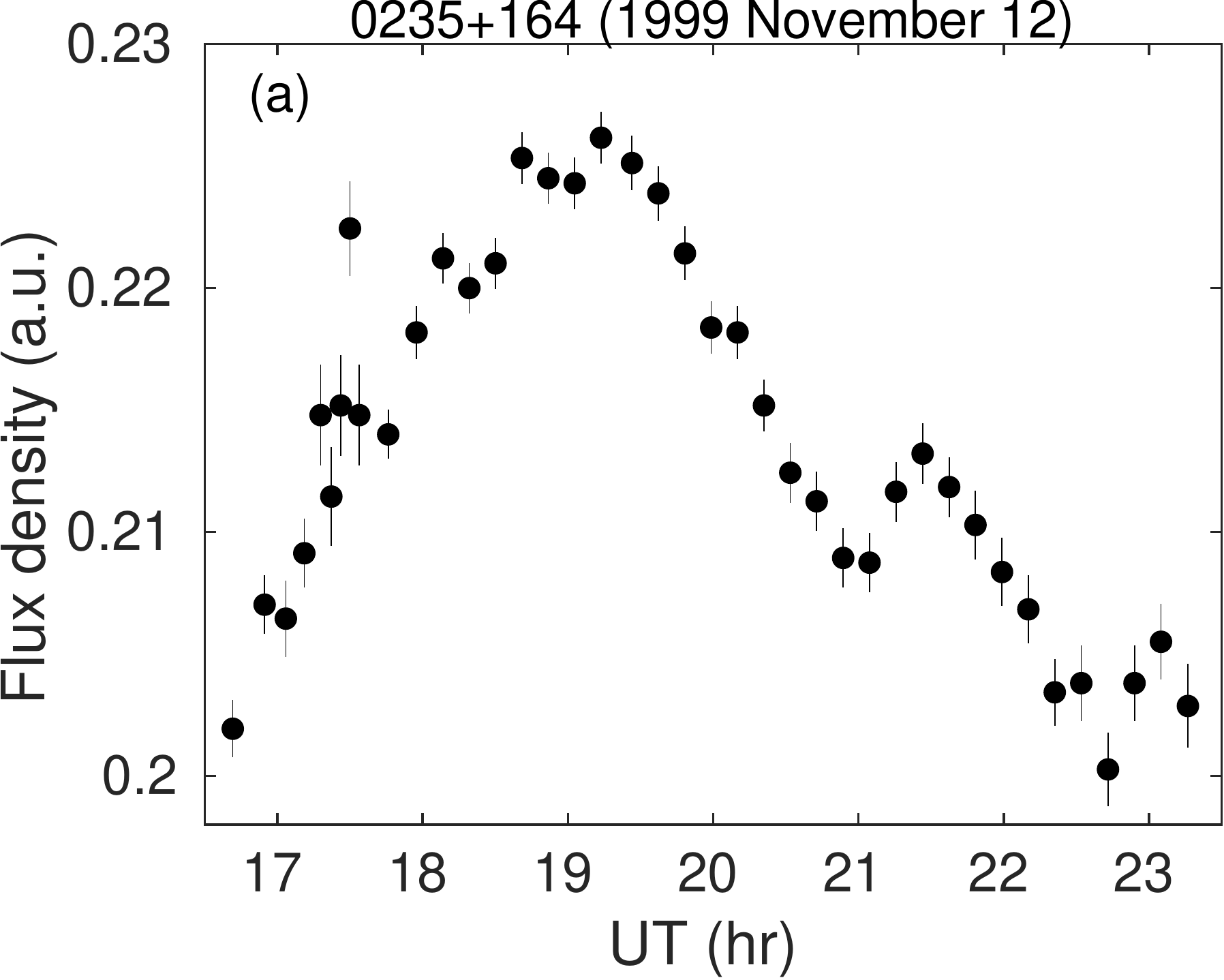}
\includegraphics[width=0.30\textwidth]{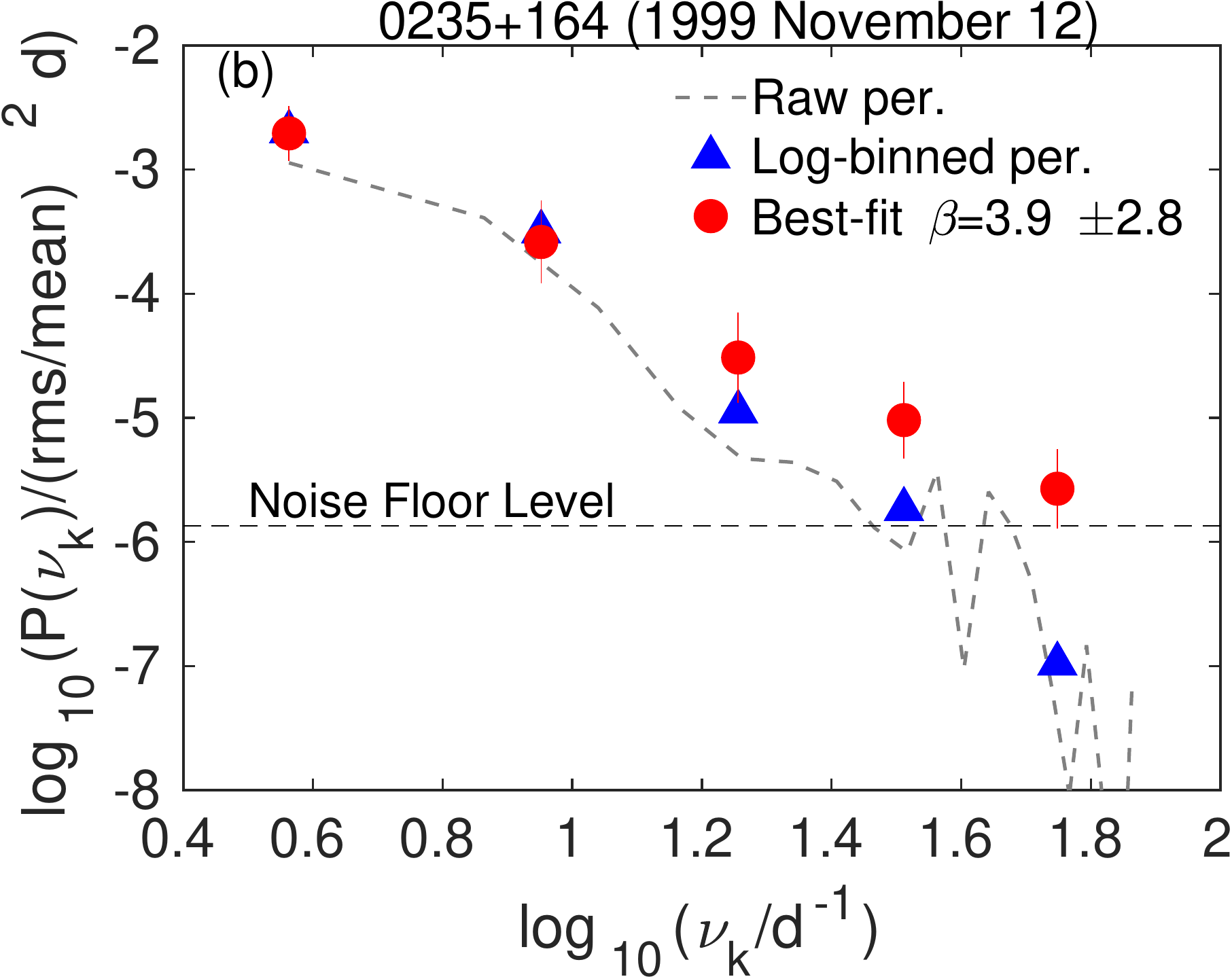}
\includegraphics[width=0.30\textwidth]{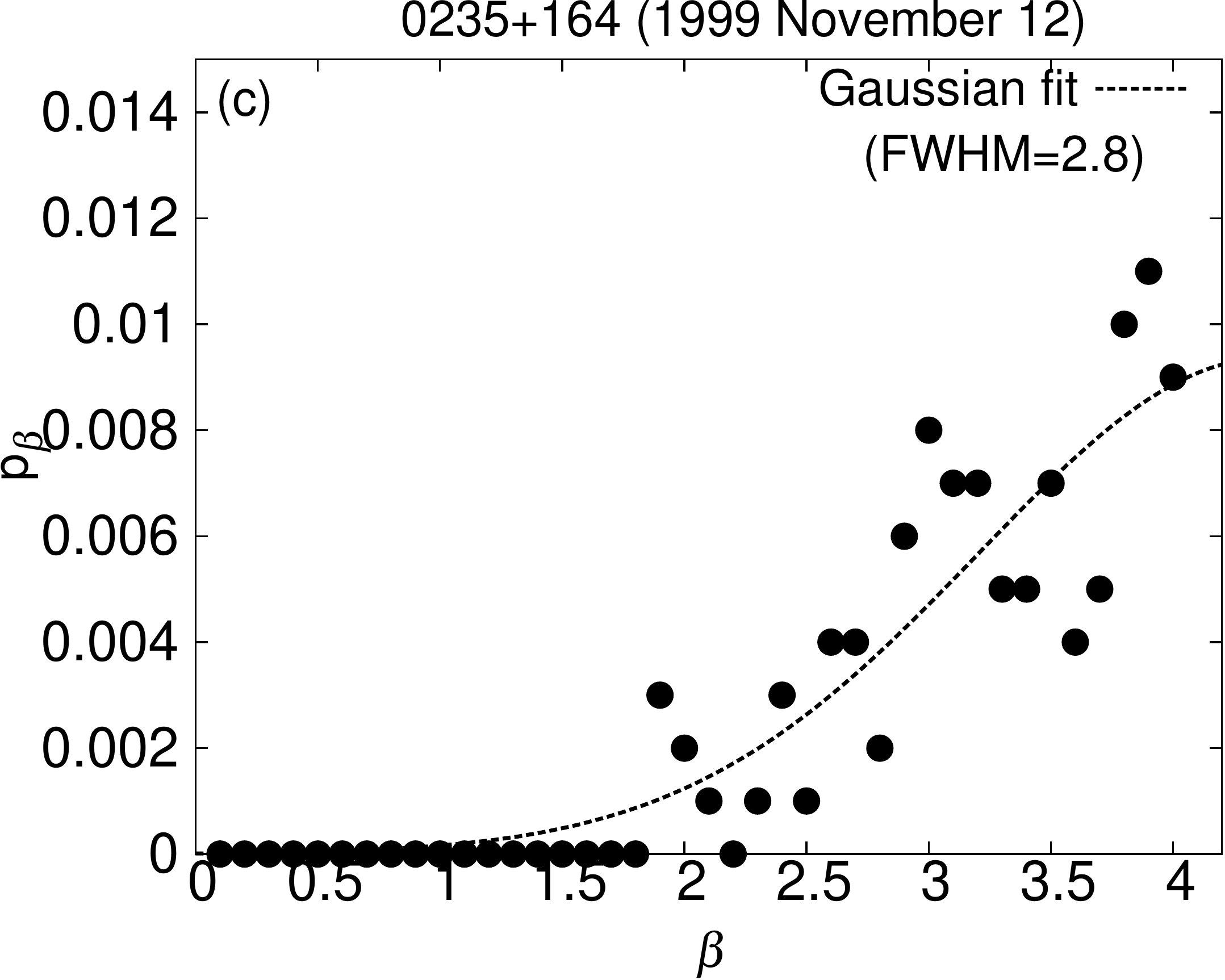}
}
\hbox{
\includegraphics[width=0.30\textwidth]{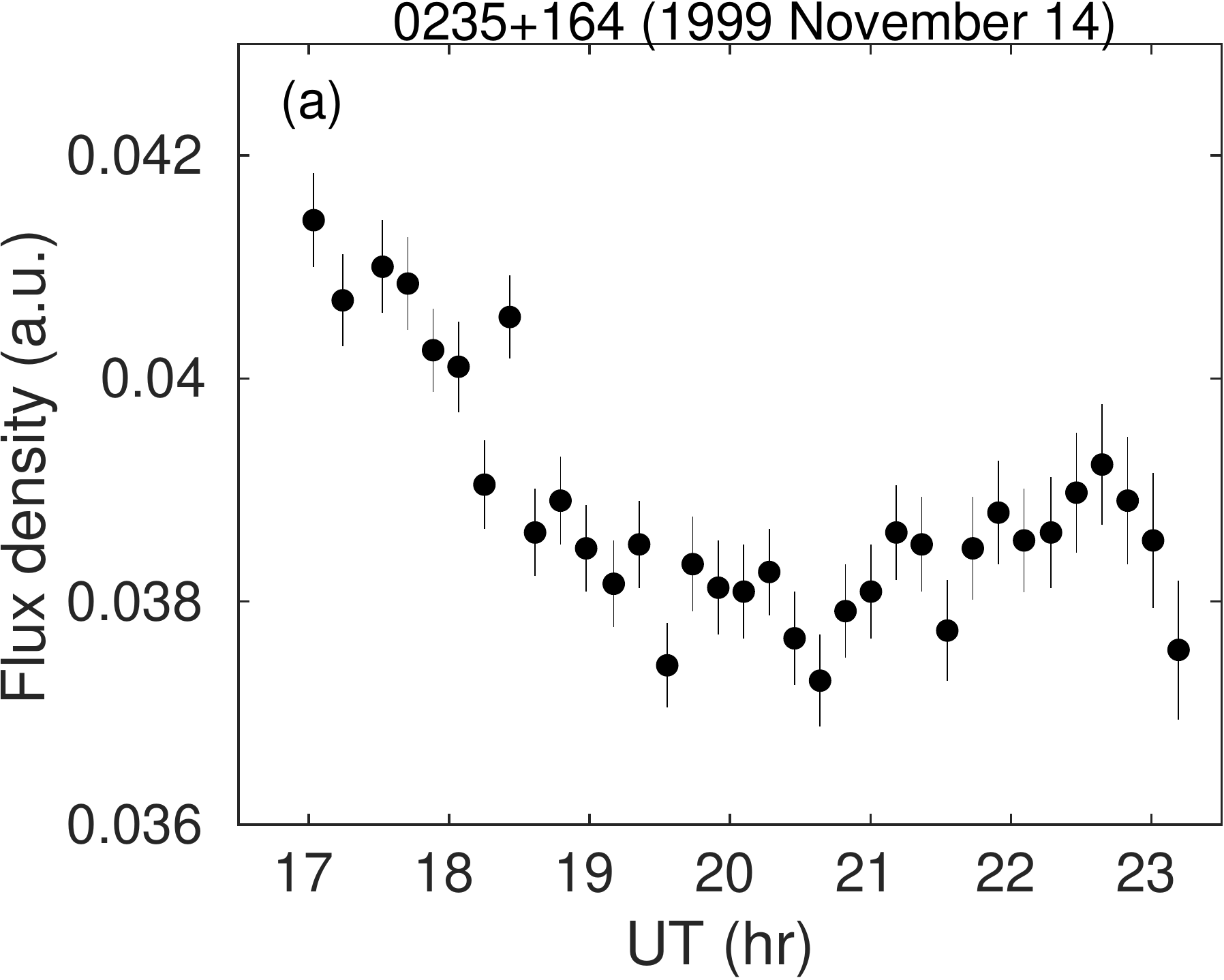}
\includegraphics[width=0.30\textwidth]{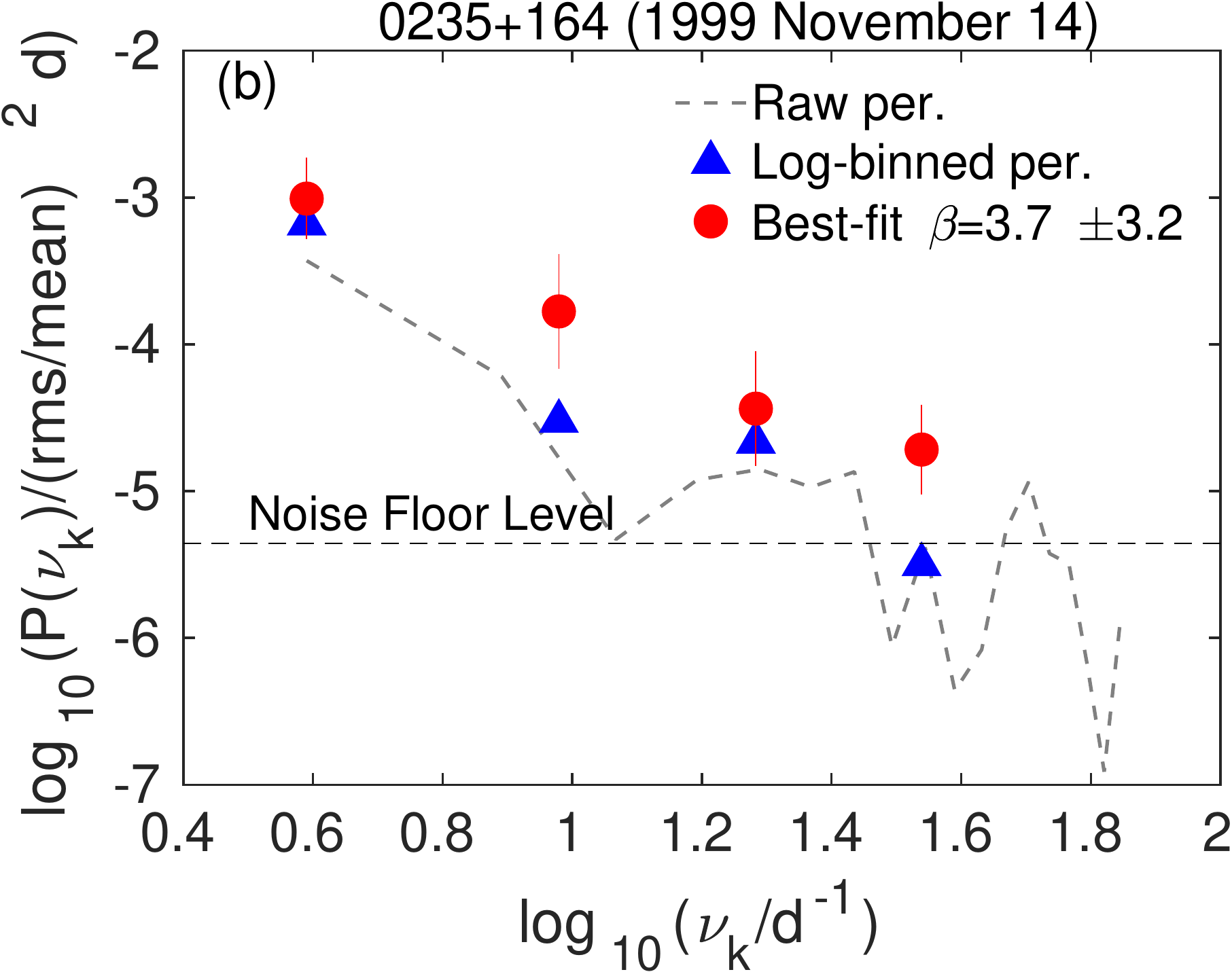}
\includegraphics[width=0.30\textwidth]{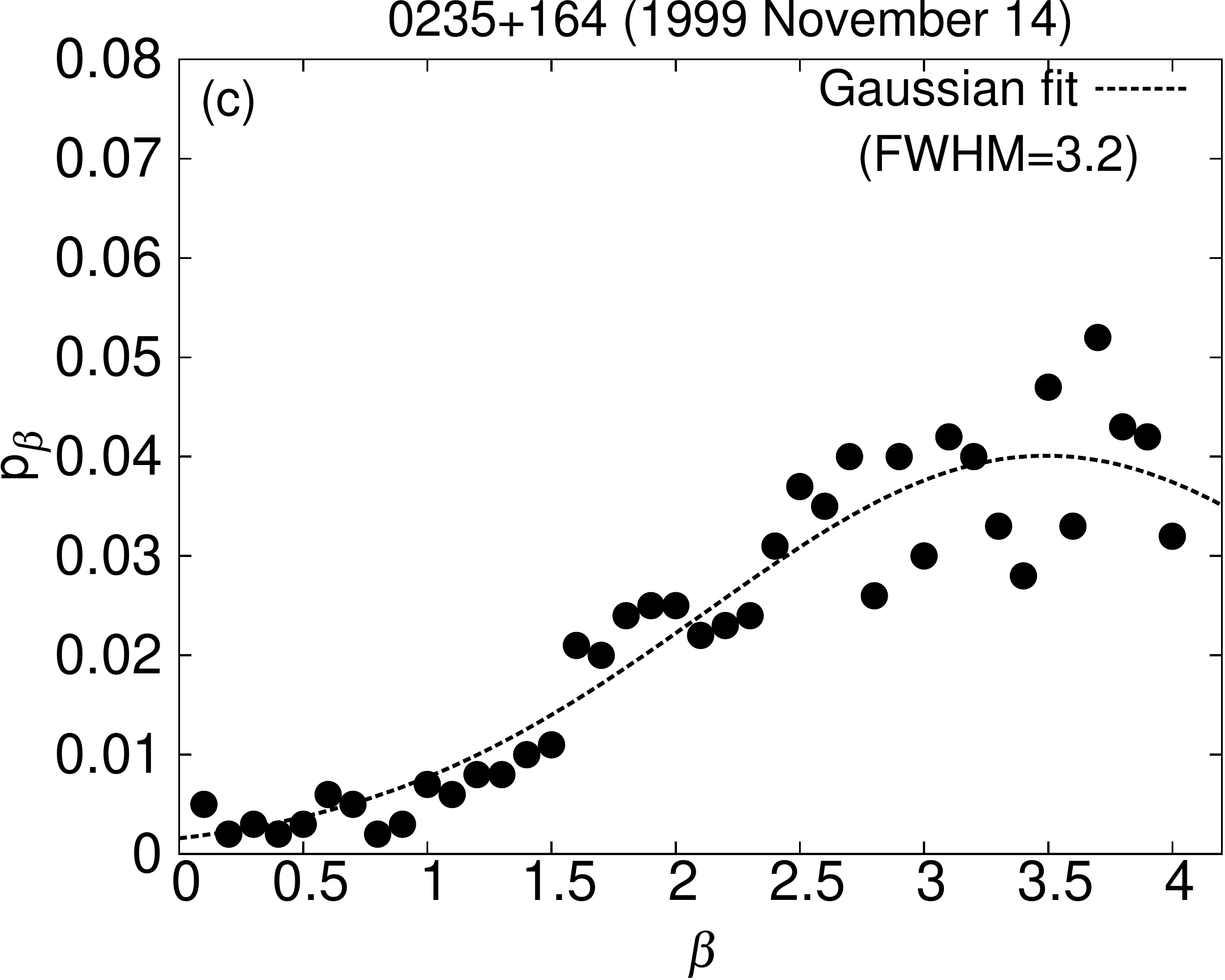}
}
\hbox{
\includegraphics[width=0.30\textwidth]{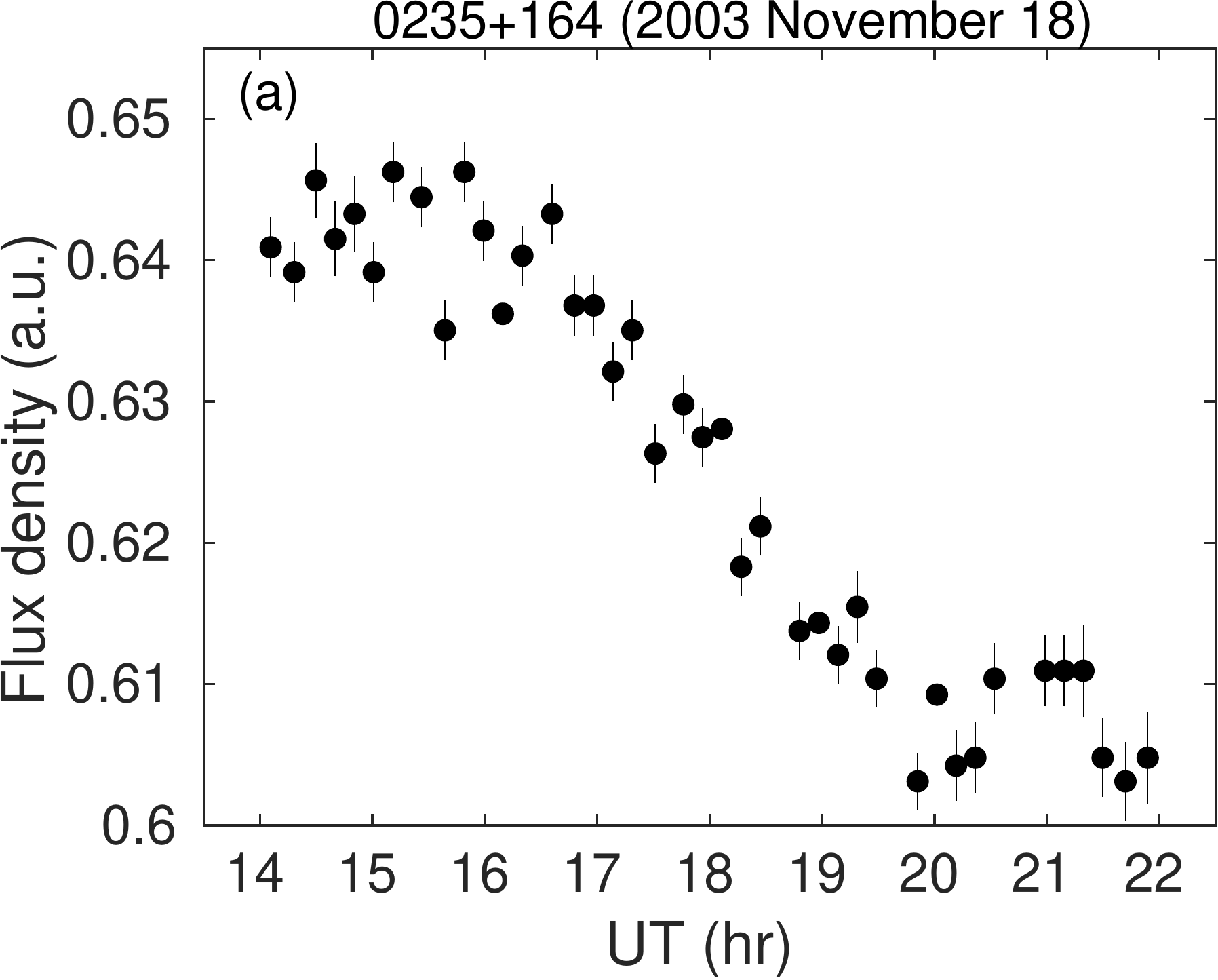}
\includegraphics[width=0.30\textwidth]{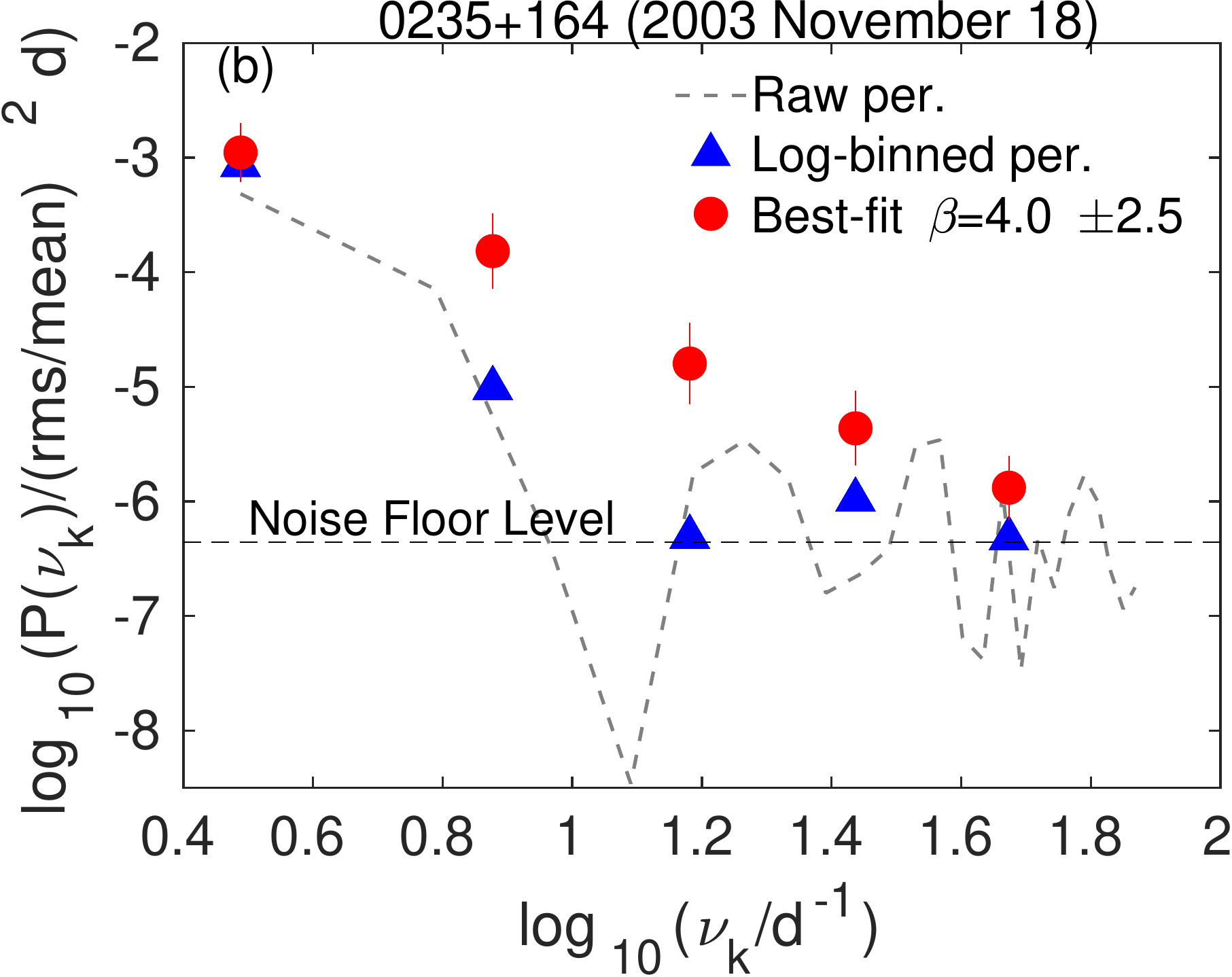}
\includegraphics[width=0.30\textwidth]{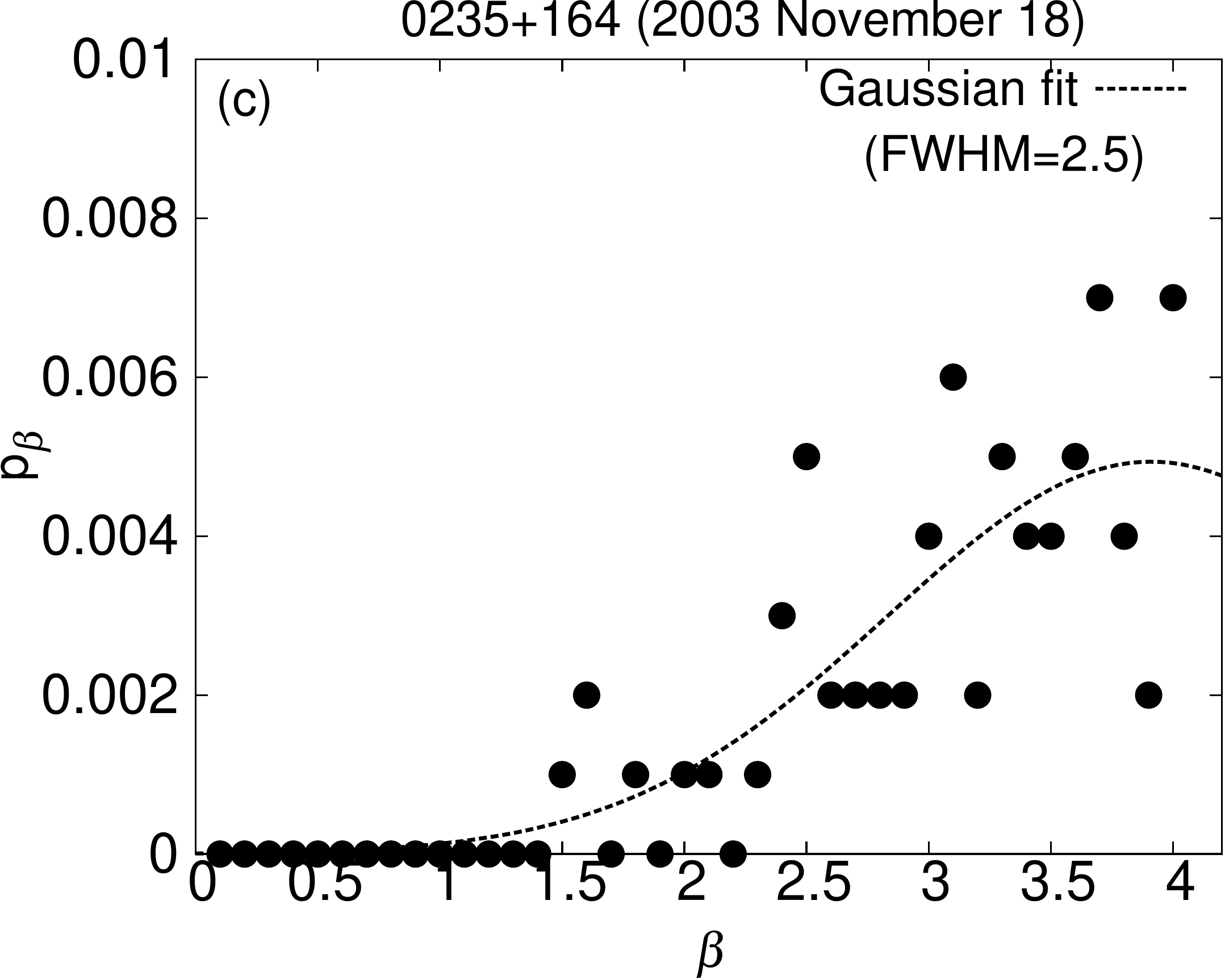}
}
\hbox{
\includegraphics[width=0.30\textwidth]{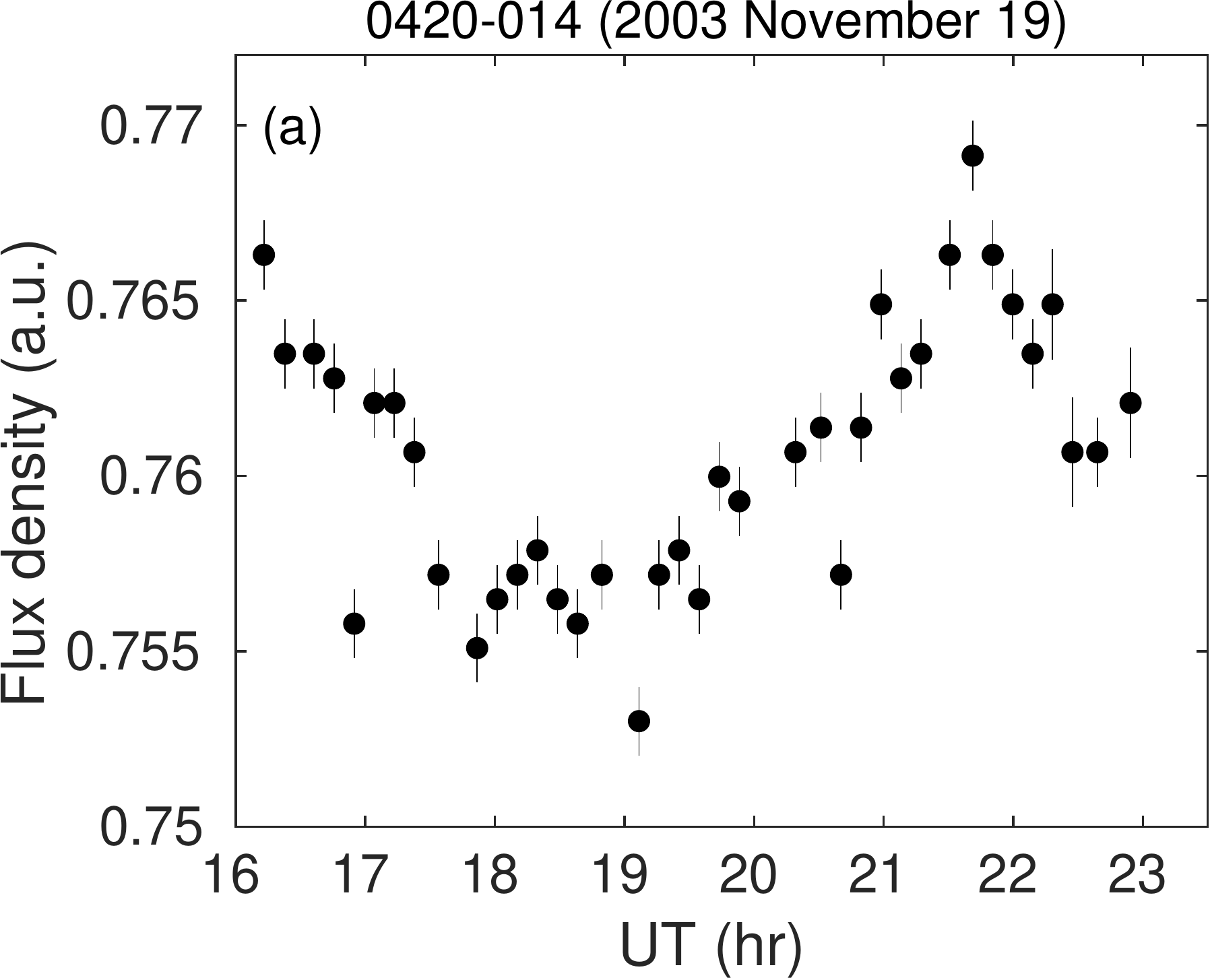}
\includegraphics[width=0.30\textwidth]{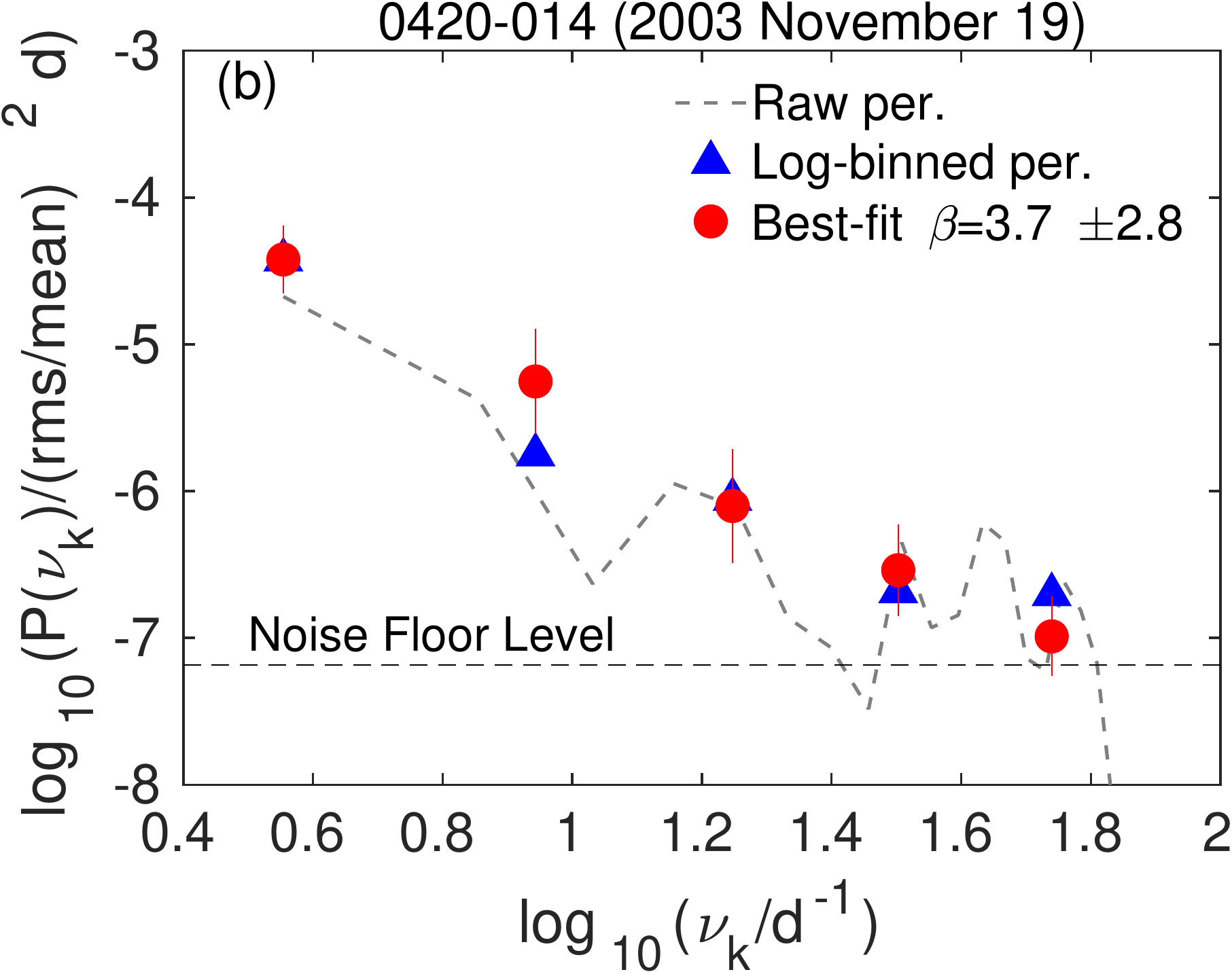}
\includegraphics[width=0.30\textwidth]{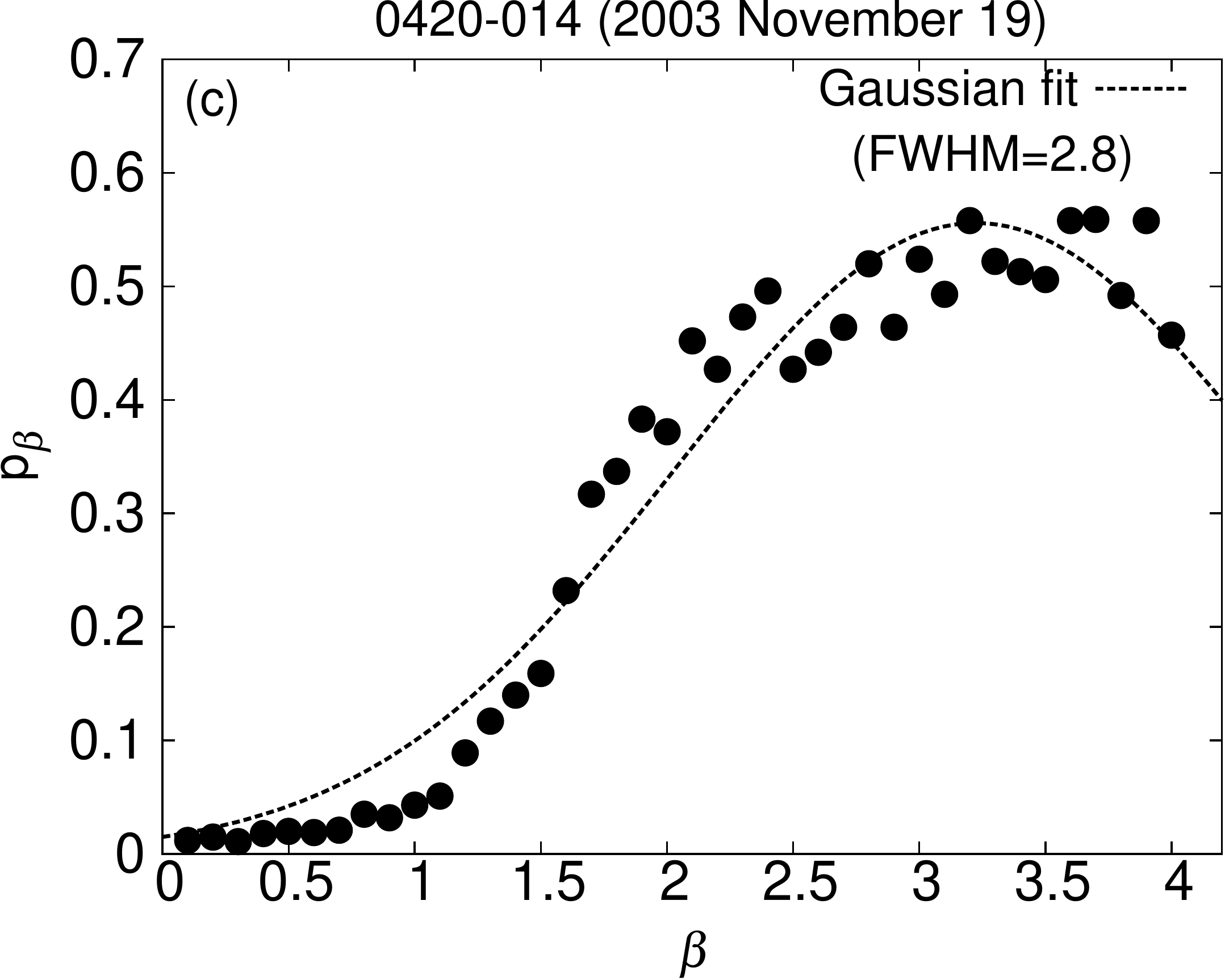}
}

\begin{minipage}{\textwidth}
\caption{
The intra-night variability power spectrum of blazar sources obtained in this study. Panel (a) presents the light curve on a linear scale (see text). Panel (b) presents the derived power spectrum down to the  Nyquist sampling frequency of the (mean) observed data. The dashed line shows the `raw' periodogram while the blue triangles and red circles give `logarithmically binned power spectrum' and the best-fit power spectrum, respectively. The error on the best-fit PSD slope corresponds to a 98\% confidence limit. The dashed horizontal line corresponds to the statistical noise floor level due to measurement noise. Panel (c) shows the probability curve as a function of the input power spectrum slope. The source name and date of monitoring are presented at the top of each panel.\label{fig:analysis}
}
\end{minipage}
\end{figure*}

\addtocounter{figure}{-1}
\begin{figure*}[ht!]
\hbox{
\includegraphics[width=0.30\textwidth]{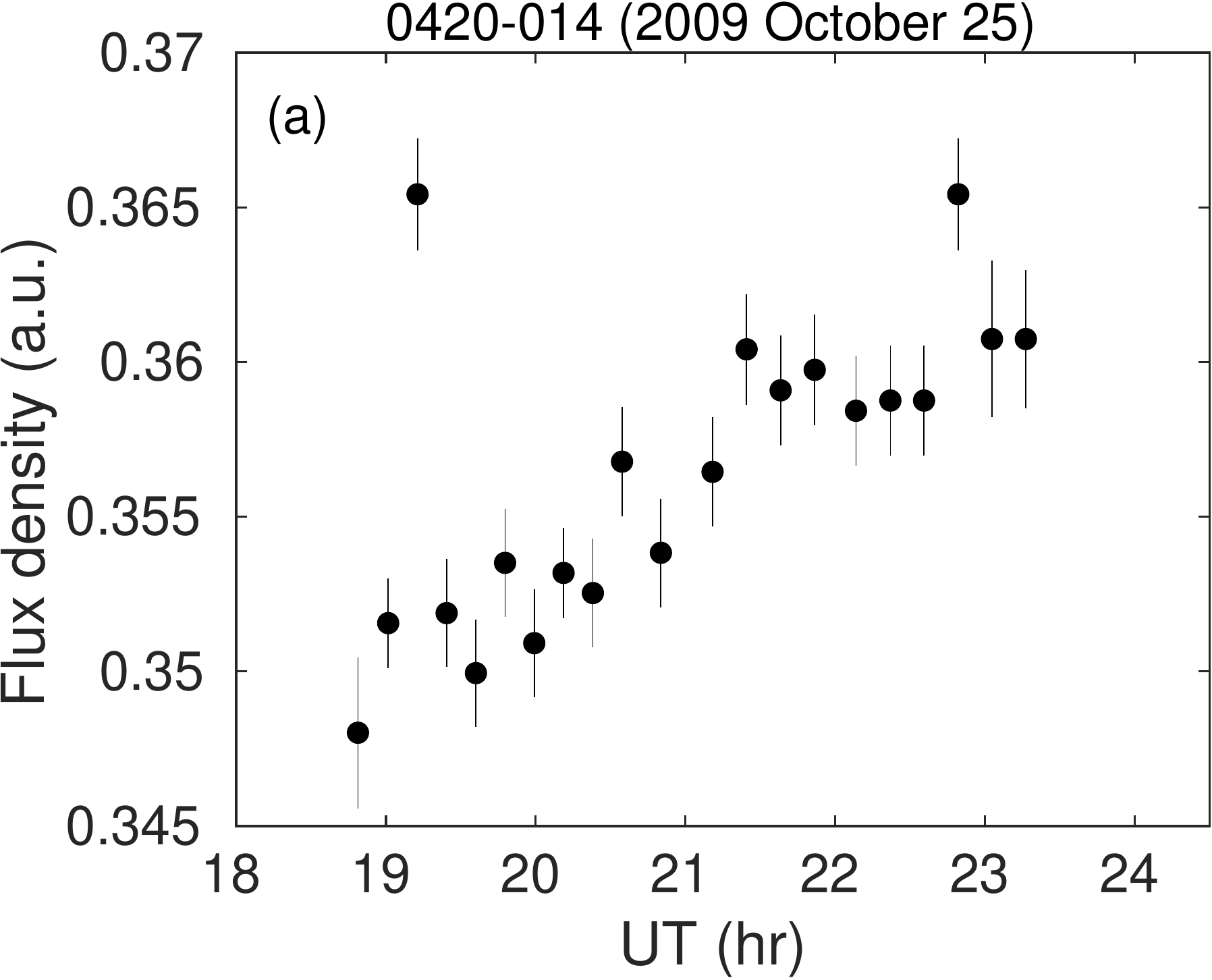}
\includegraphics[width=0.30\textwidth]{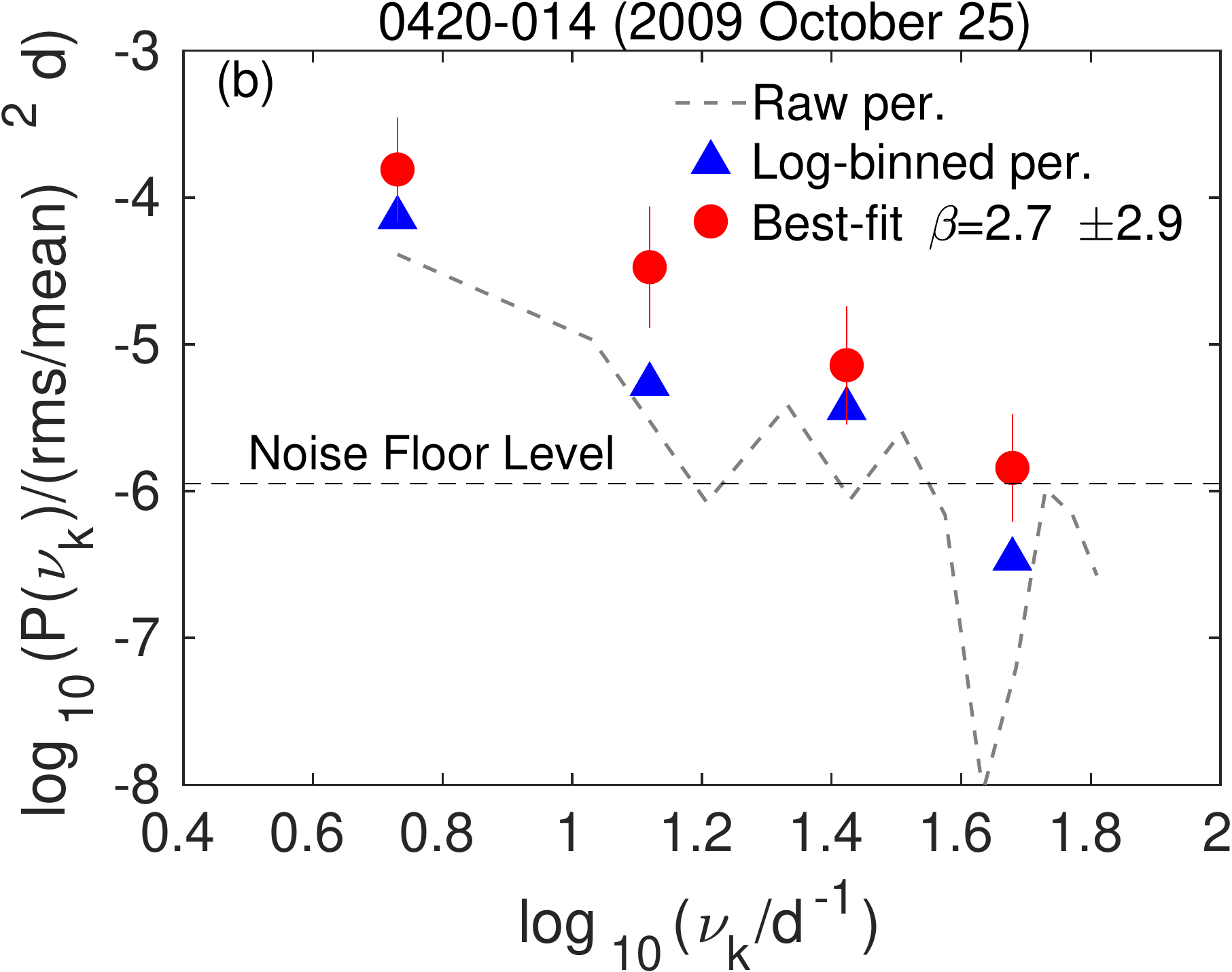}
\includegraphics[width=0.30\textwidth]{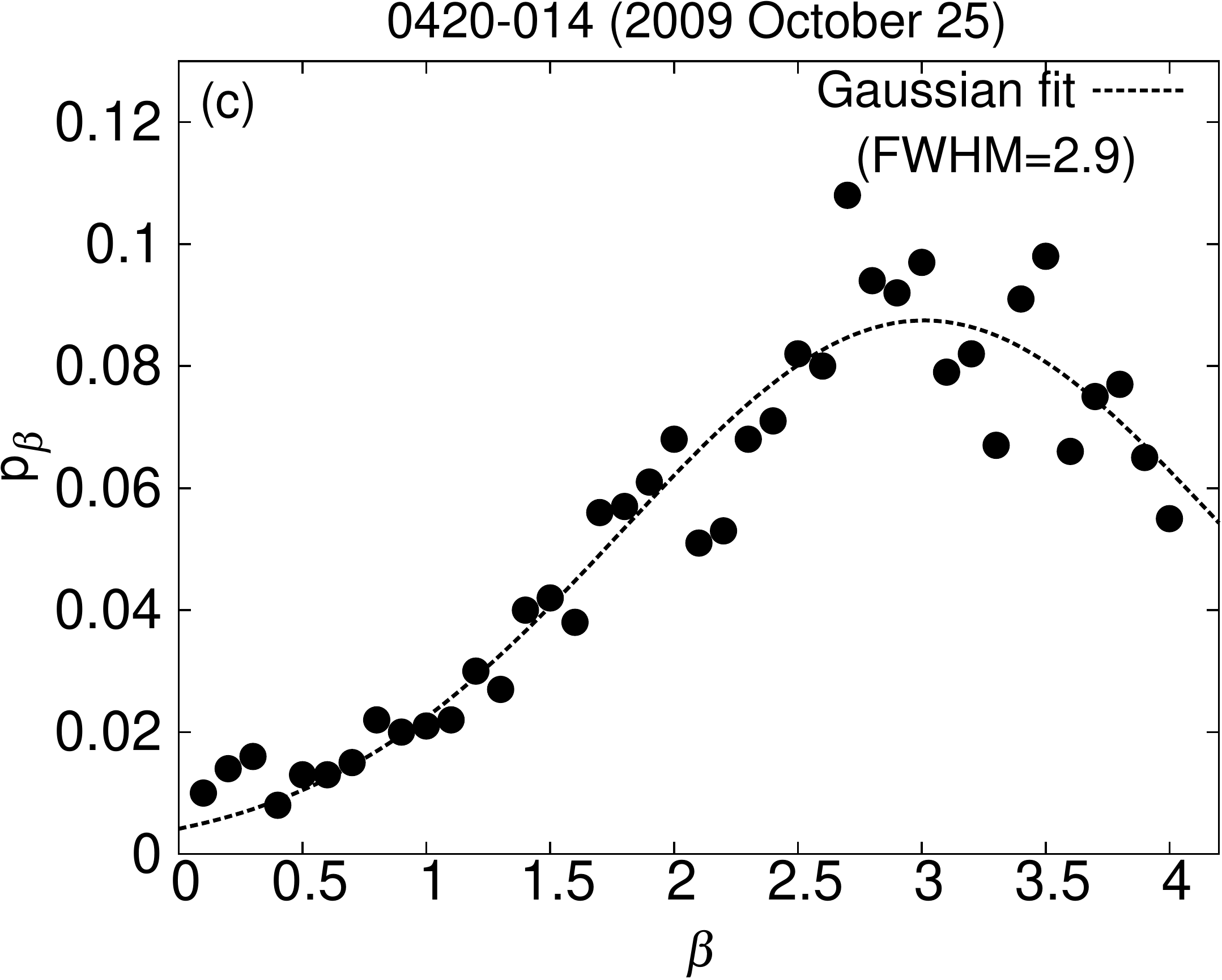}
}
\hbox{
\includegraphics[width=0.30\textwidth]{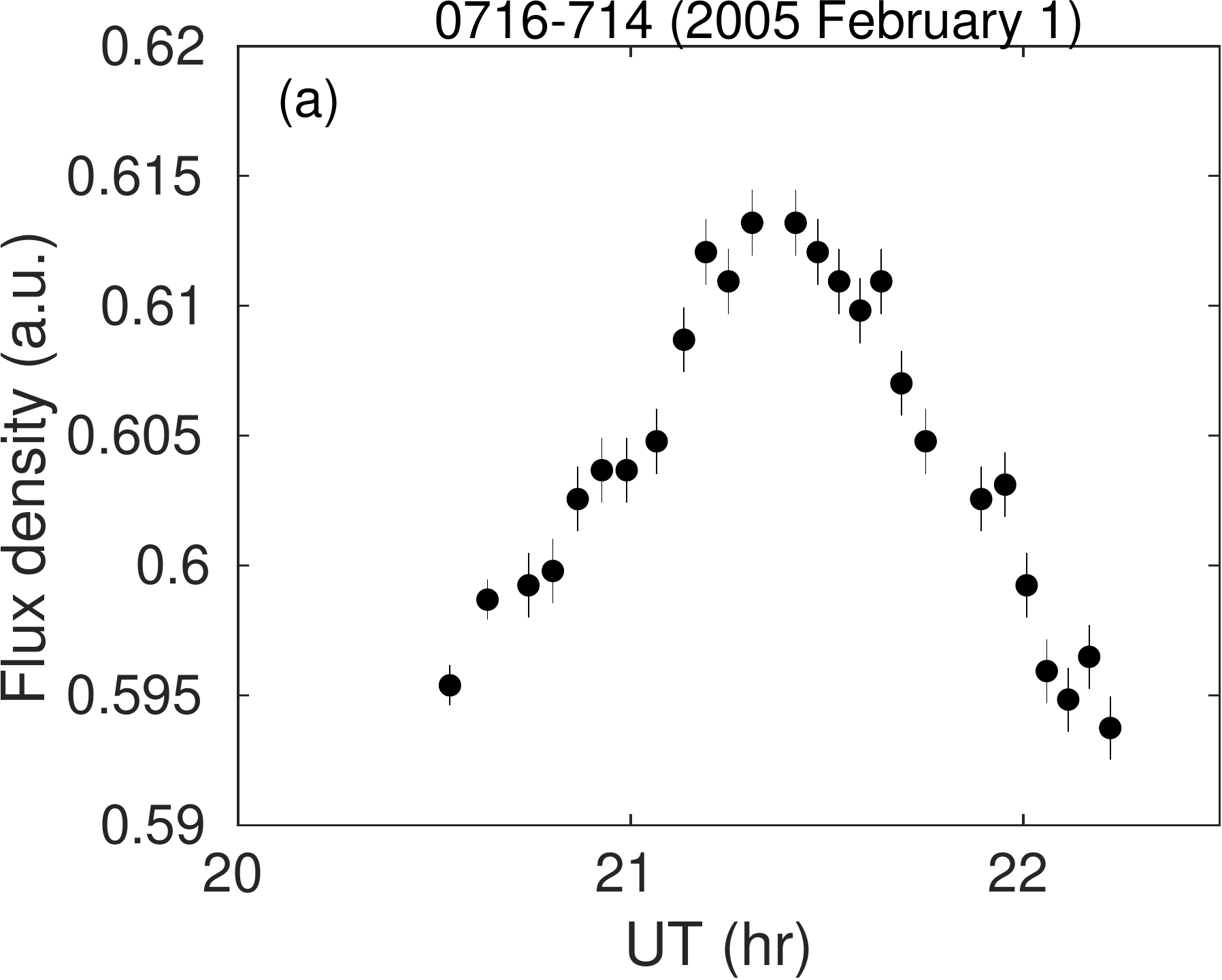}
\includegraphics[width=0.30\textwidth]{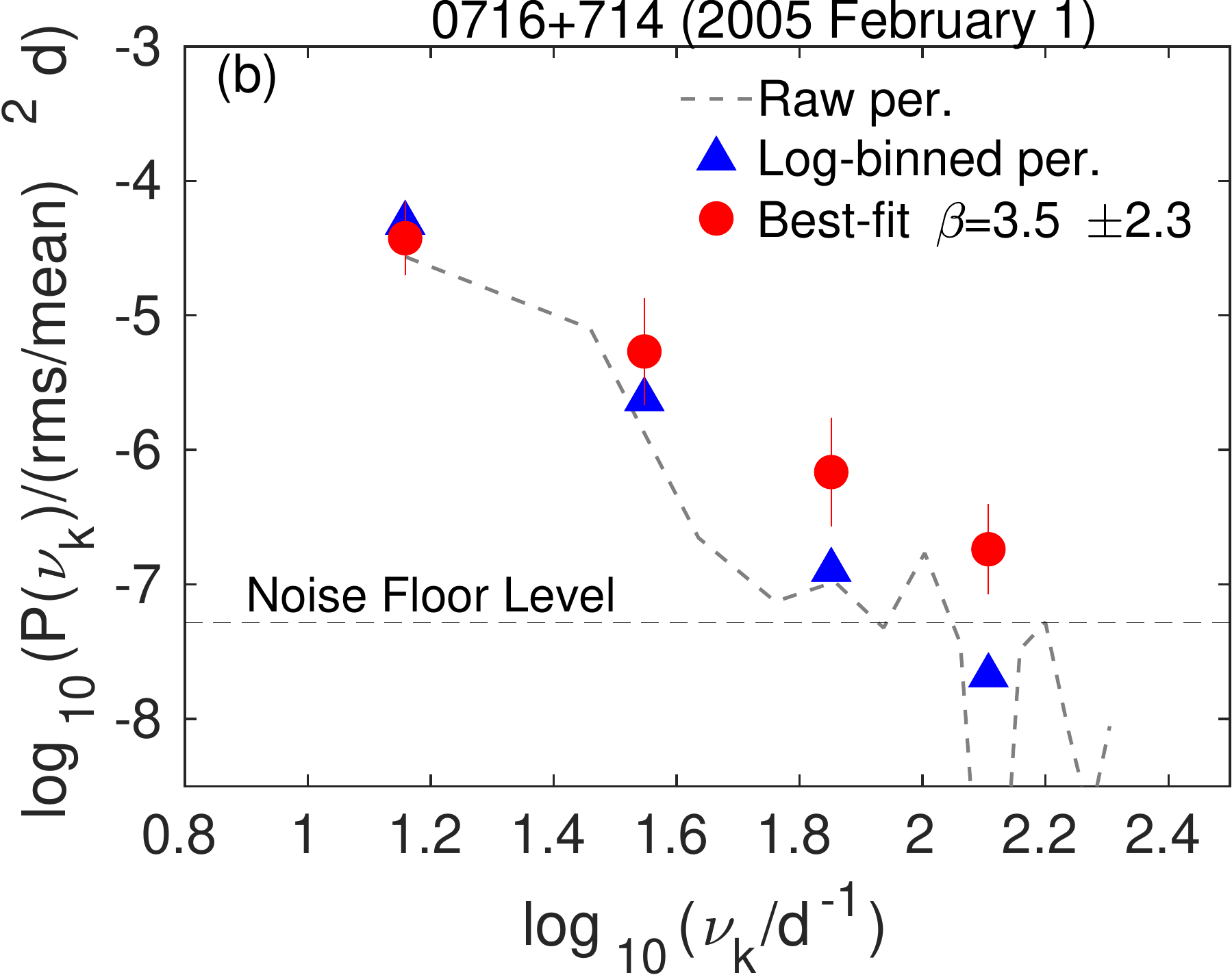}
\includegraphics[width=0.30\textwidth]{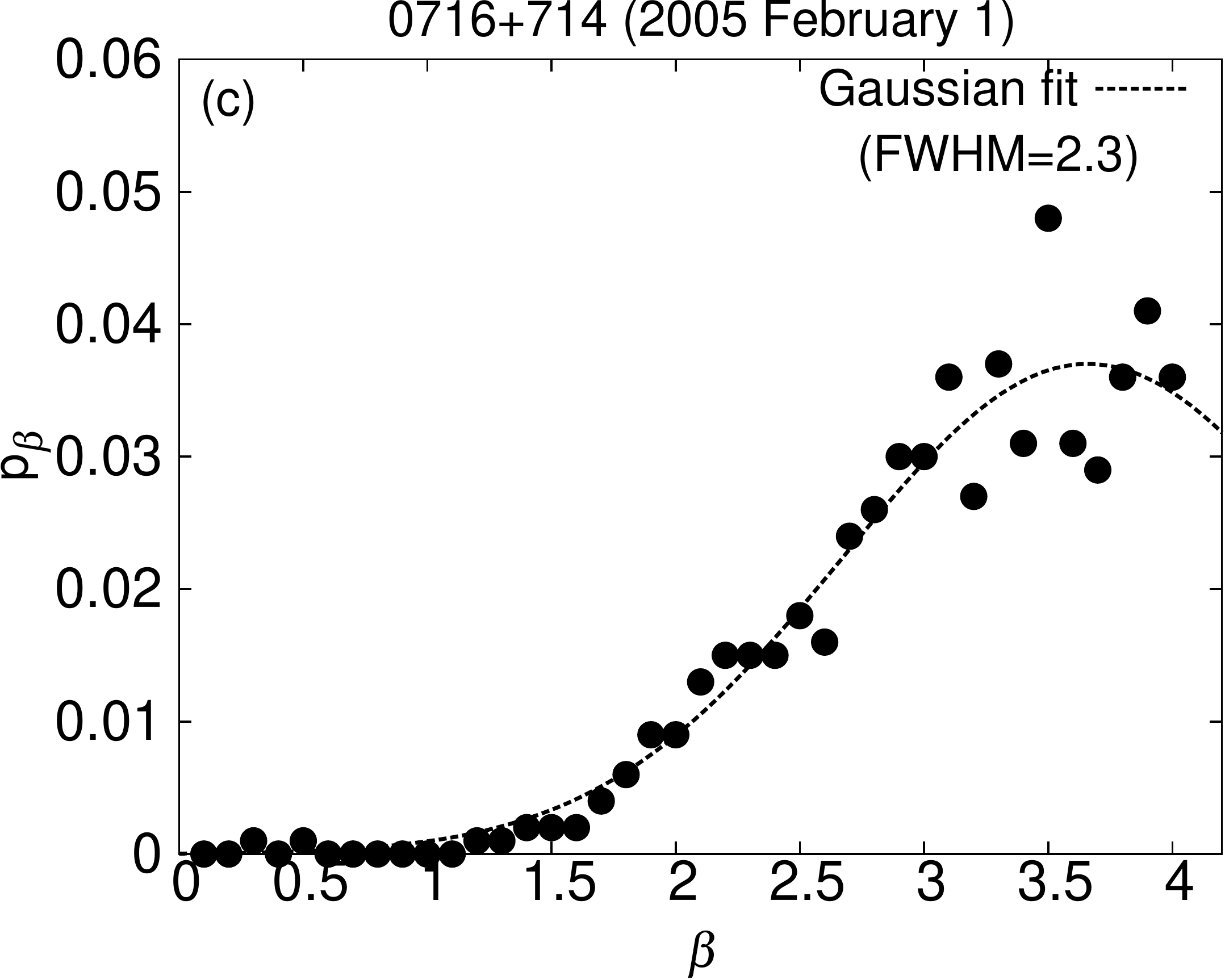}
}
\hbox{
\includegraphics[width=0.30\textwidth]{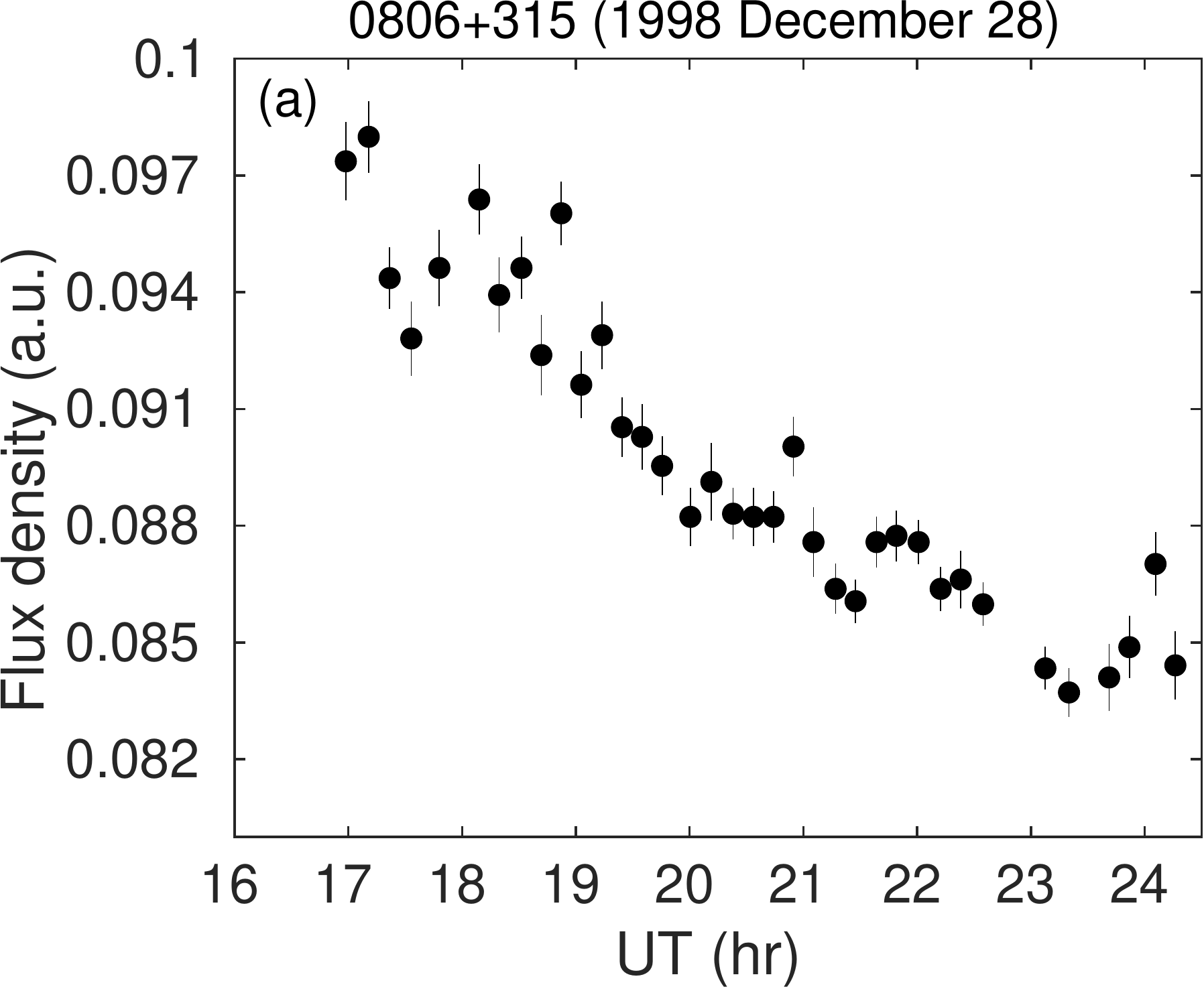}
\includegraphics[width=0.30\textwidth]{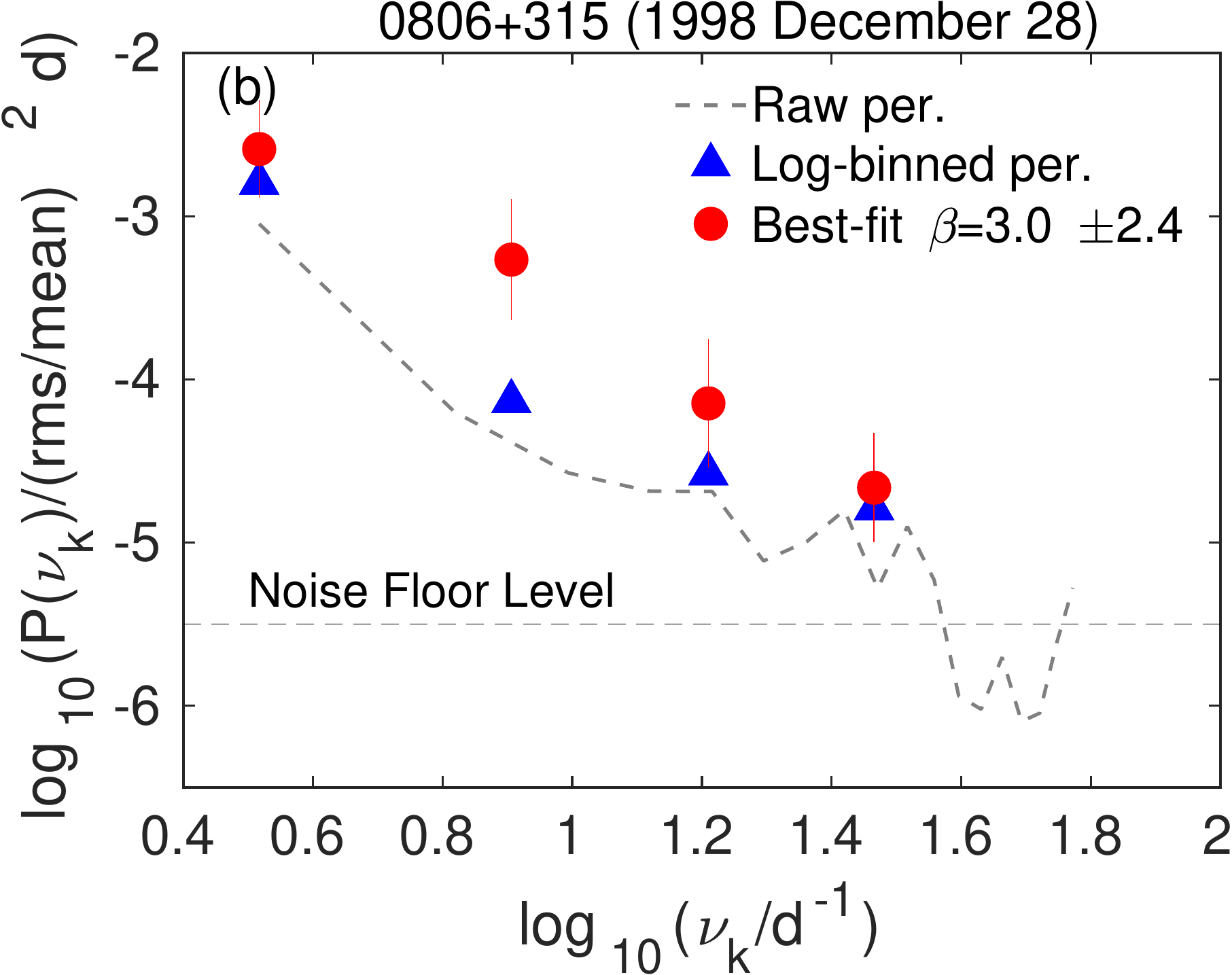}
\includegraphics[width=0.30\textwidth]{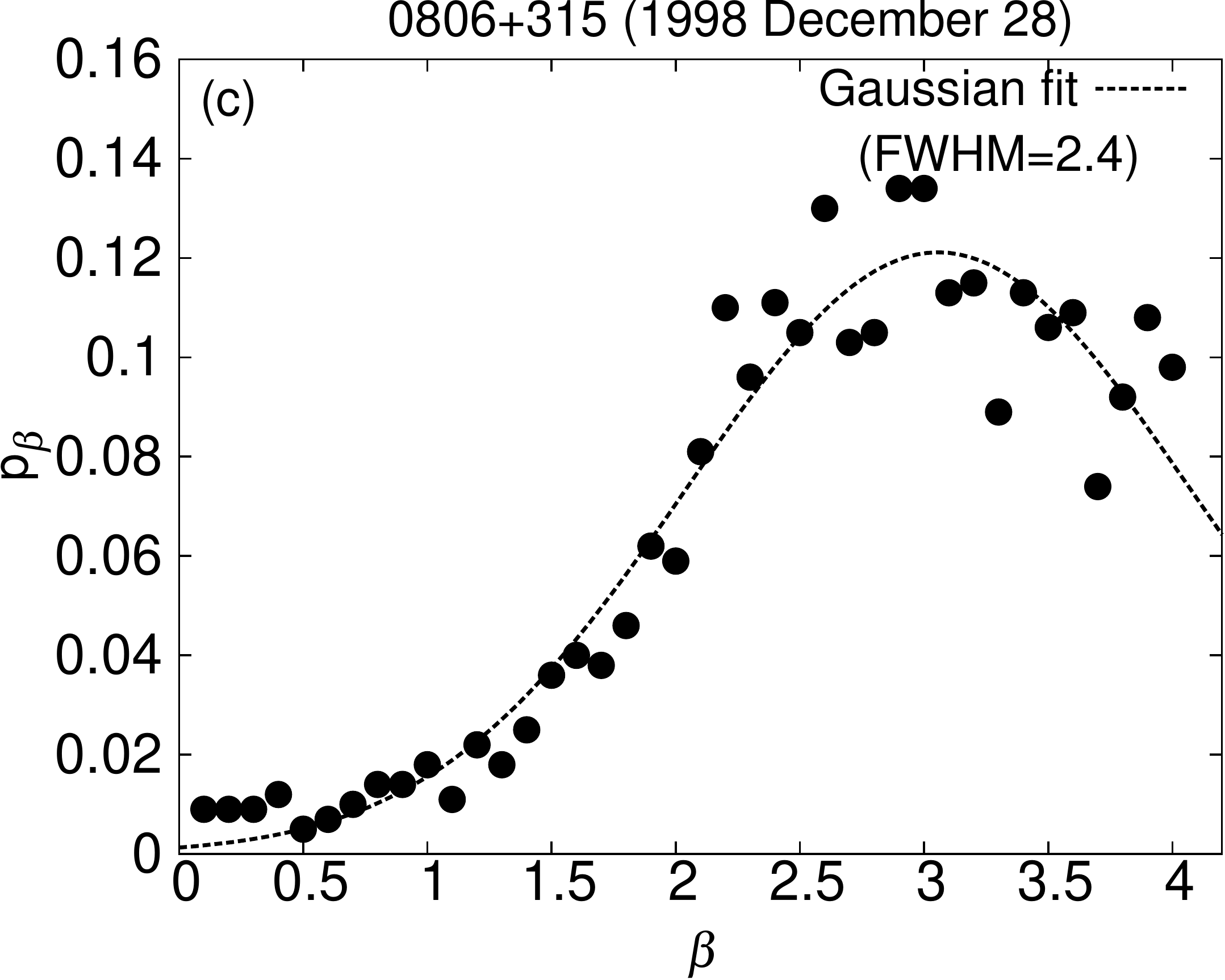}
}
\hbox{
\includegraphics[width=0.30\textwidth]{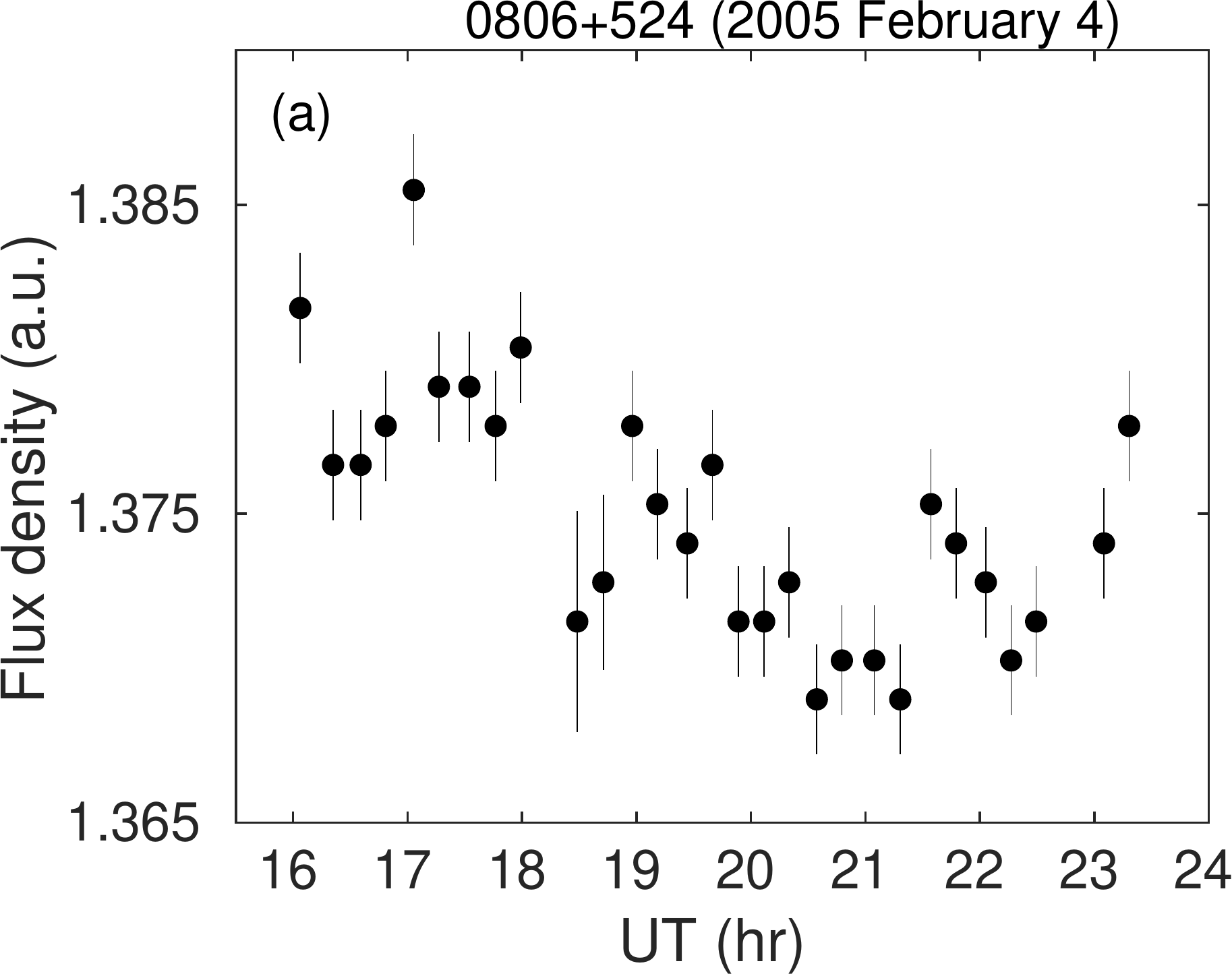}
\includegraphics[width=0.30\textwidth]{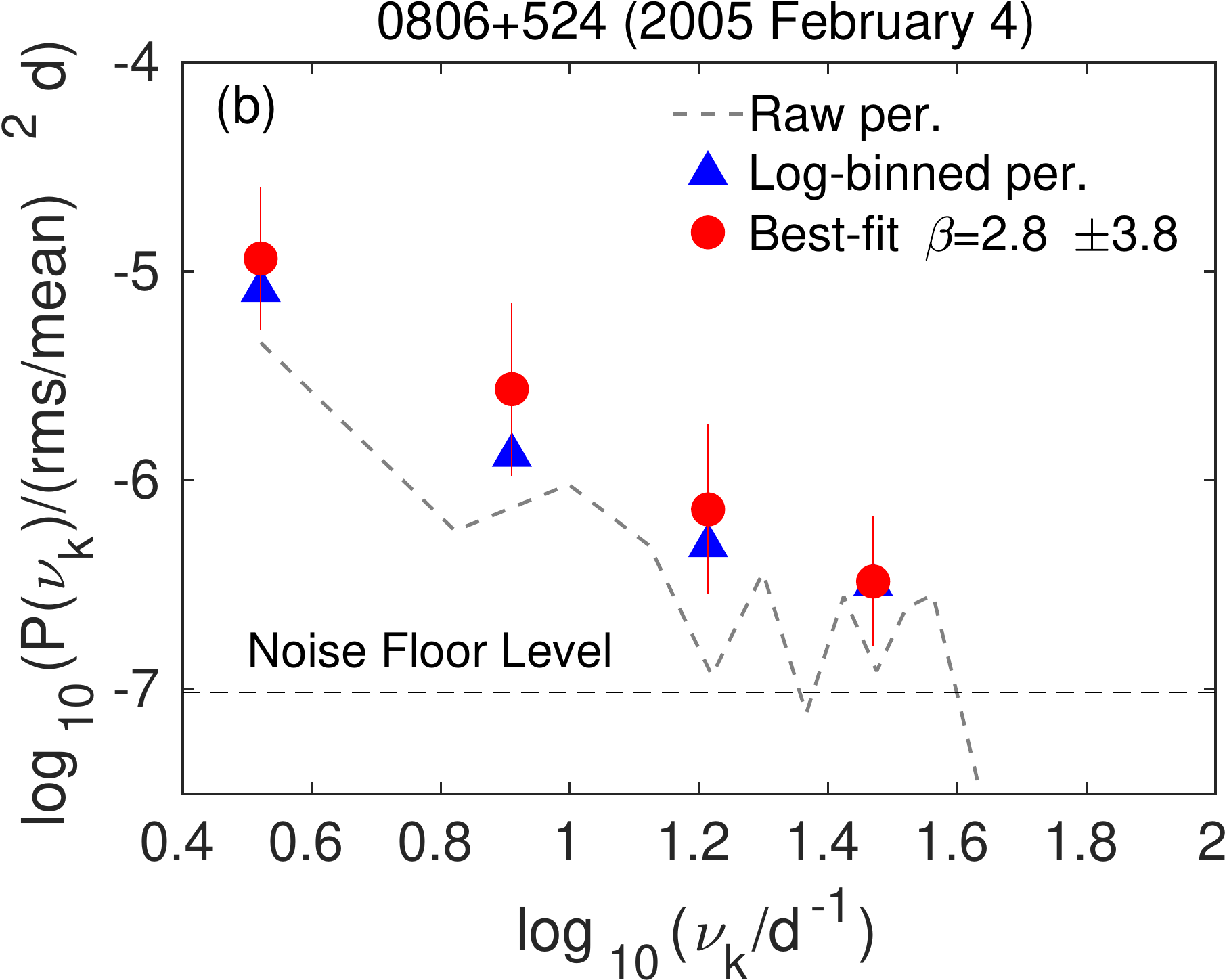}
\includegraphics[width=0.30\textwidth]{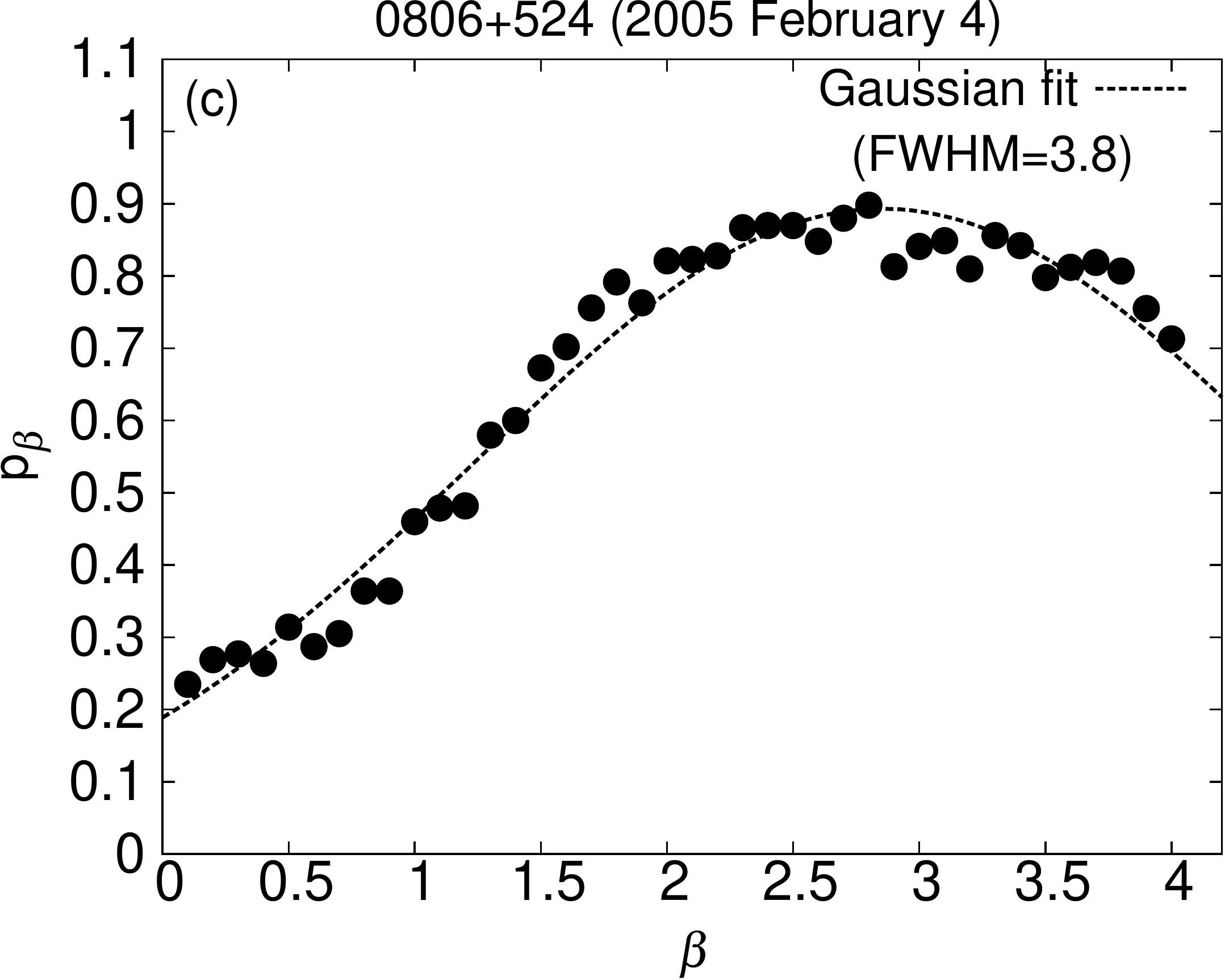}
}
\hbox{
\includegraphics[width=0.30\textwidth]{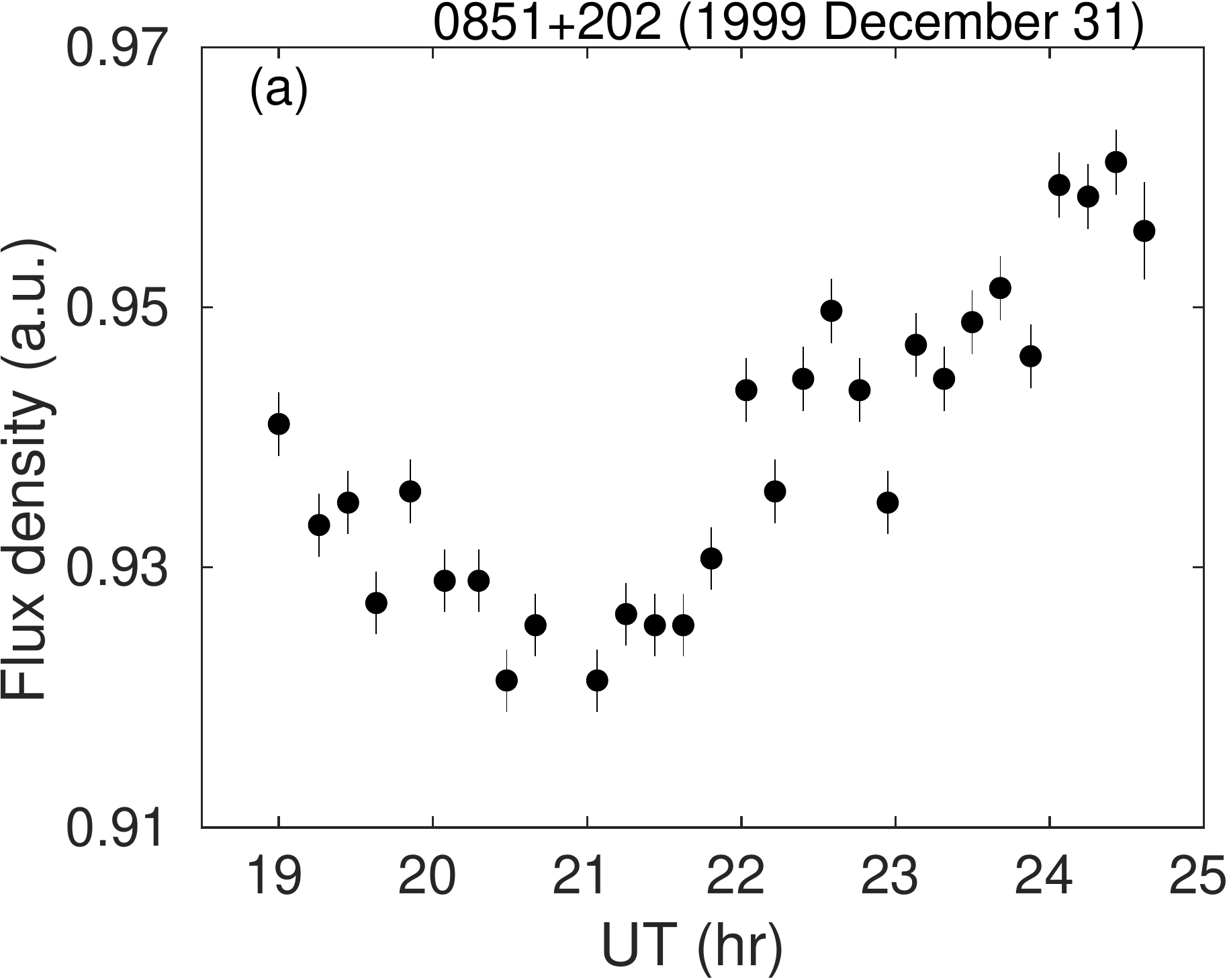}
\includegraphics[width=0.30\textwidth]{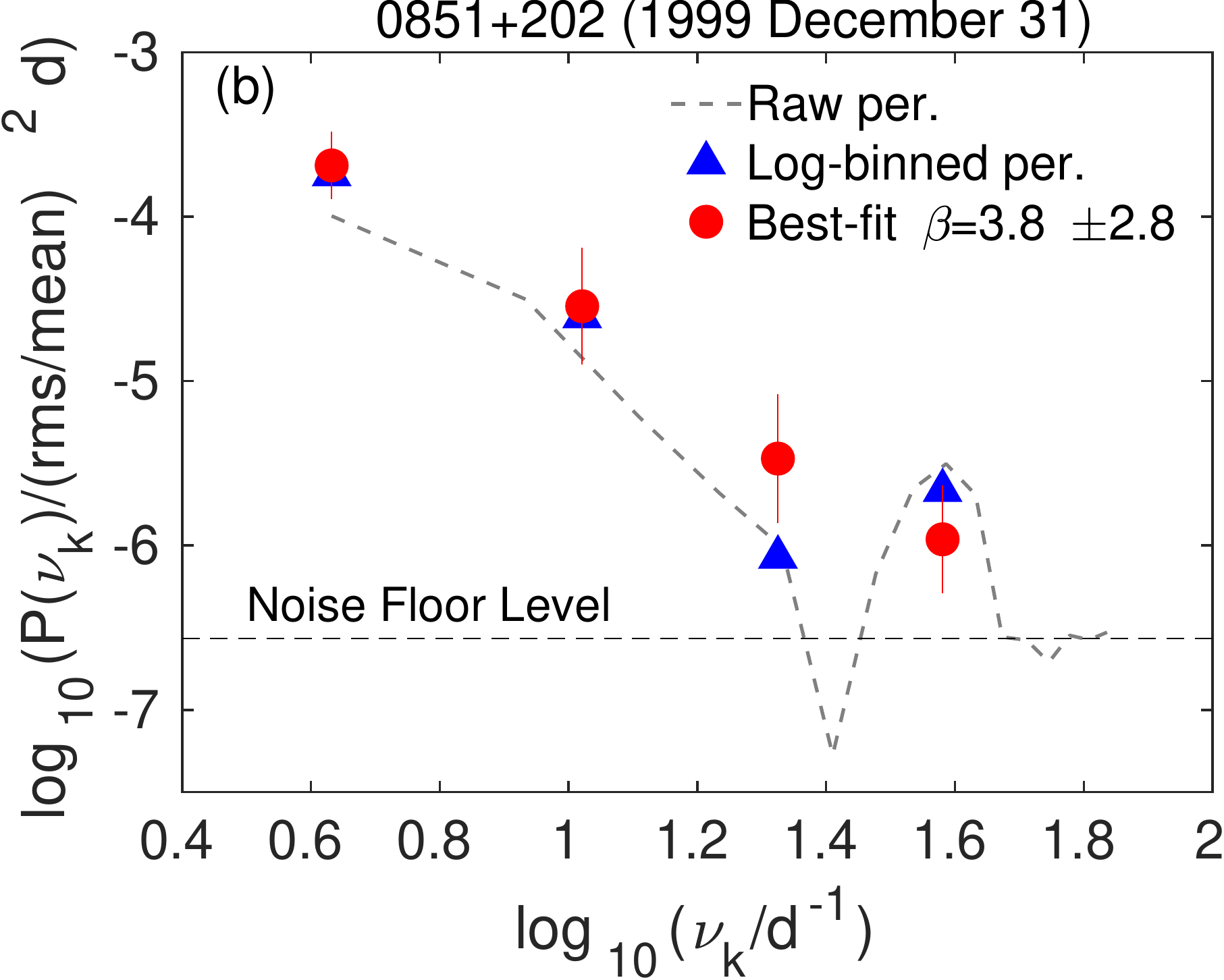}
\includegraphics[width=0.30\textwidth]{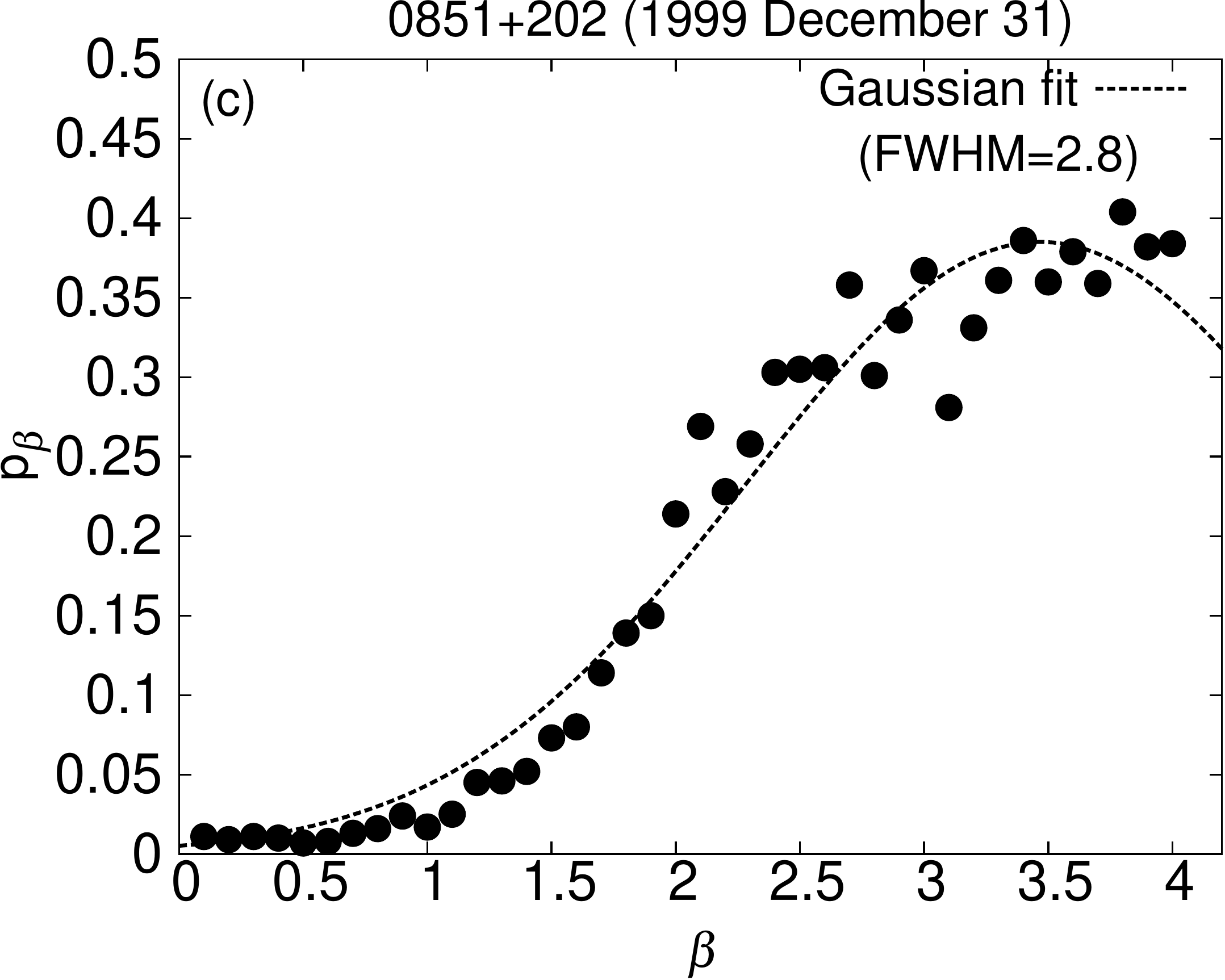}
}
\begin{minipage}{\textwidth}
\caption{(continued) }
\end{minipage}
\end{figure*}

\addtocounter{figure}{-1}
\begin{figure*}[ht!]

\hbox{
\includegraphics[width=0.30\textwidth]{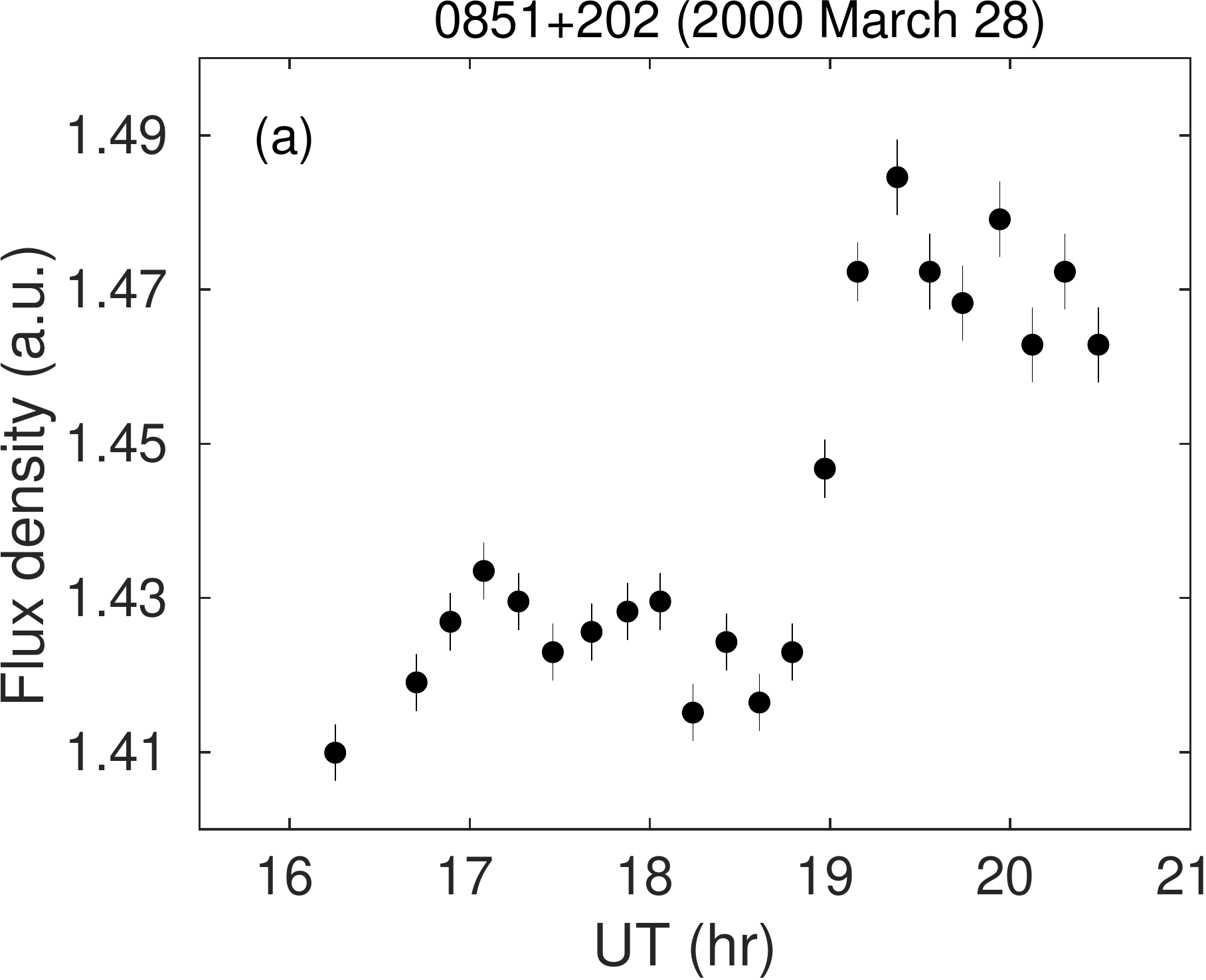}
\includegraphics[width=0.30\textwidth]{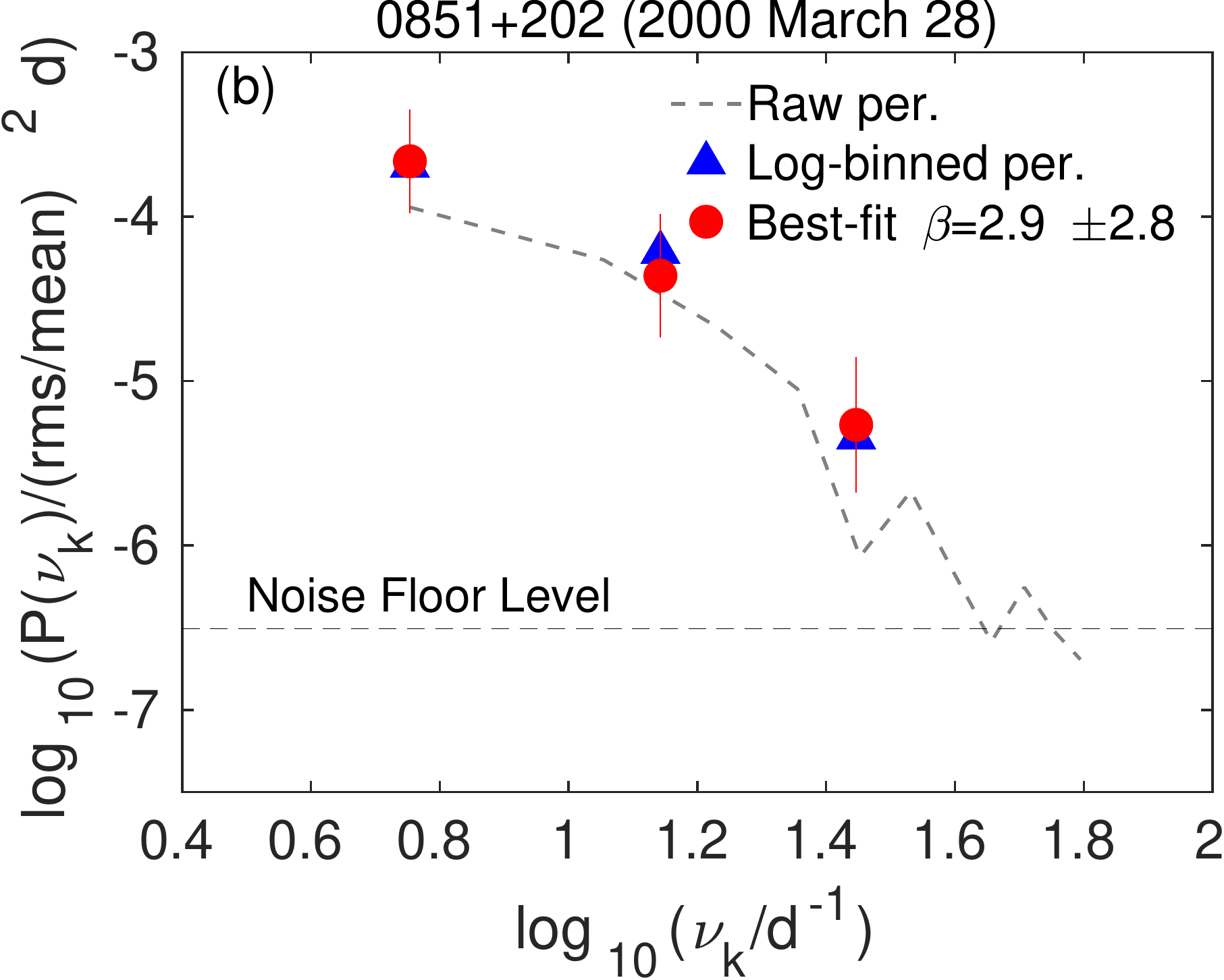}
\includegraphics[width=0.30\textwidth]{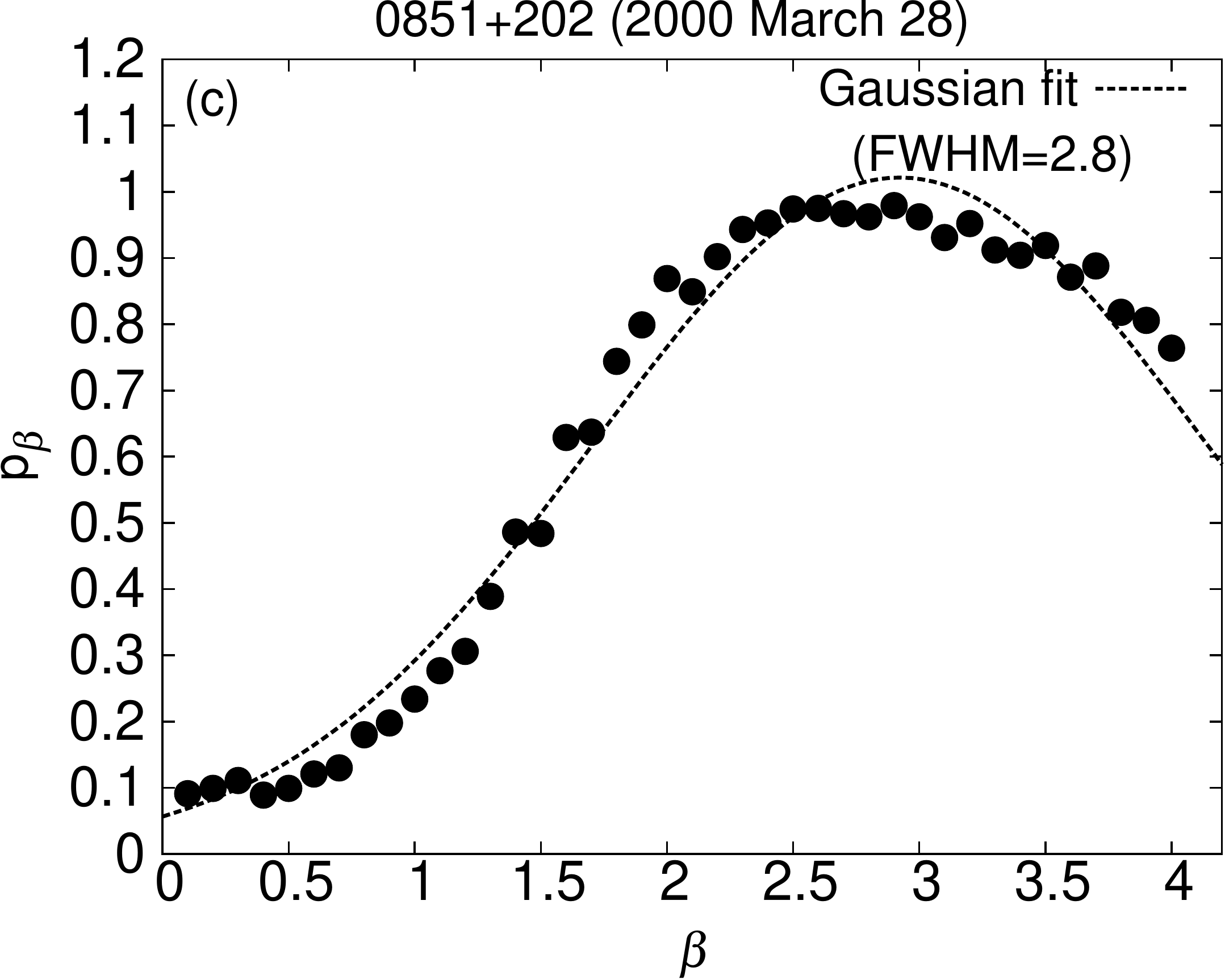}
}
\hbox{
\includegraphics[width=0.30\textwidth]{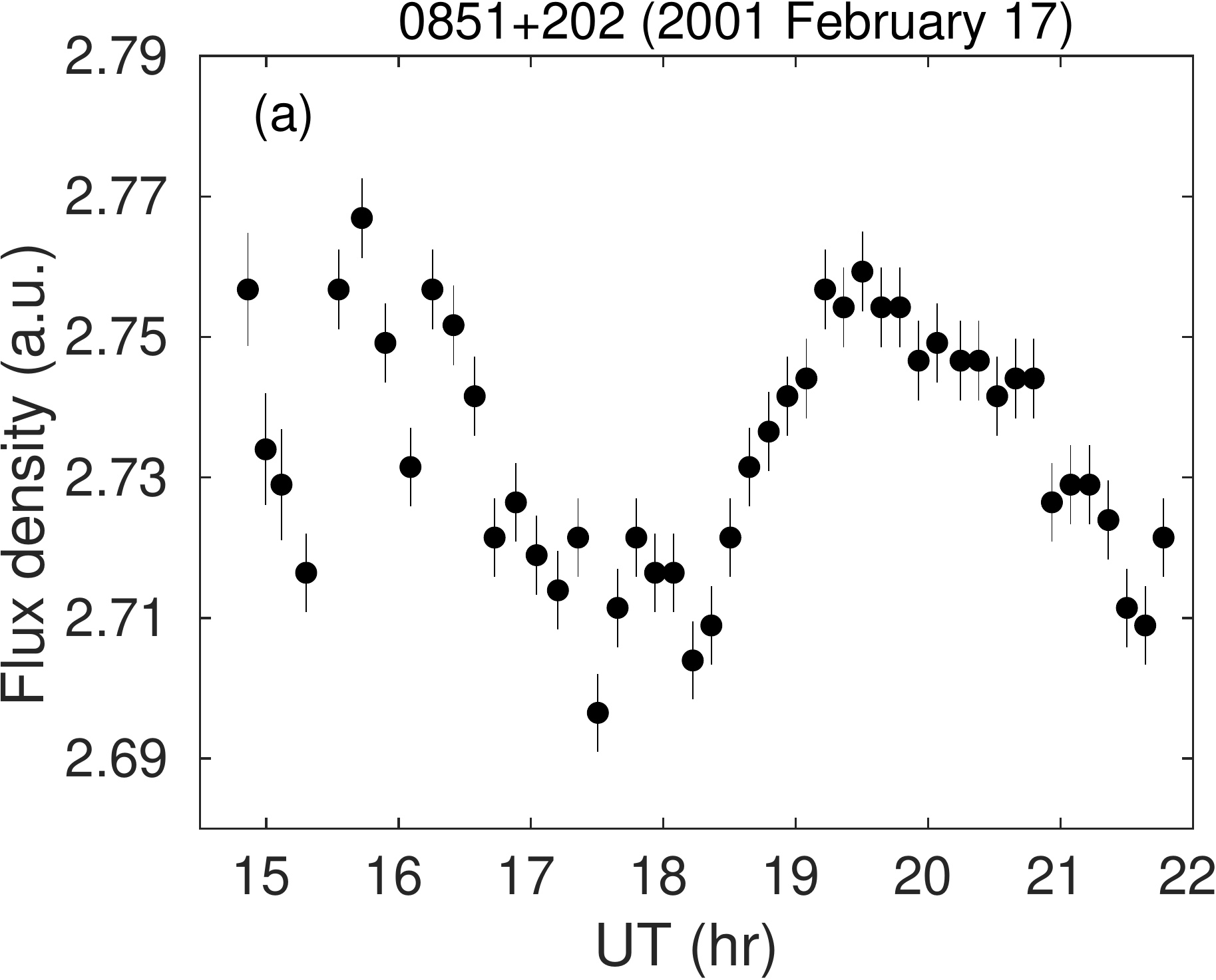}
\includegraphics[width=0.30\textwidth]{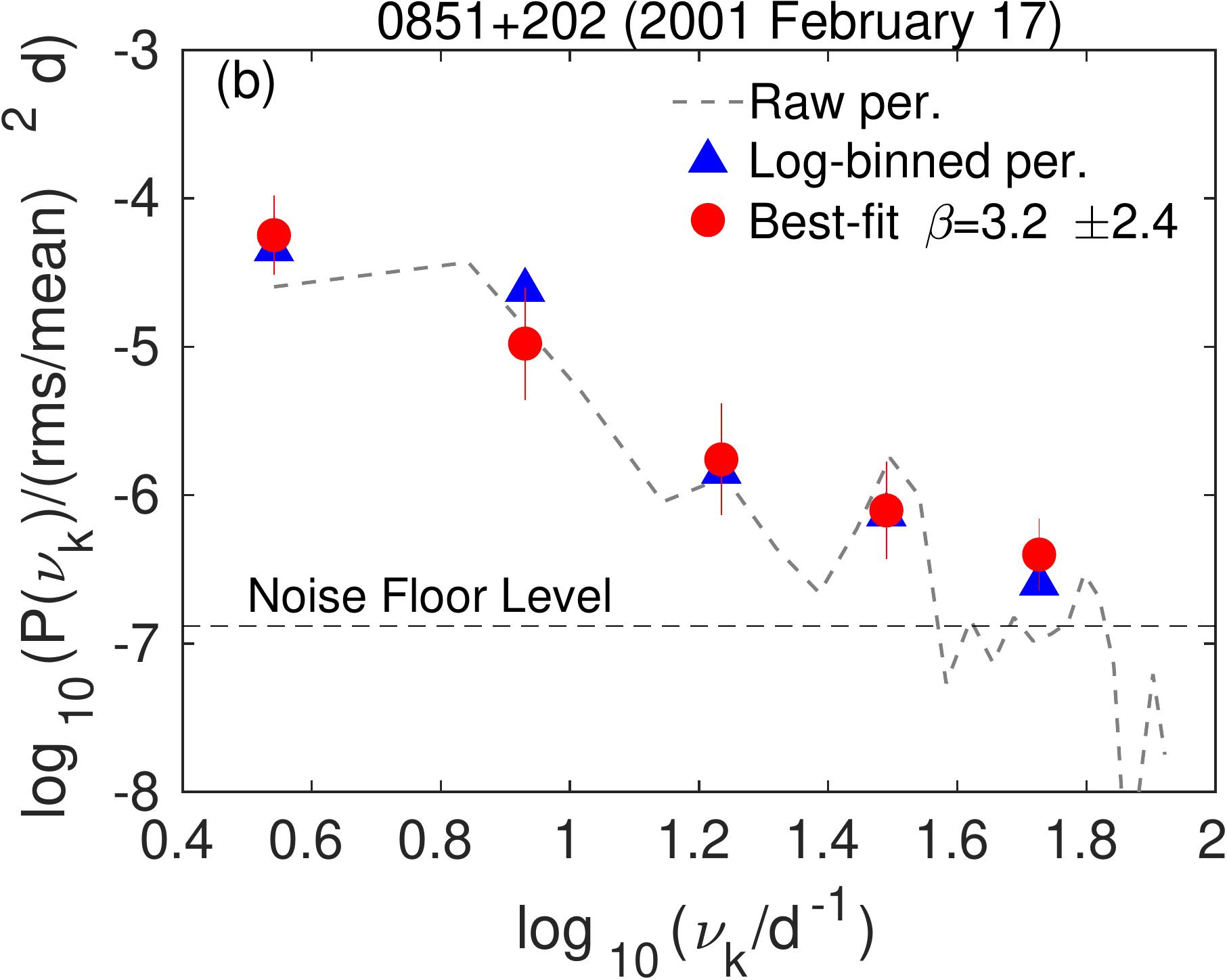}
\includegraphics[width=0.30\textwidth]{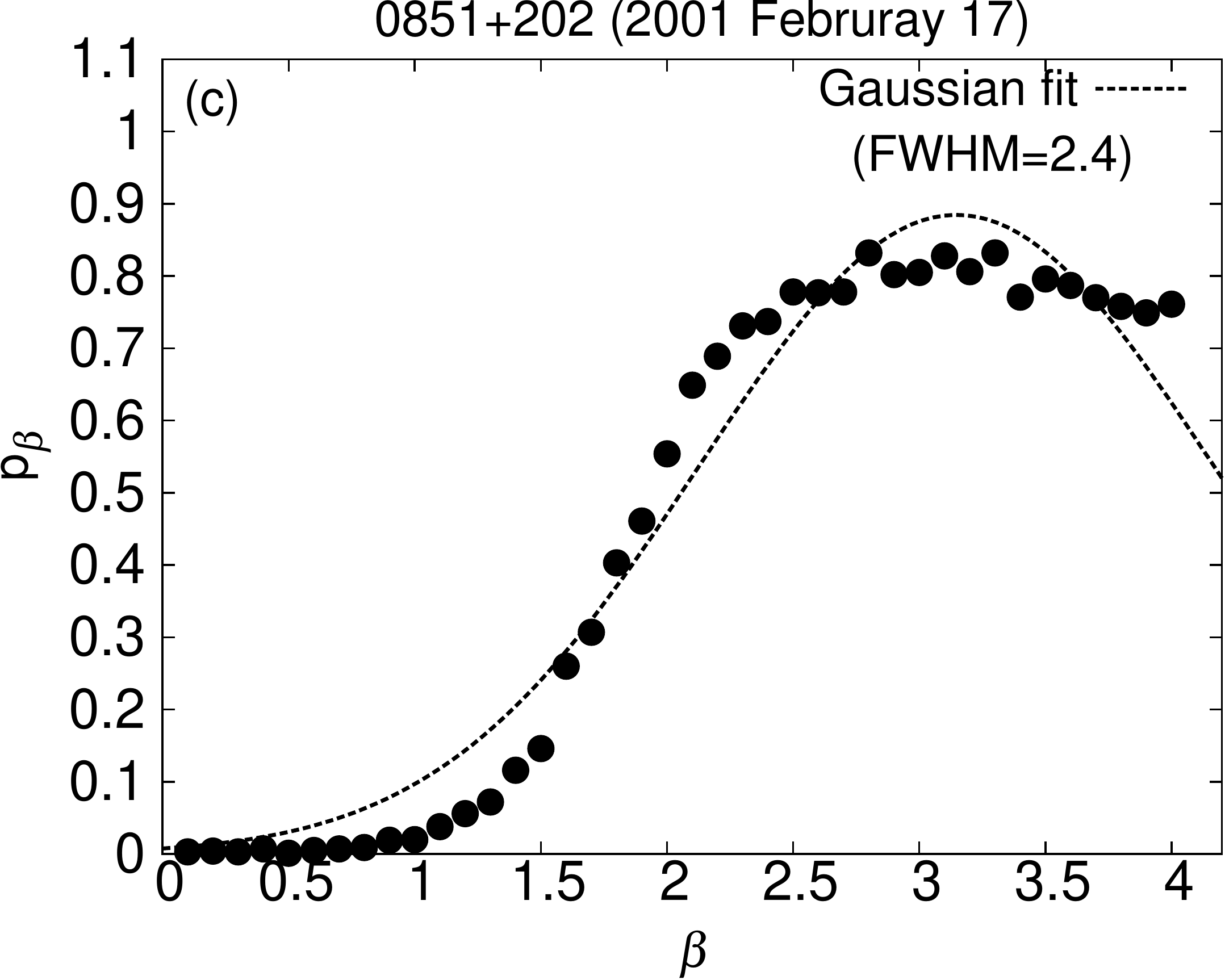}
}
\hbox{
\includegraphics[width=0.30\textwidth]{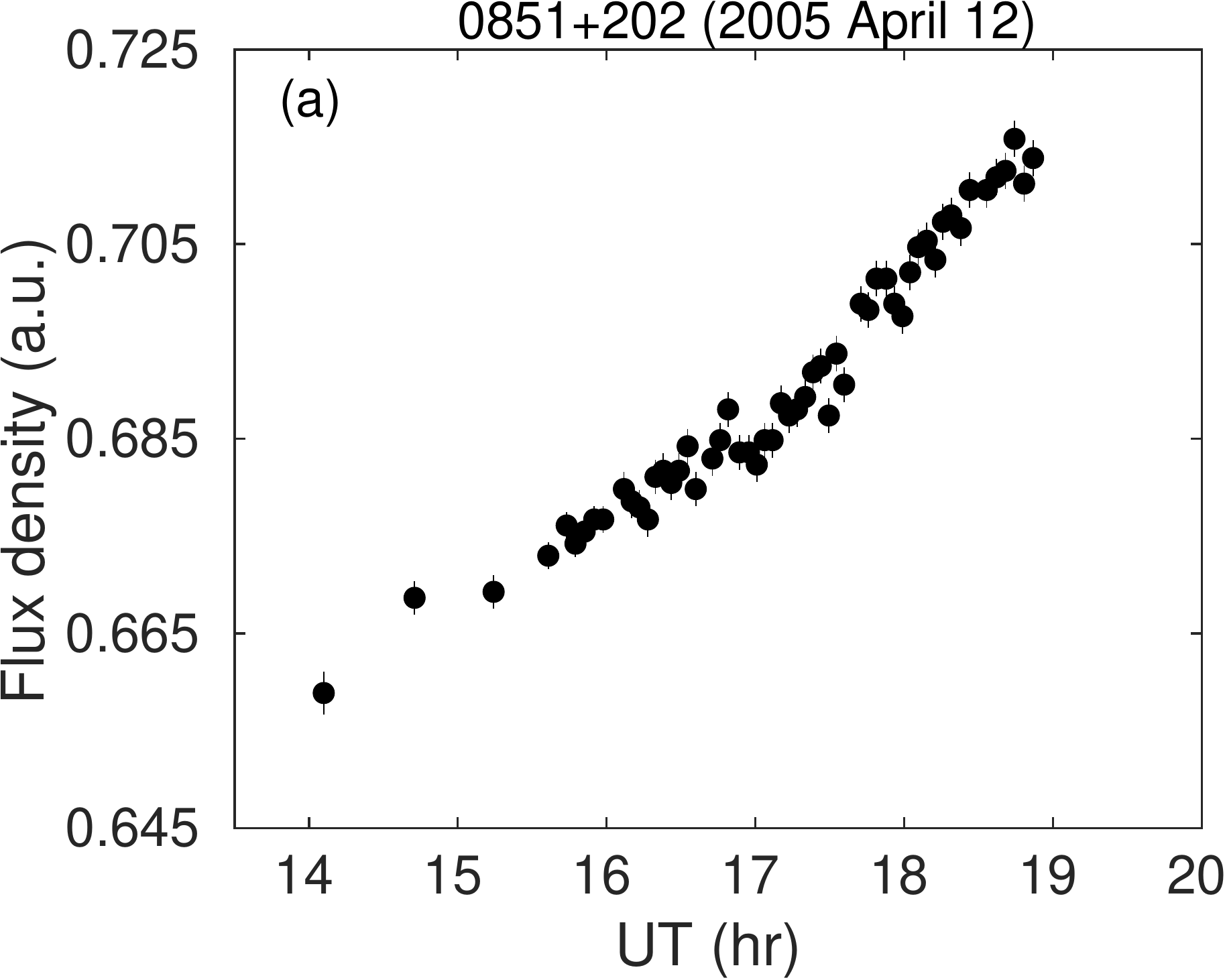}
\includegraphics[width=0.30\textwidth]{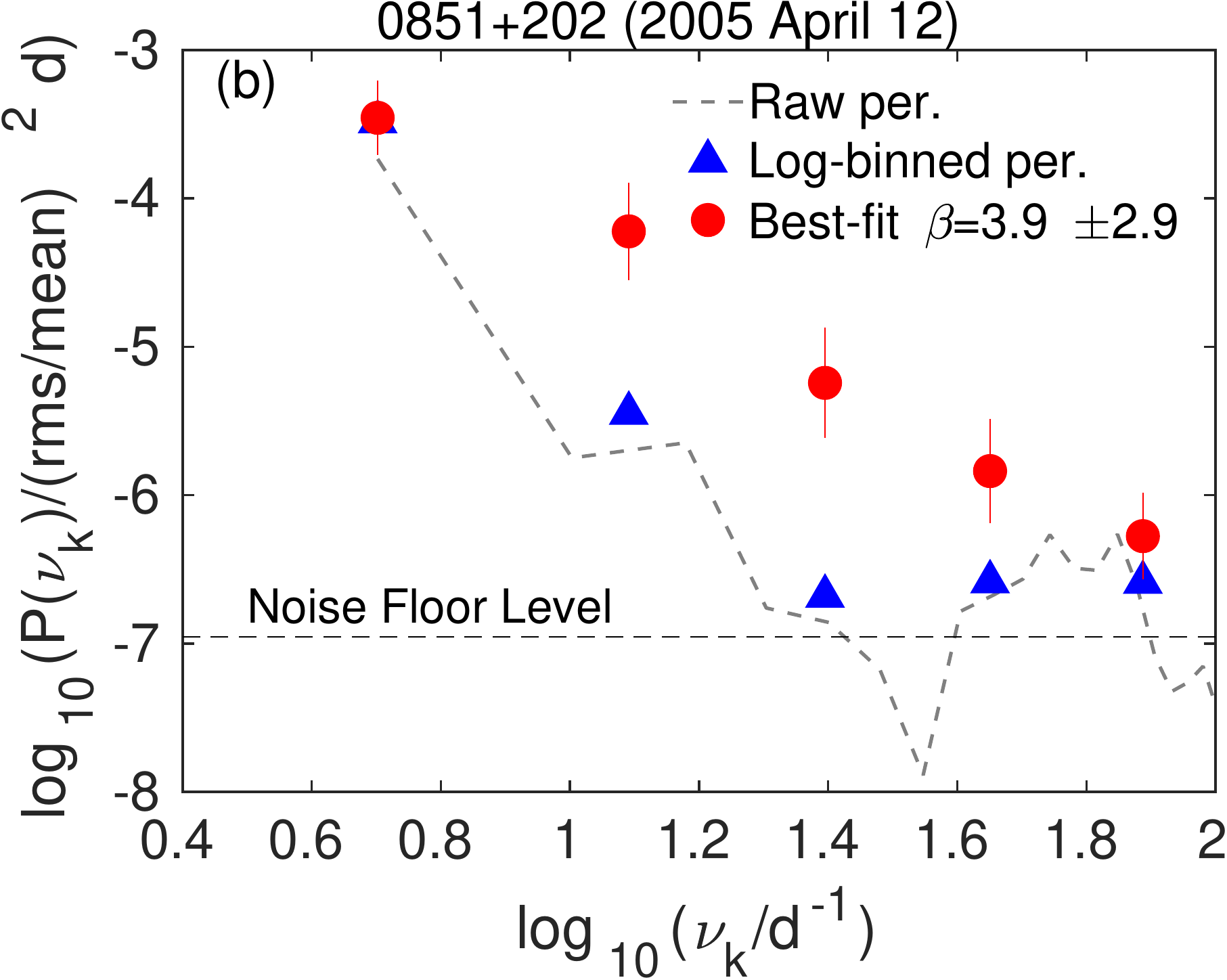}
\includegraphics[width=0.30\textwidth]{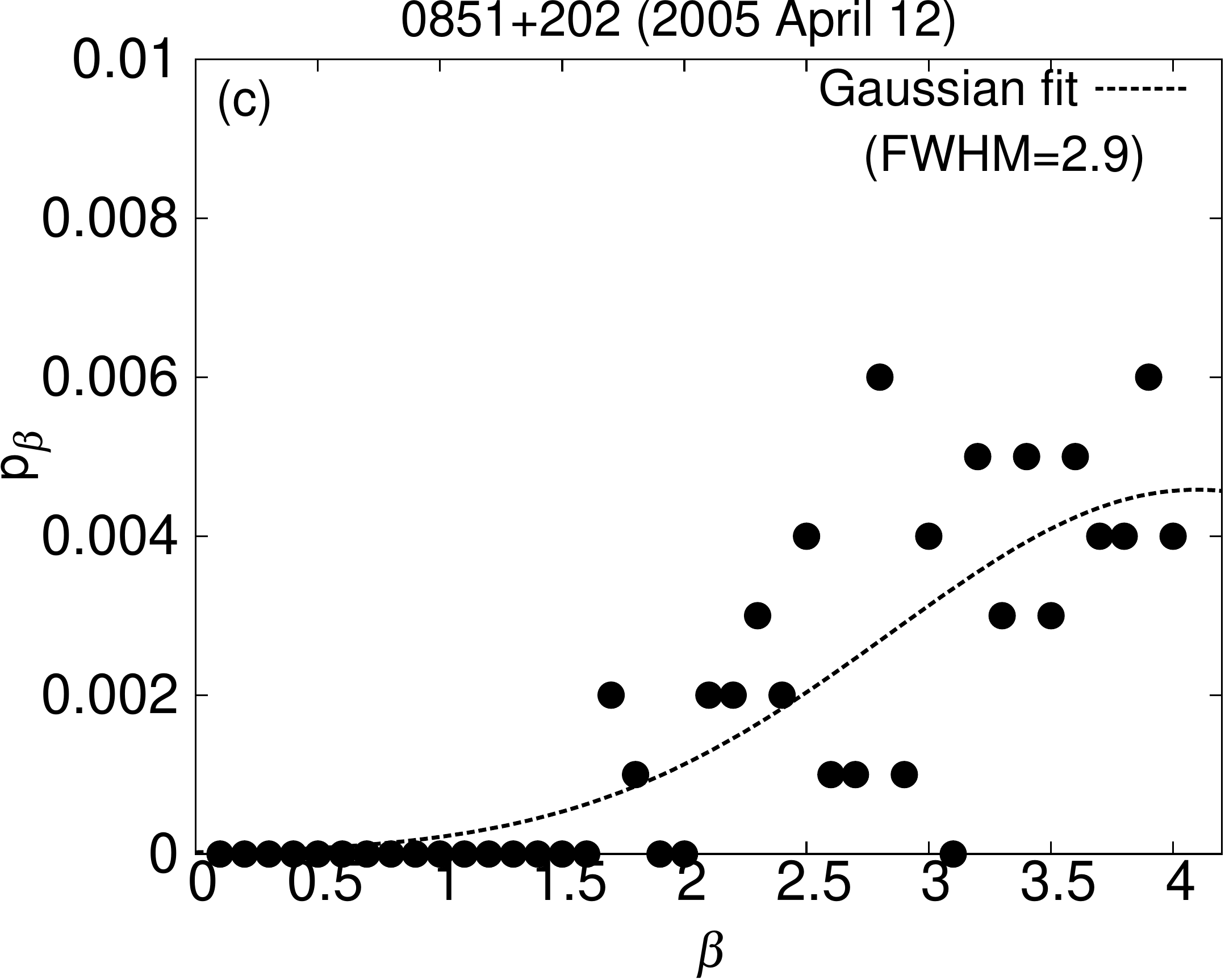}
}
\hbox{
\includegraphics[width=0.30\textwidth]{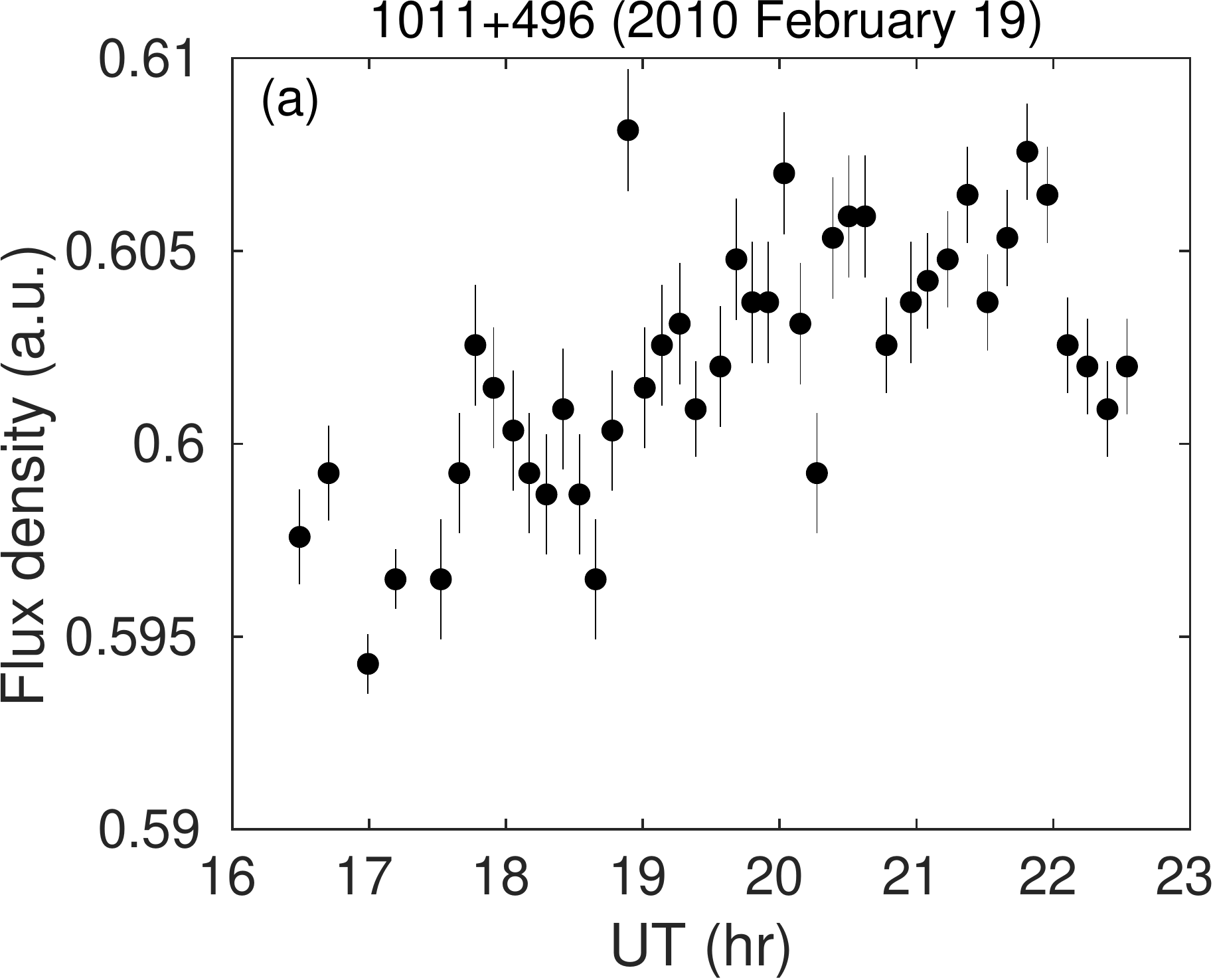}
\includegraphics[width=0.30\textwidth]{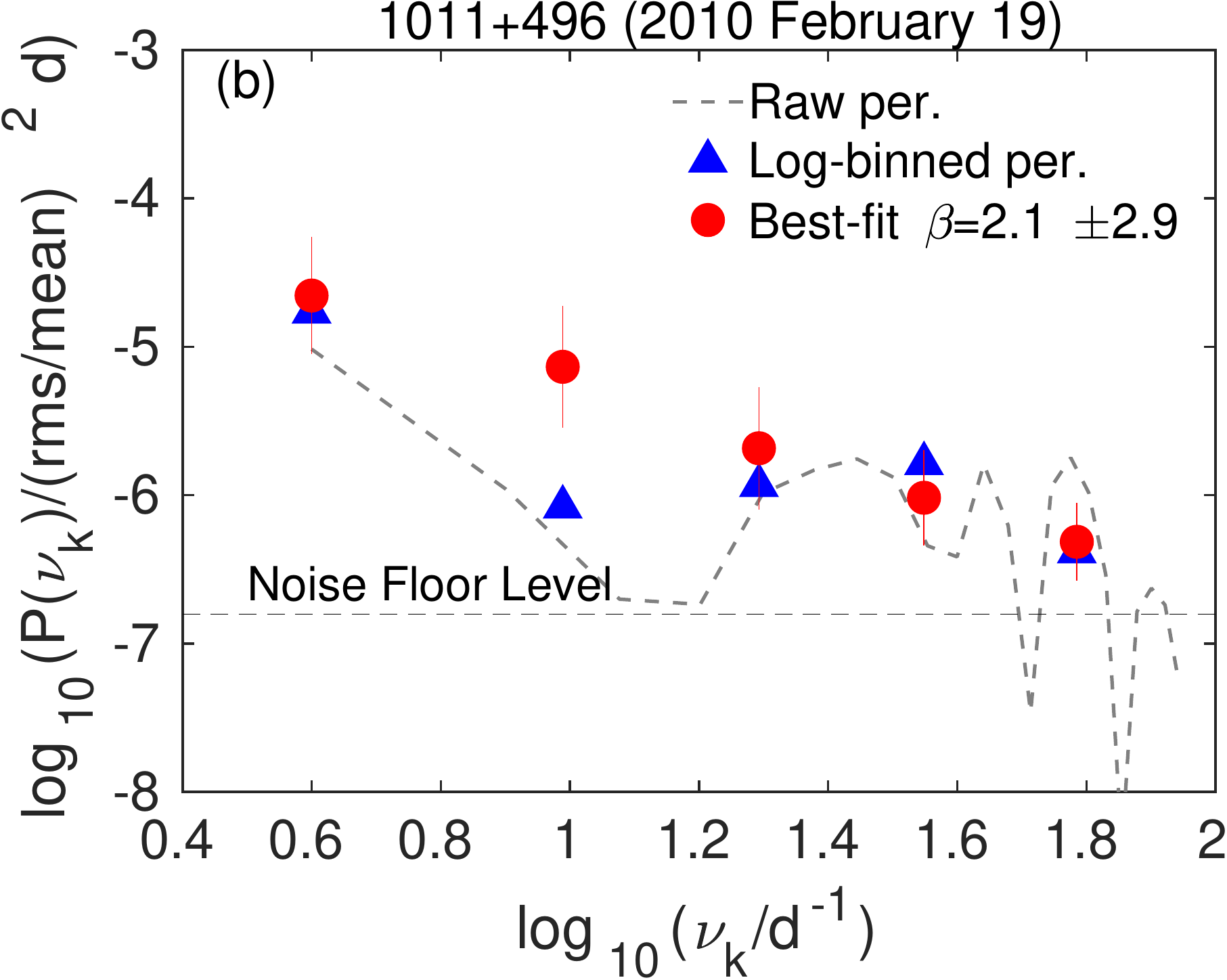}
\includegraphics[width=0.30\textwidth]{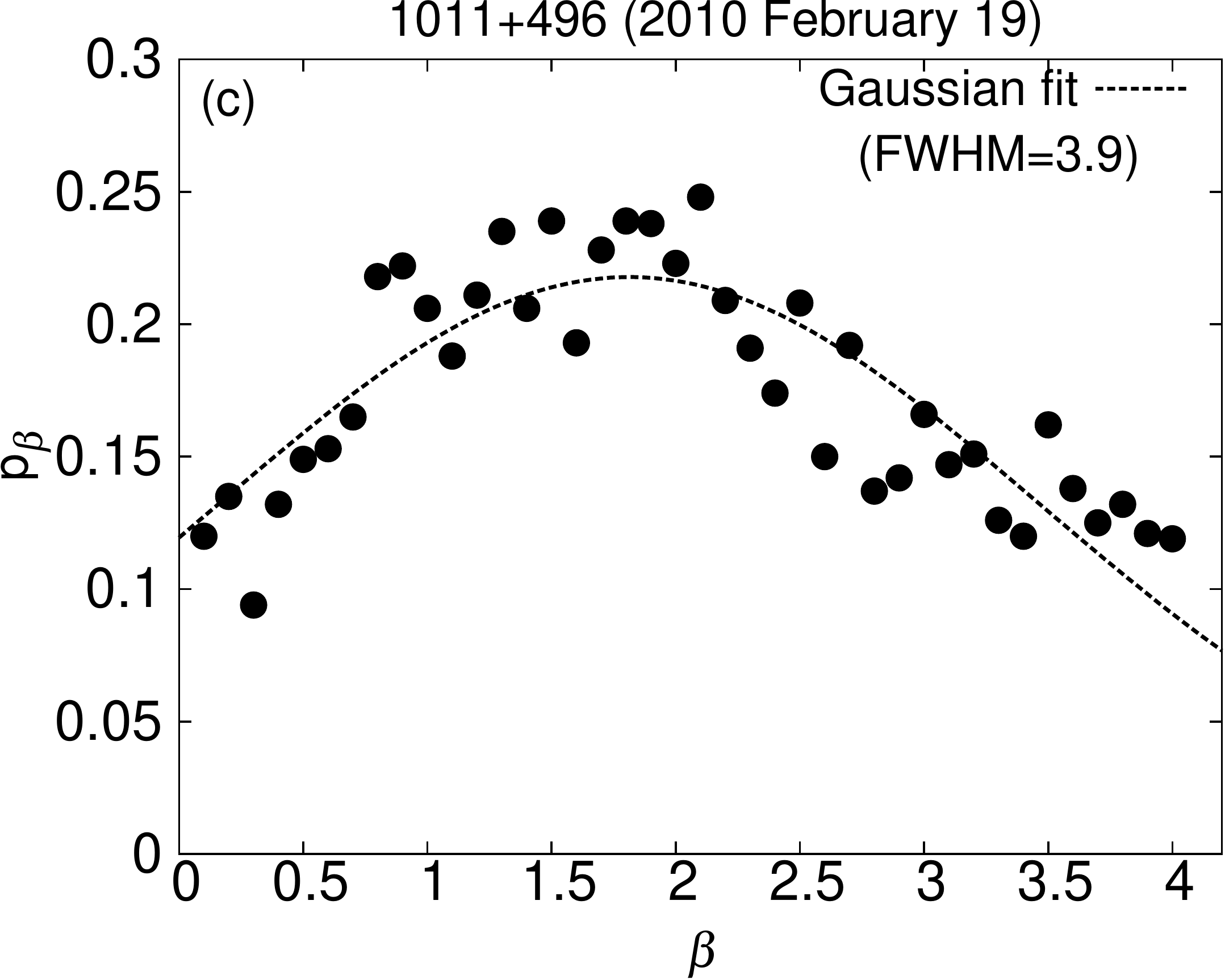}
}
\hbox{
\includegraphics[width=0.30\textwidth]{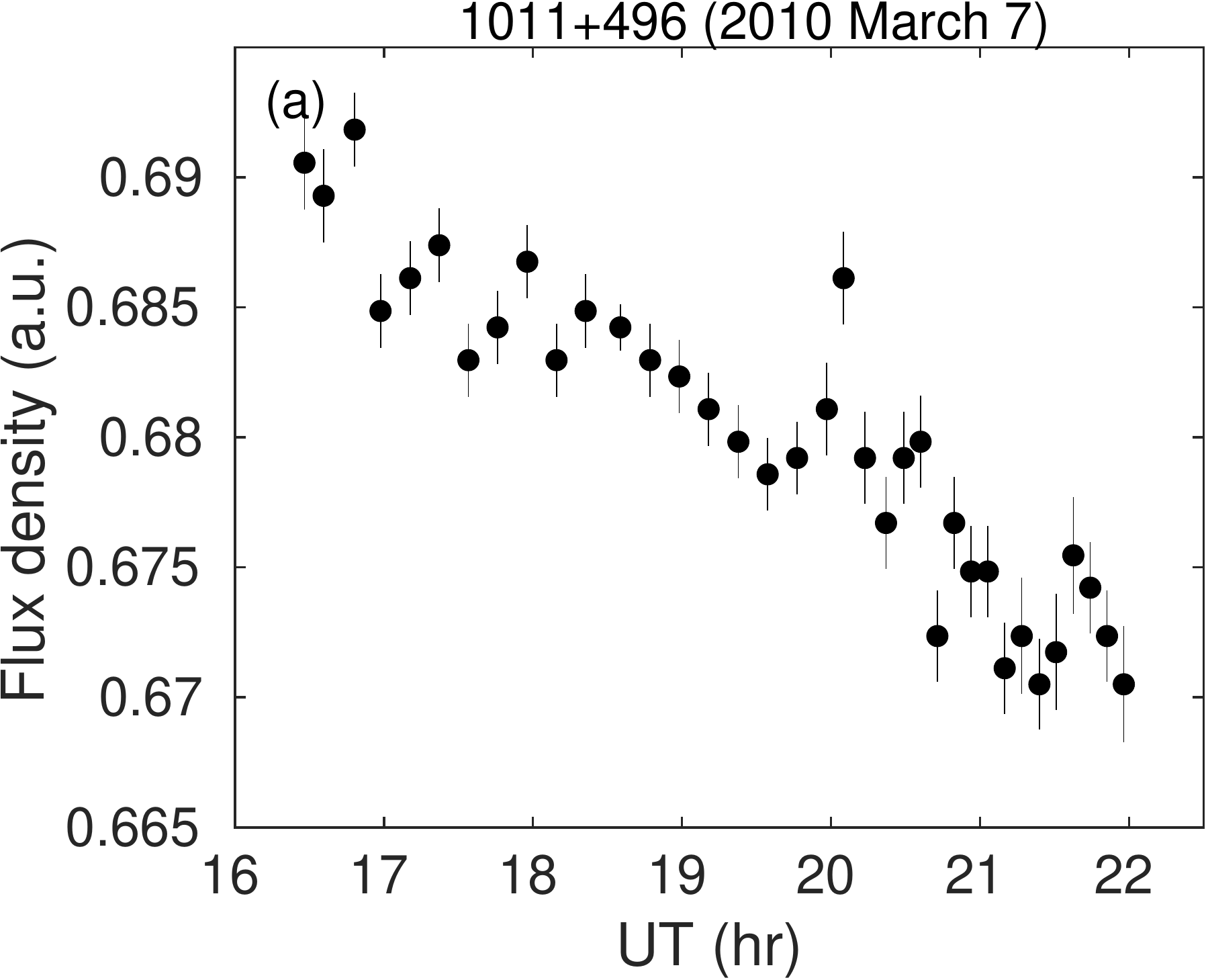}
\includegraphics[width=0.30\textwidth]{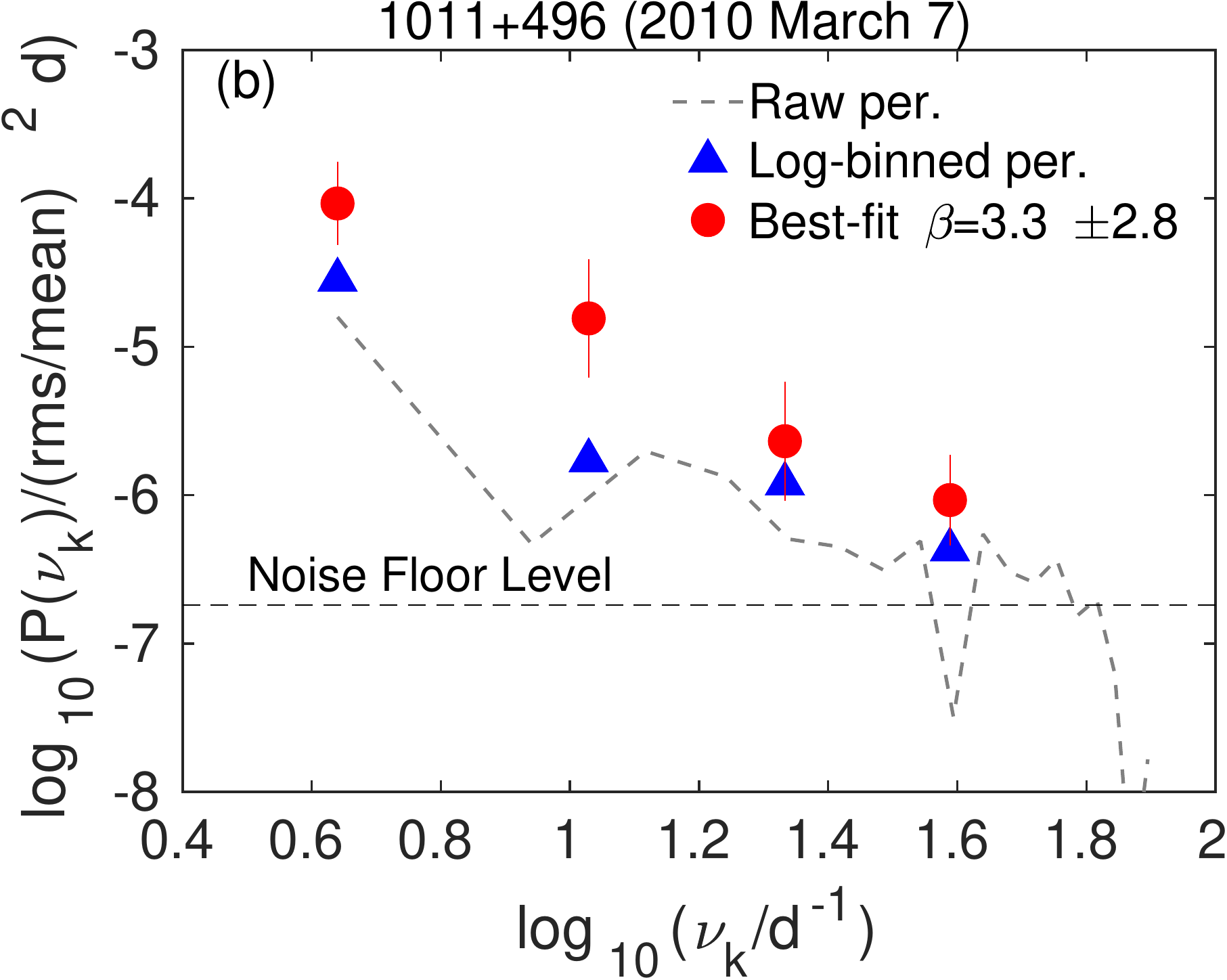}
\includegraphics[width=0.30\textwidth]{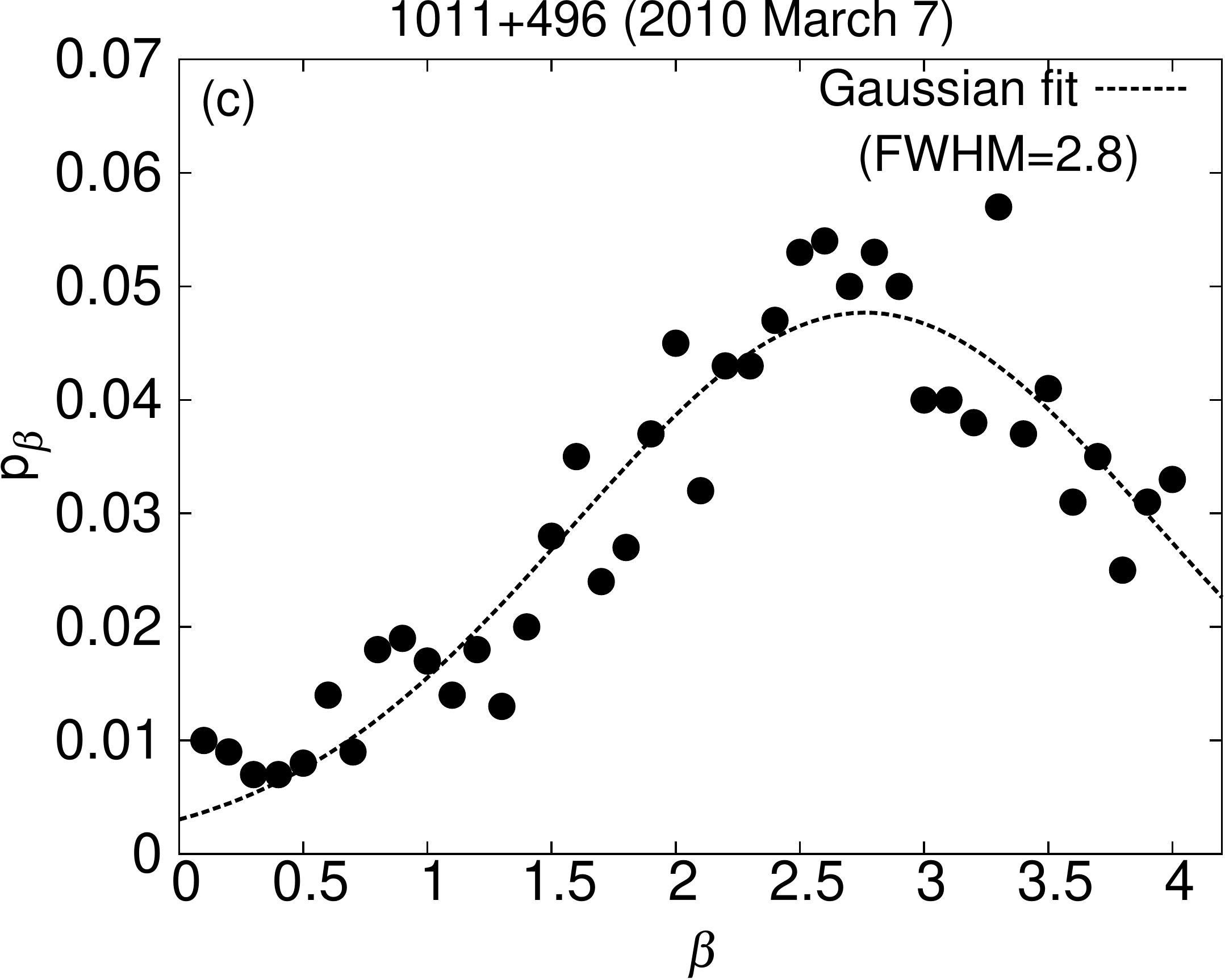}
}

\begin{minipage}{\textwidth}
\caption{(continued) }
\end{minipage}
\end{figure*}

\addtocounter{figure}{-1}
\begin{figure*}[ht!]

\hbox{
\includegraphics[width=0.30\textwidth]{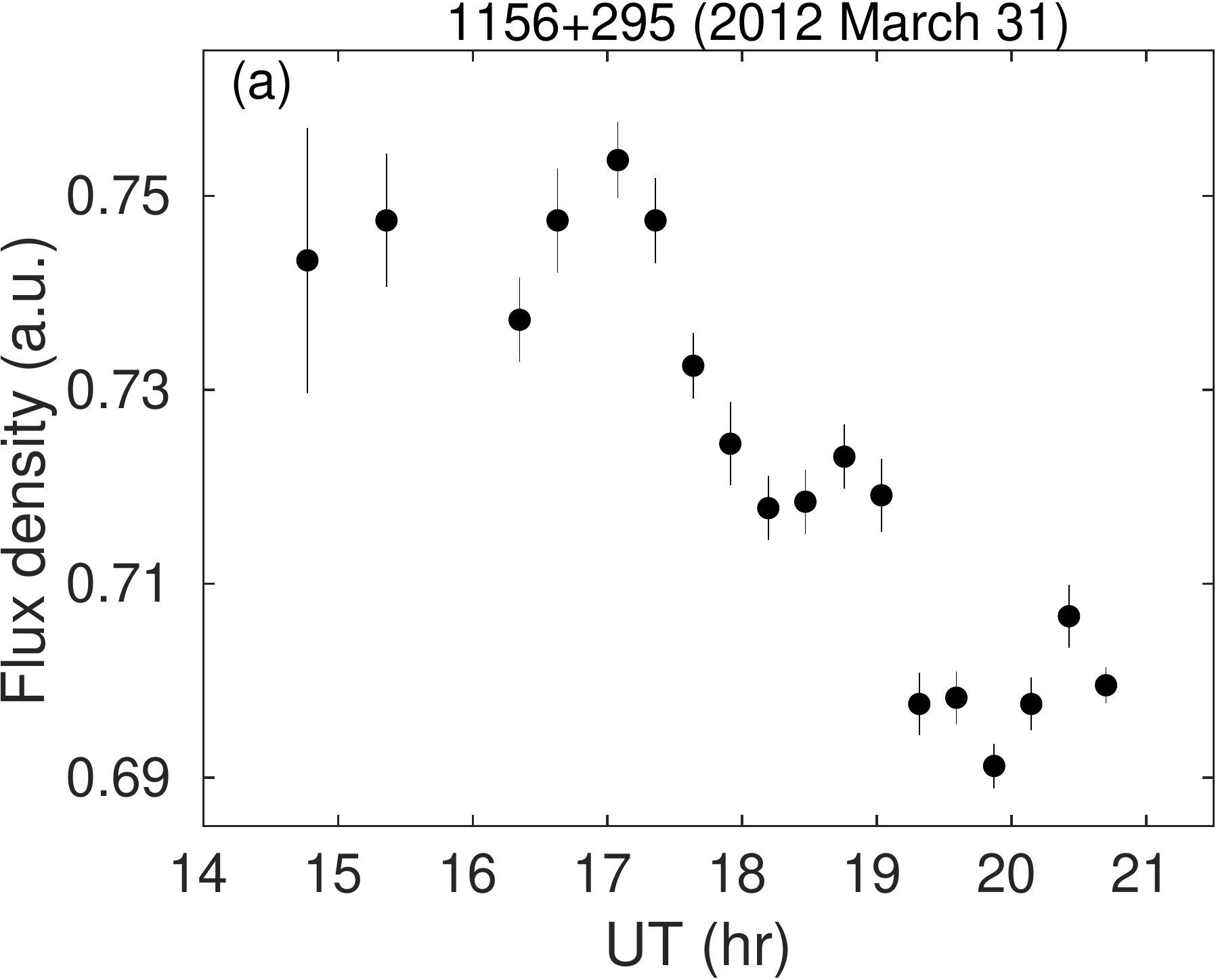}
\includegraphics[width=0.30\textwidth]{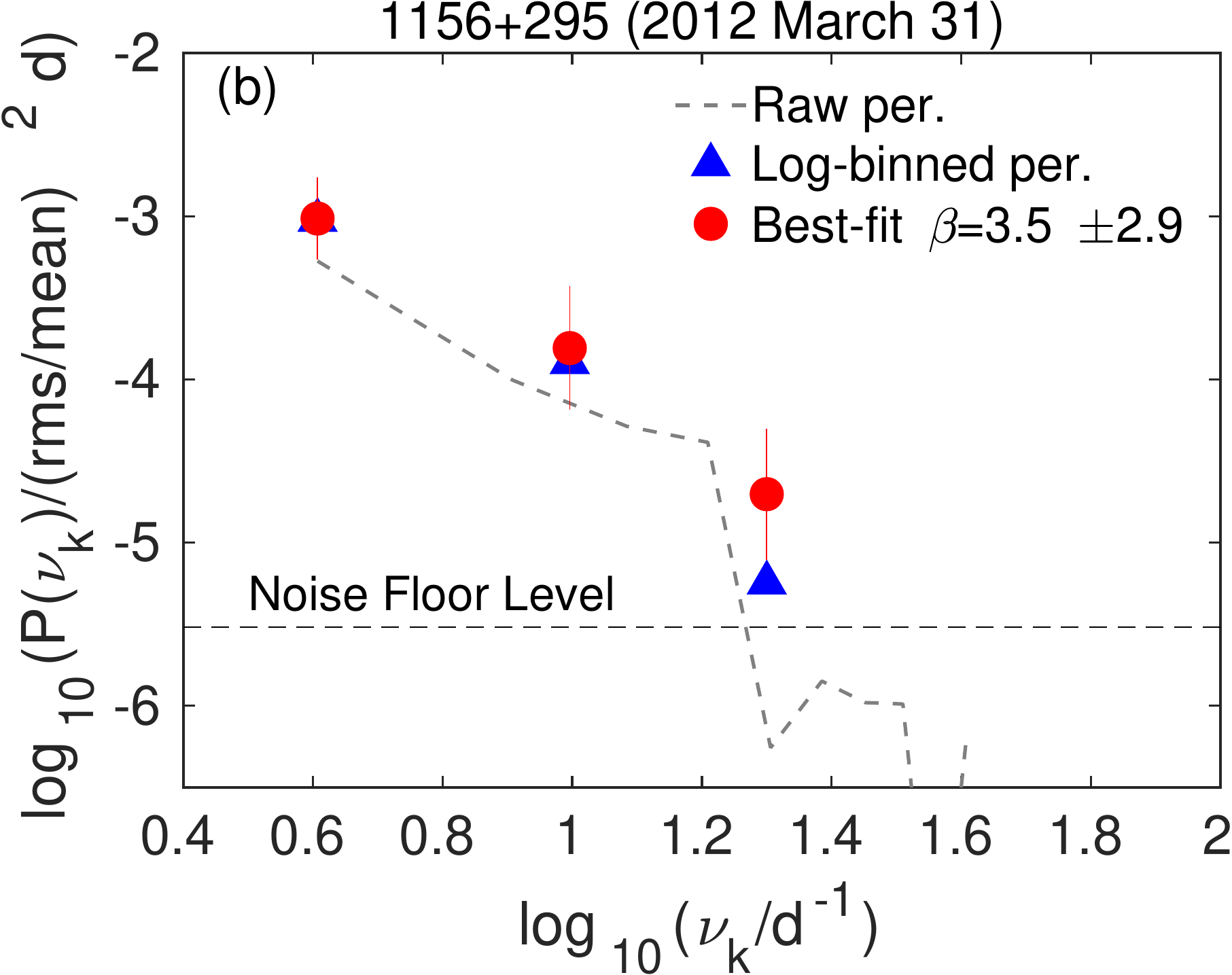}
\includegraphics[width=0.30\textwidth]{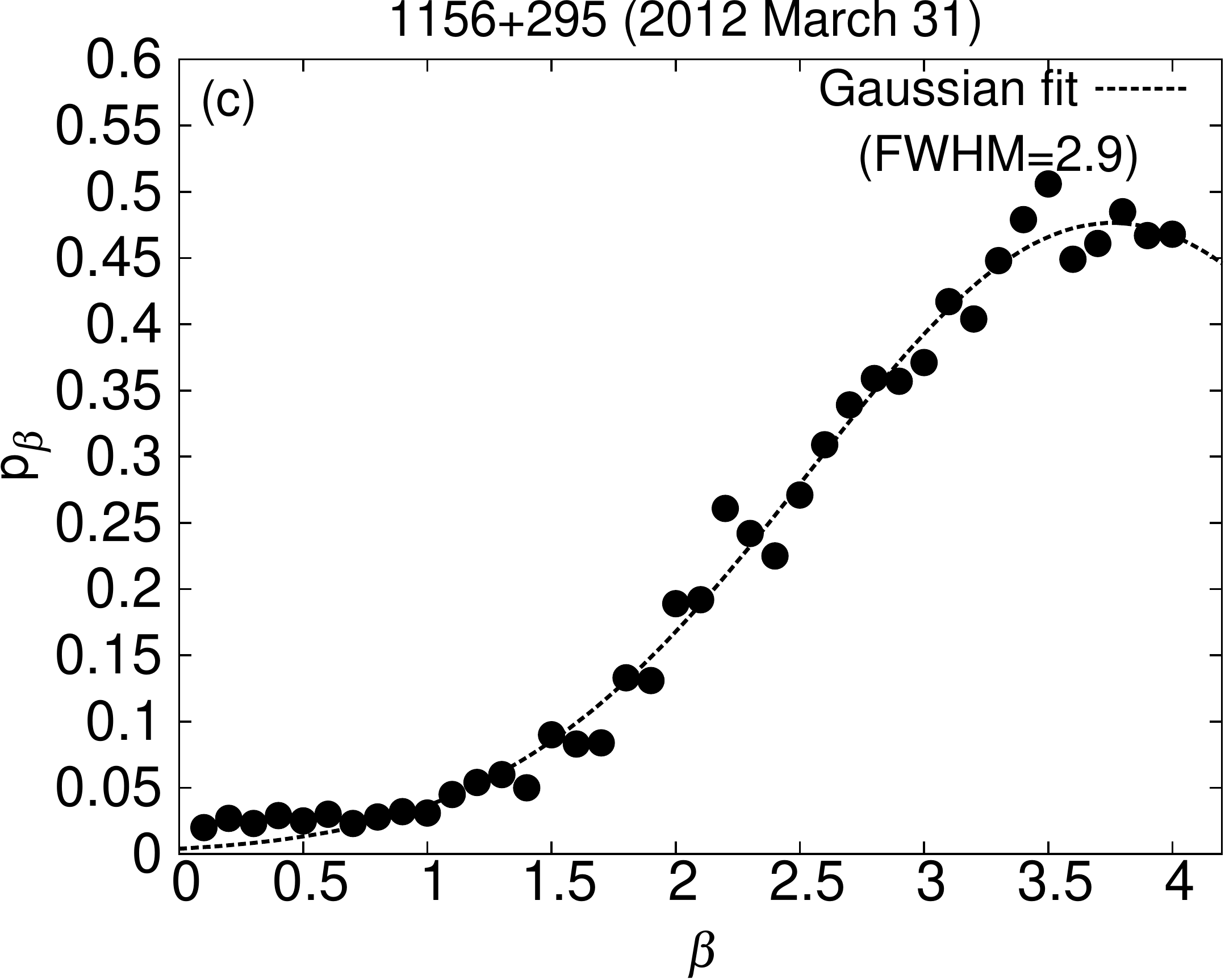}
}
\hbox{
\includegraphics[width=0.30\textwidth]{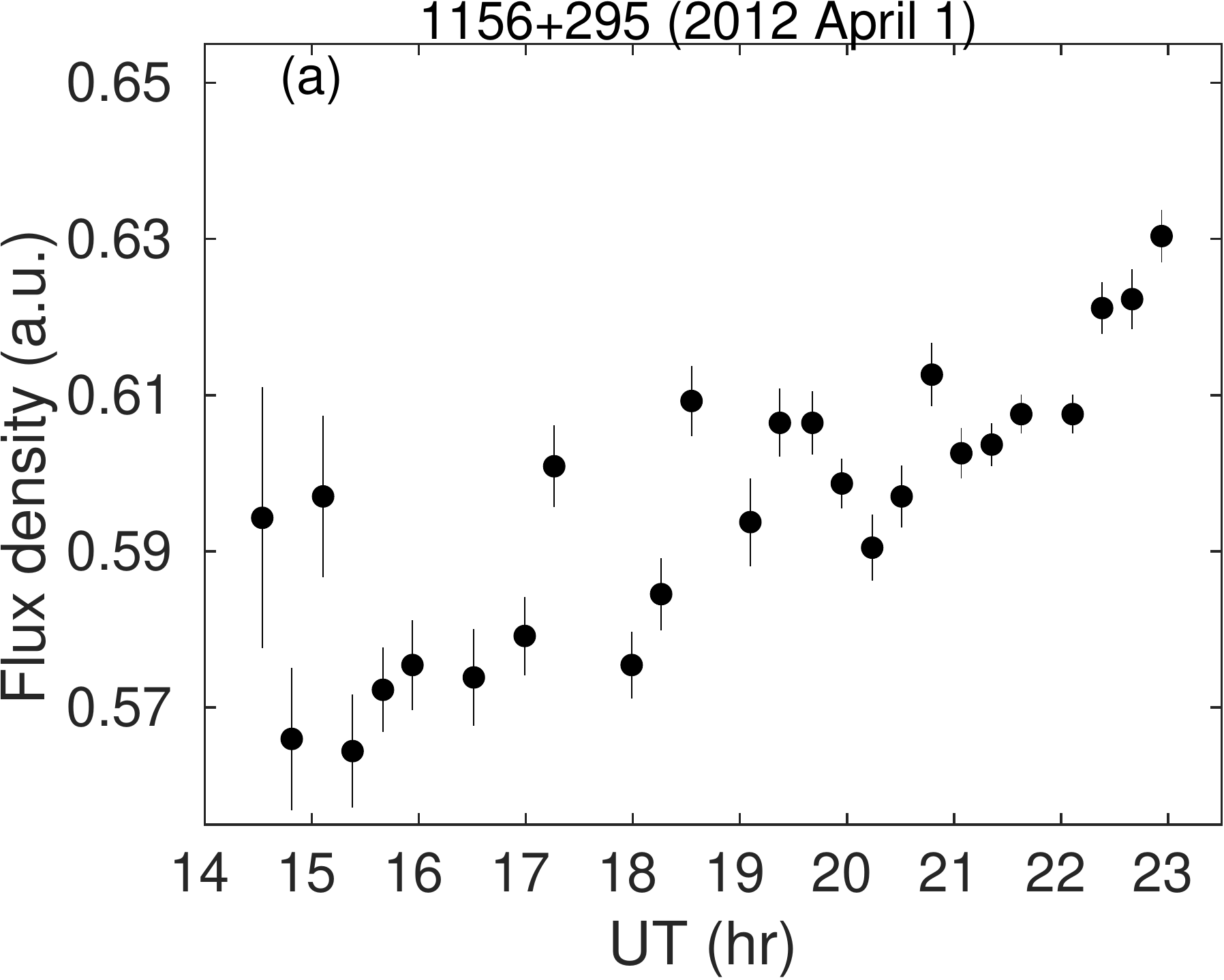}
\includegraphics[width=0.30\textwidth]{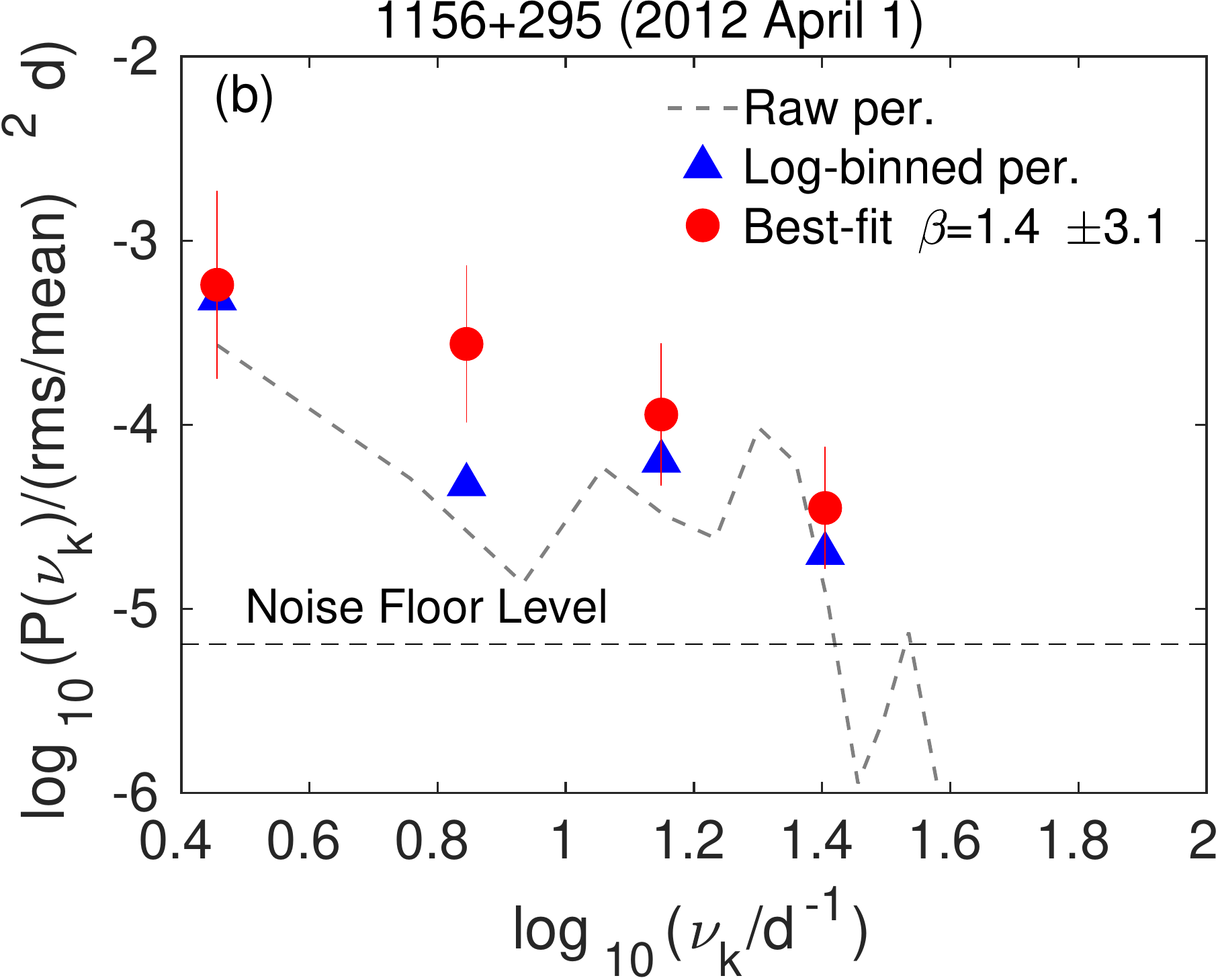}
\includegraphics[width=0.30\textwidth]{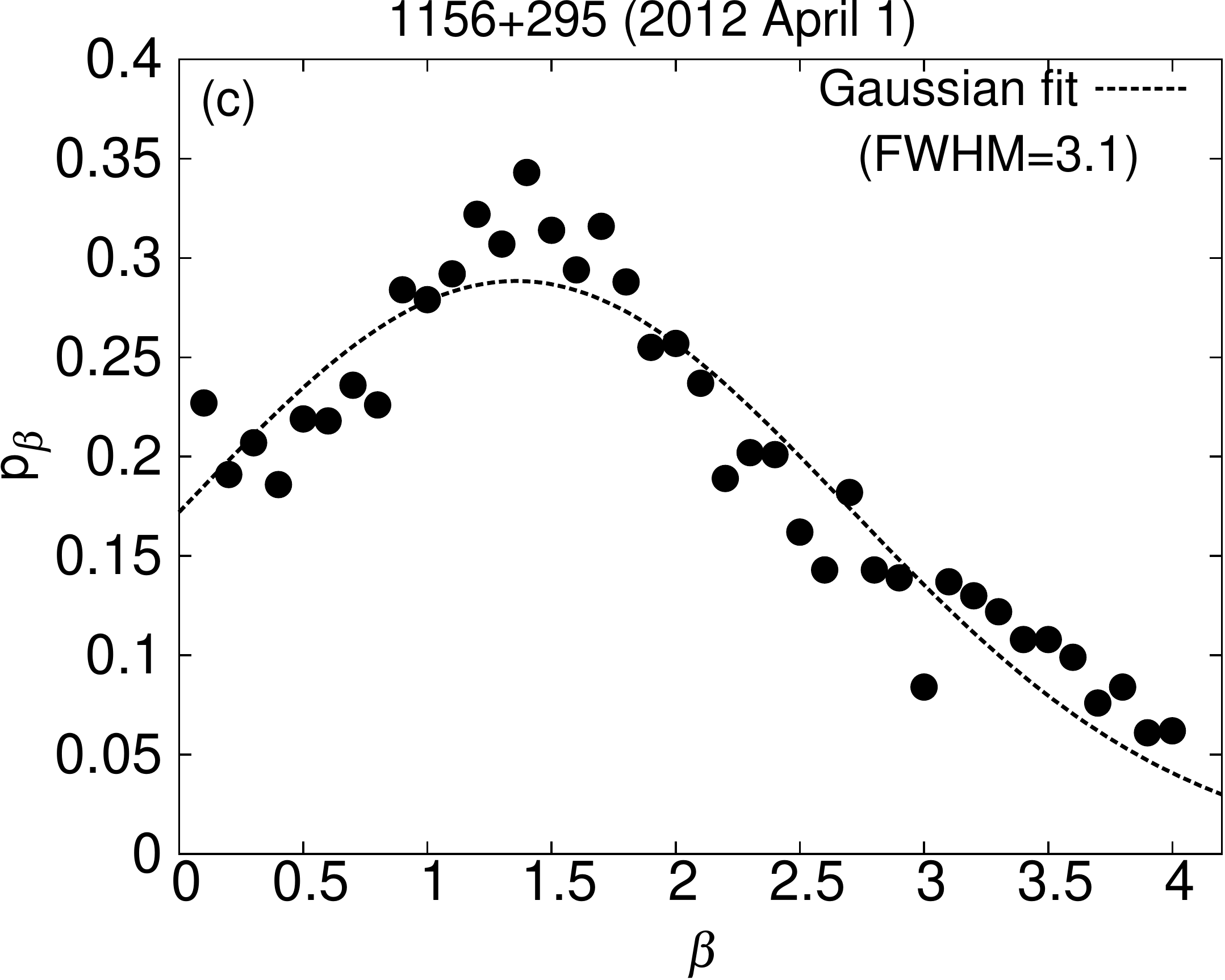}
}
\hbox{
\includegraphics[width=0.30\textwidth]{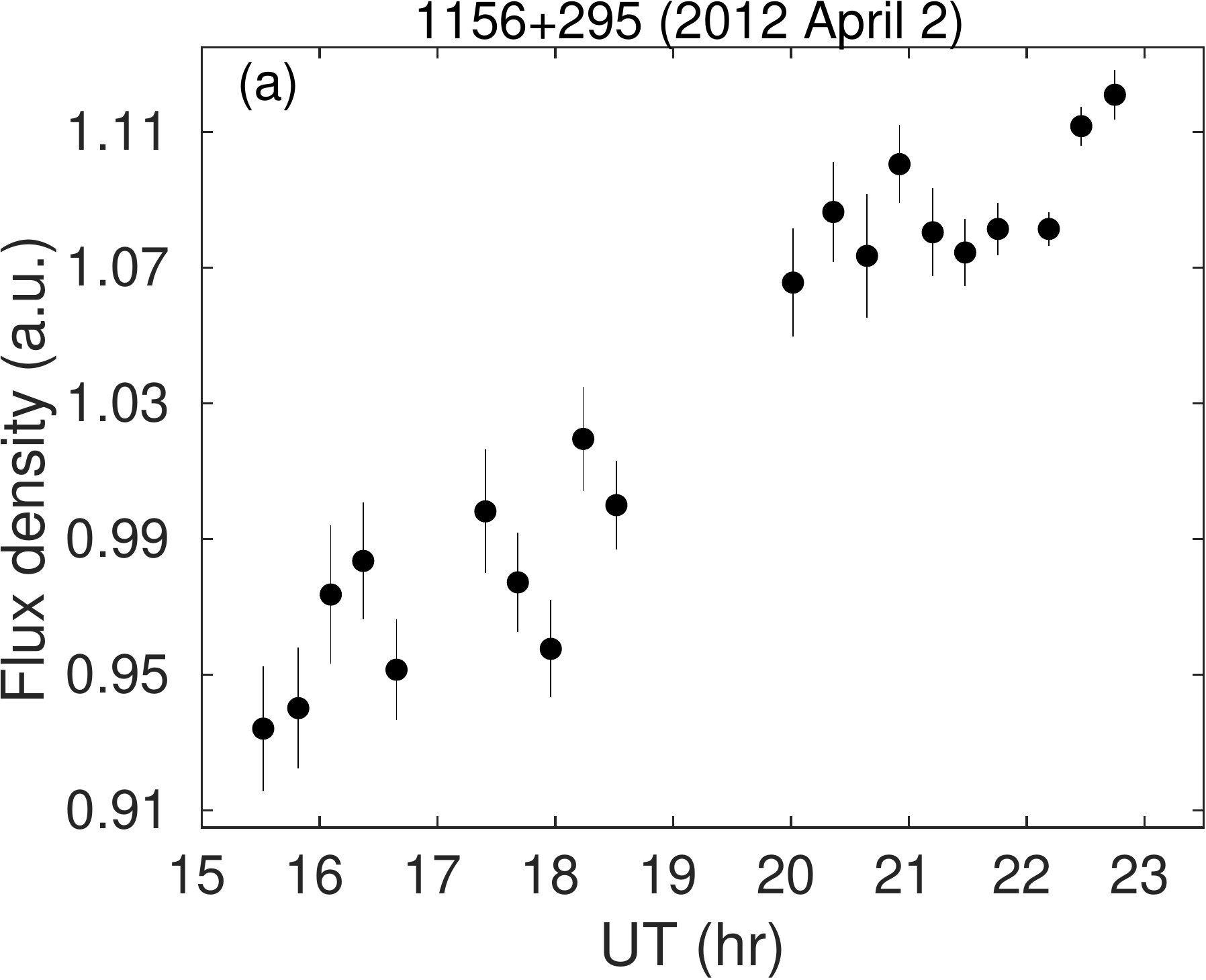}
\includegraphics[width=0.30\textwidth]{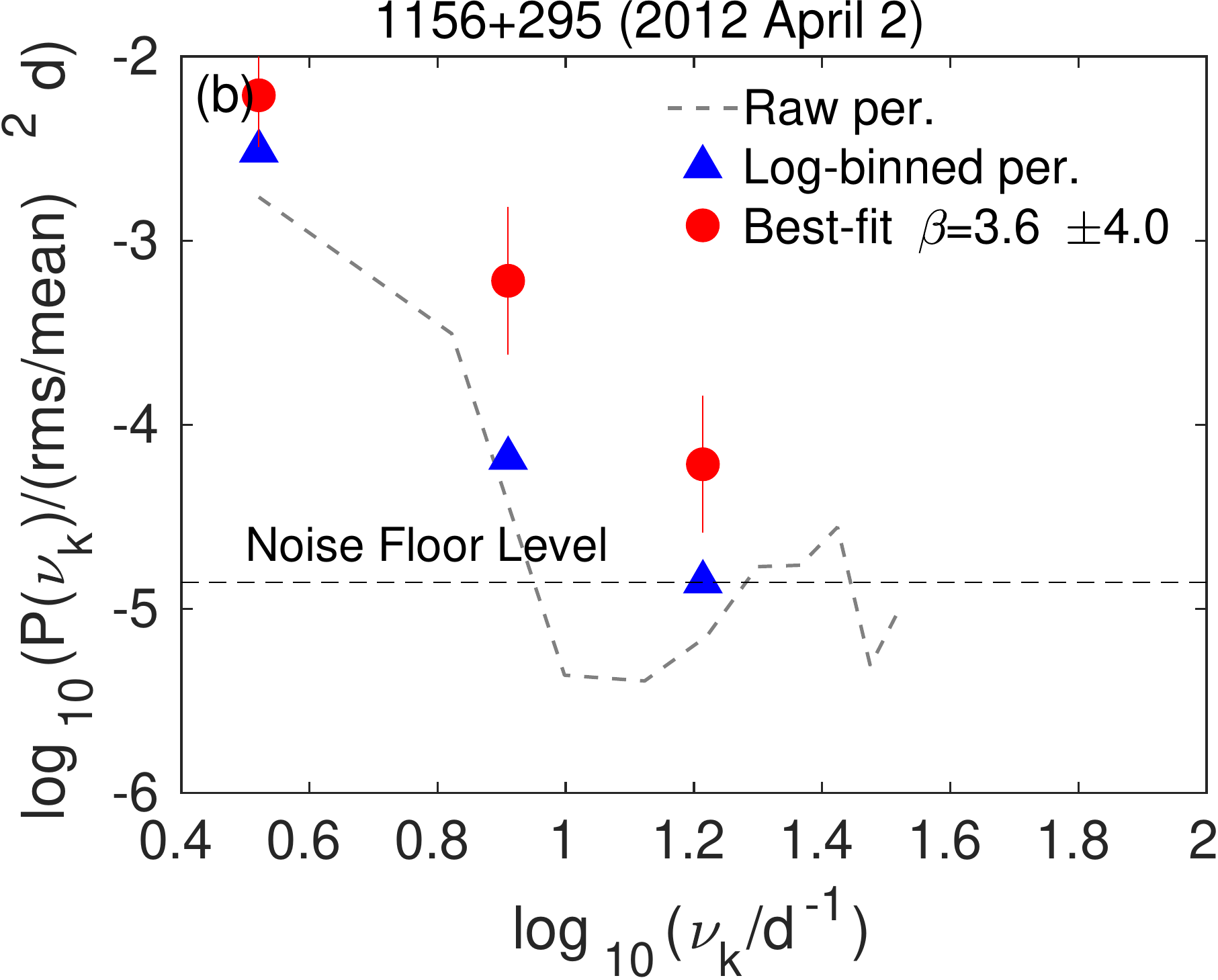}
\includegraphics[width=0.30\textwidth]{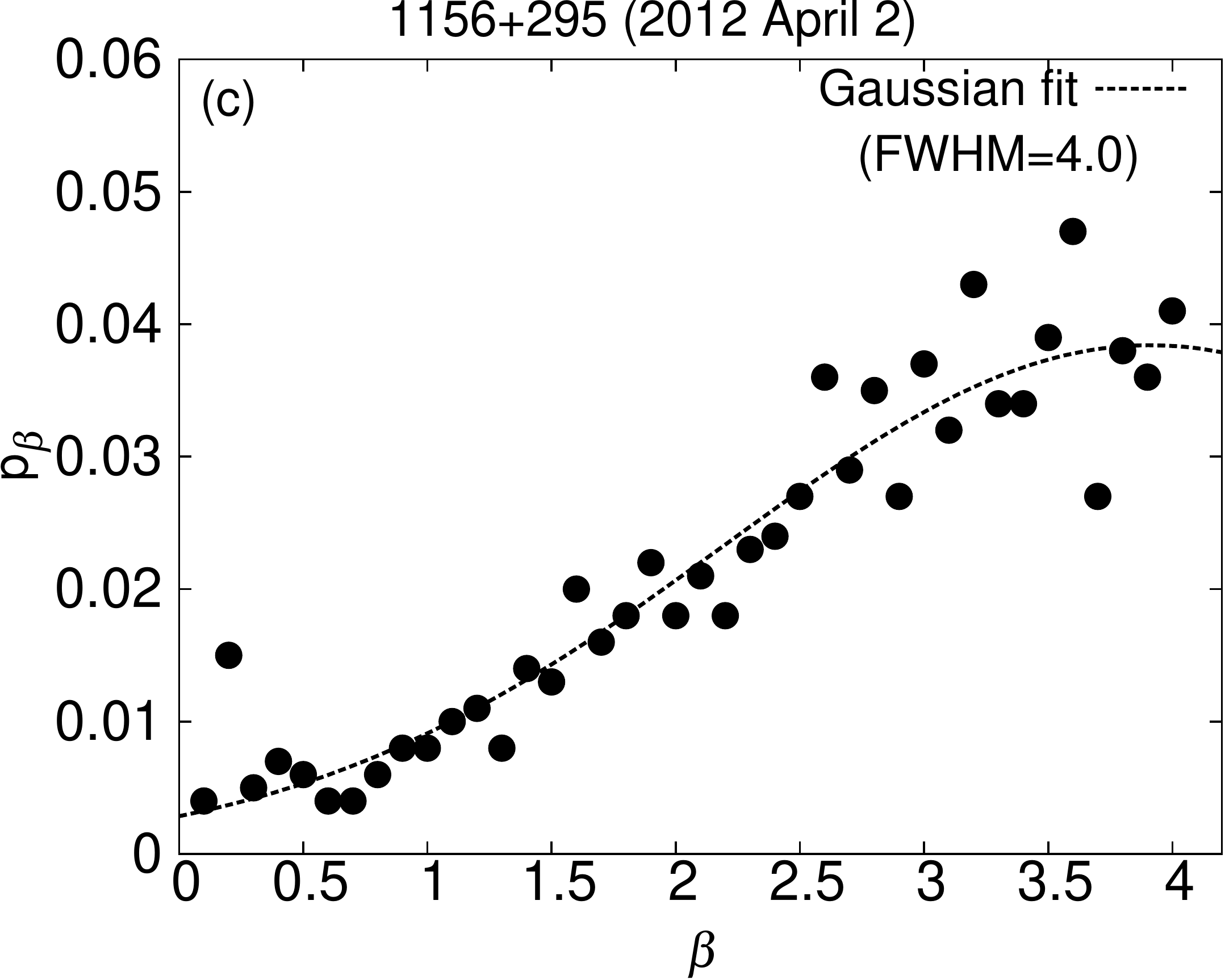}
}
\hbox{
\includegraphics[width=0.30\textwidth]{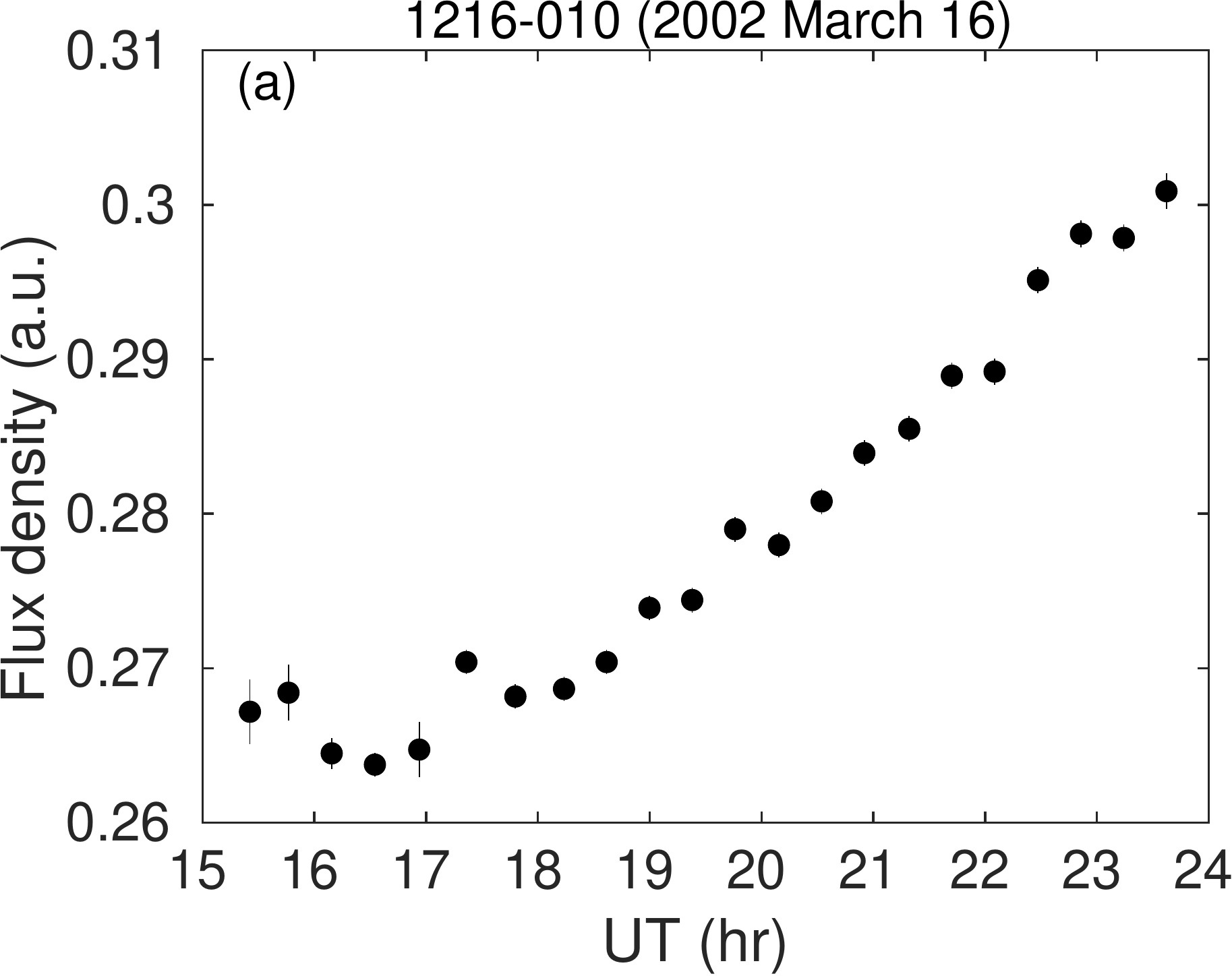}
\includegraphics[width=0.30\textwidth]{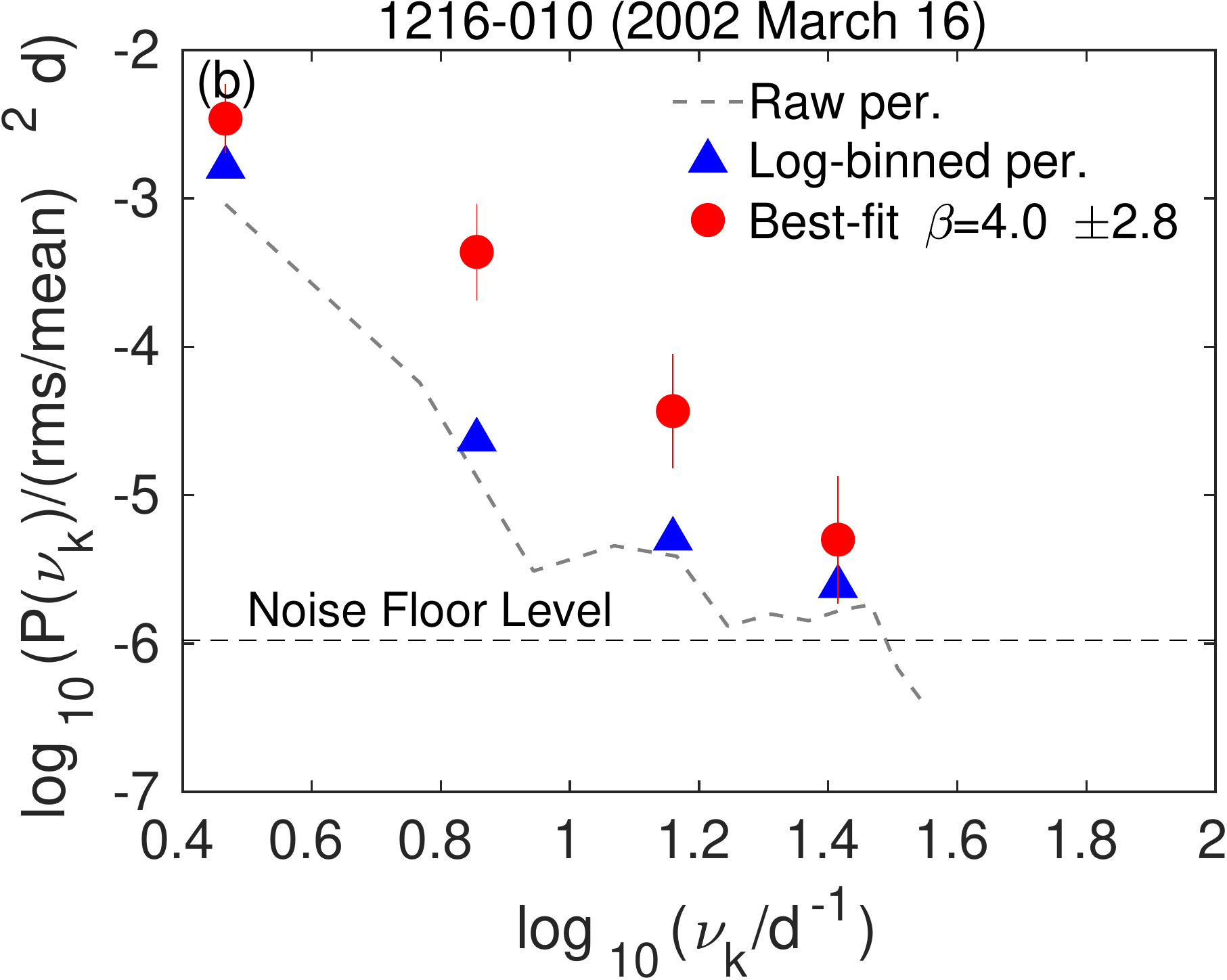}
\includegraphics[width=0.30\textwidth]{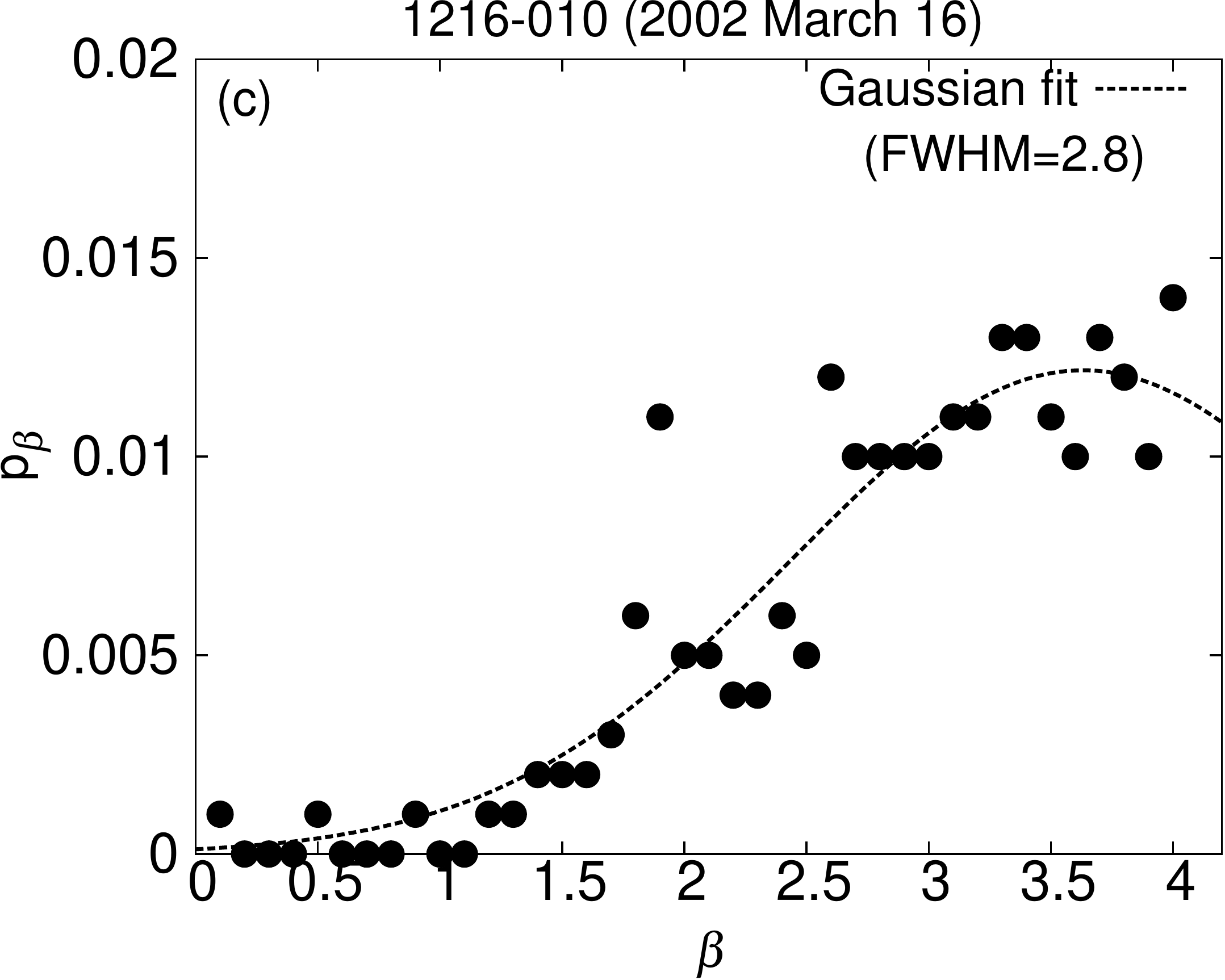}
}
\hbox{
\includegraphics[width=0.30\textwidth]{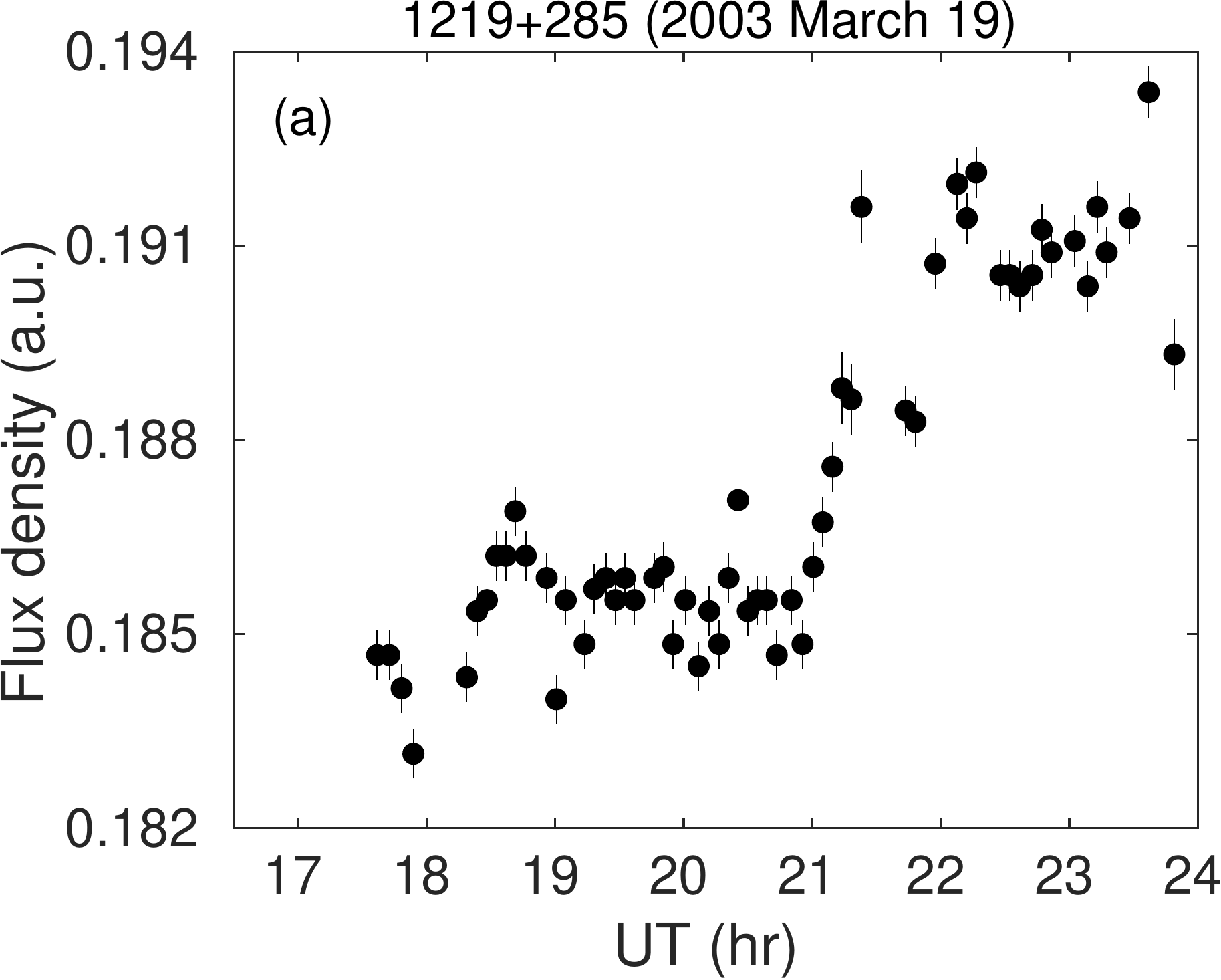}
\includegraphics[width=0.30\textwidth]{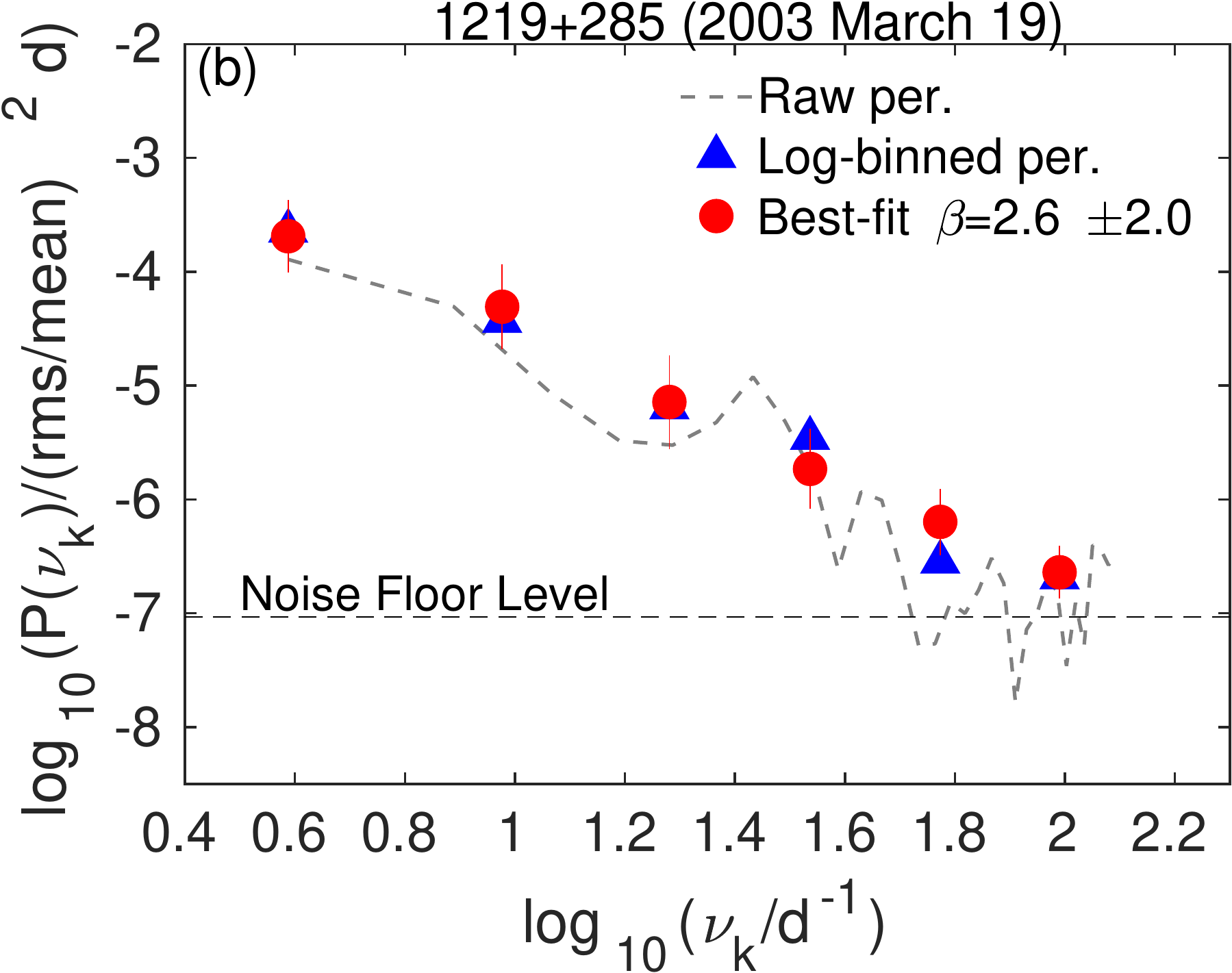}
\includegraphics[width=0.30\textwidth]{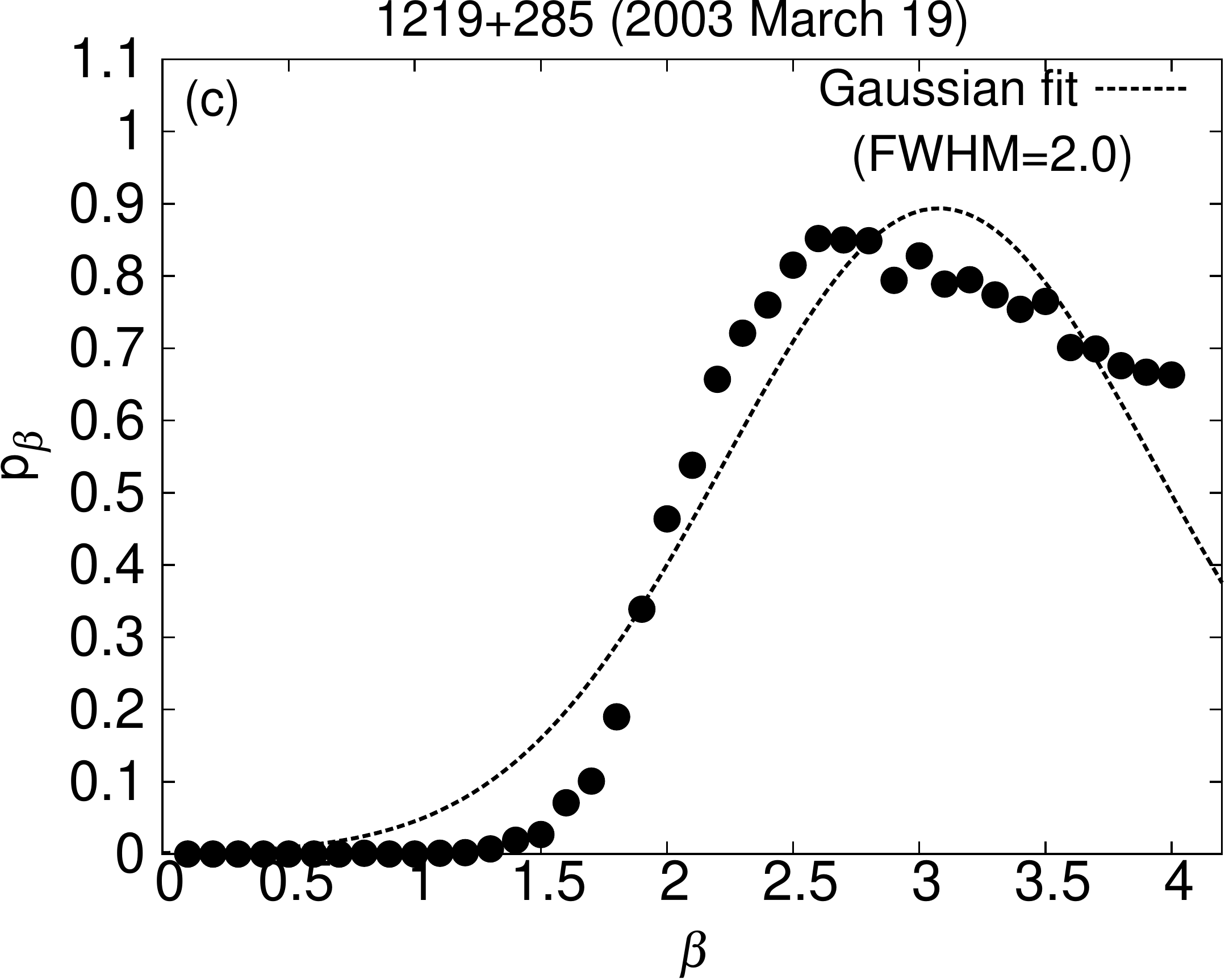}
}
\begin{minipage}{\textwidth}
\caption{(continued) }
\end{minipage}
\end{figure*}

\addtocounter{figure}{-1}
\begin{figure*}[ht!]
\hbox{
\includegraphics[width=0.30\textwidth]{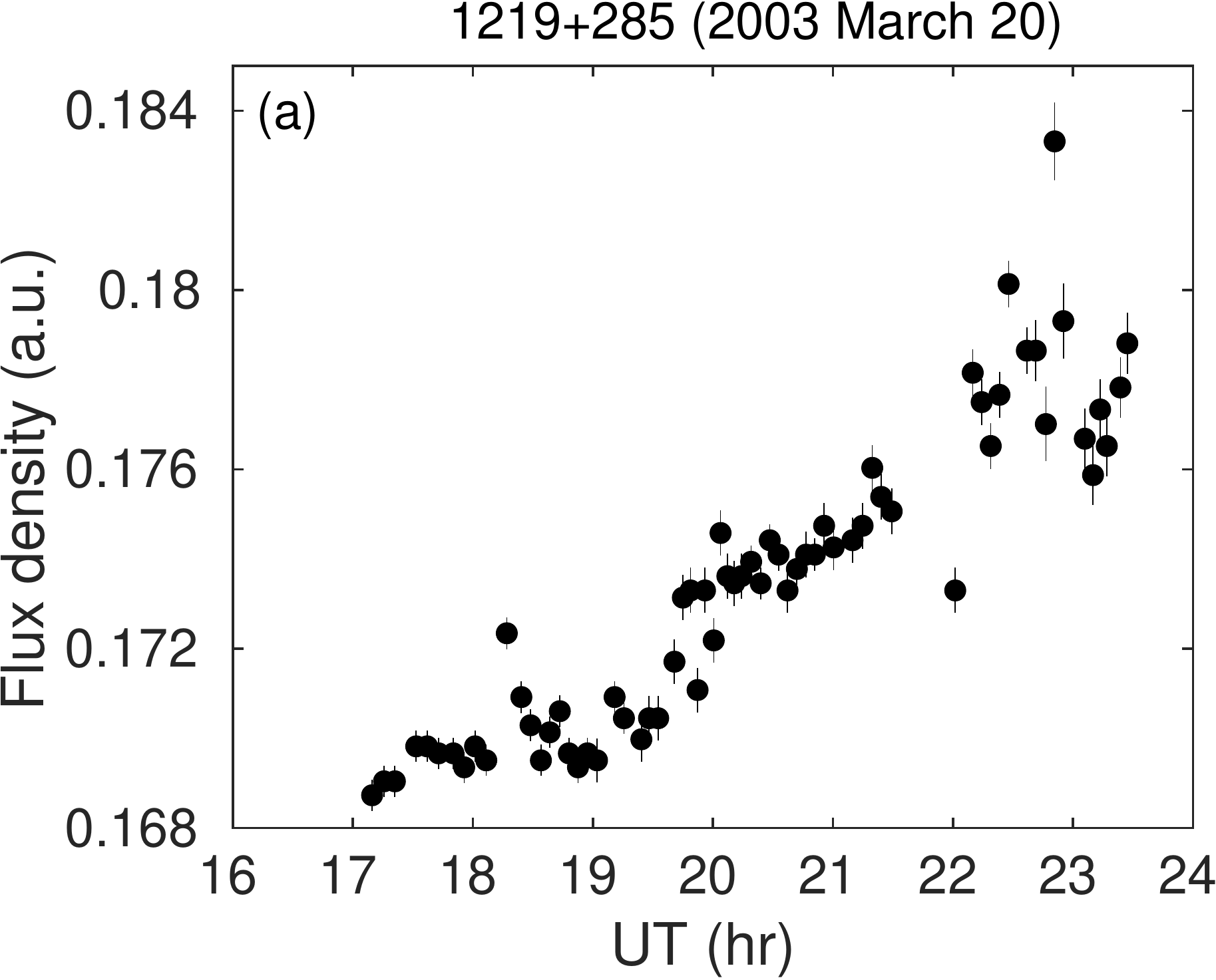}
\includegraphics[width=0.30\textwidth]{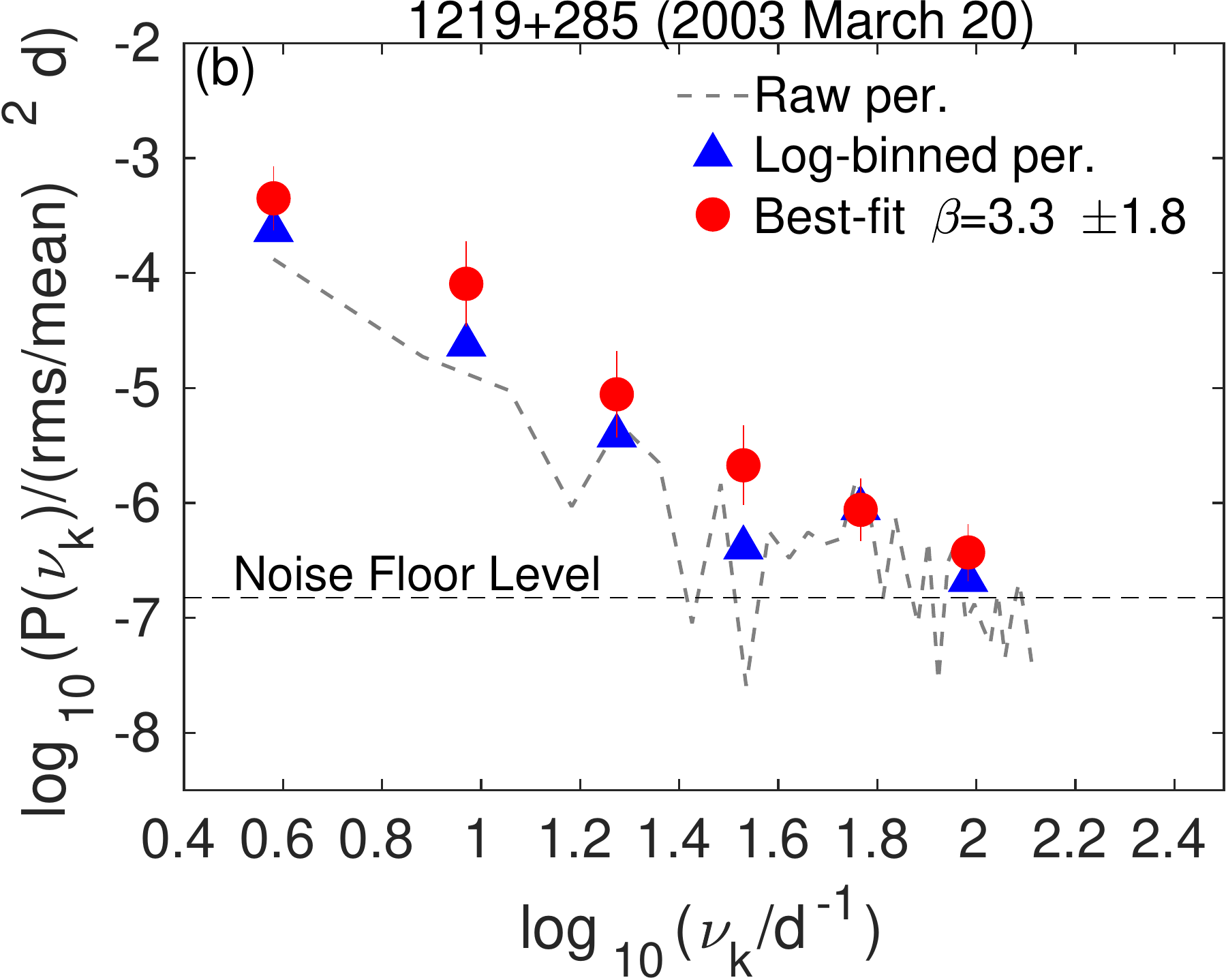}
\includegraphics[width=0.30\textwidth]{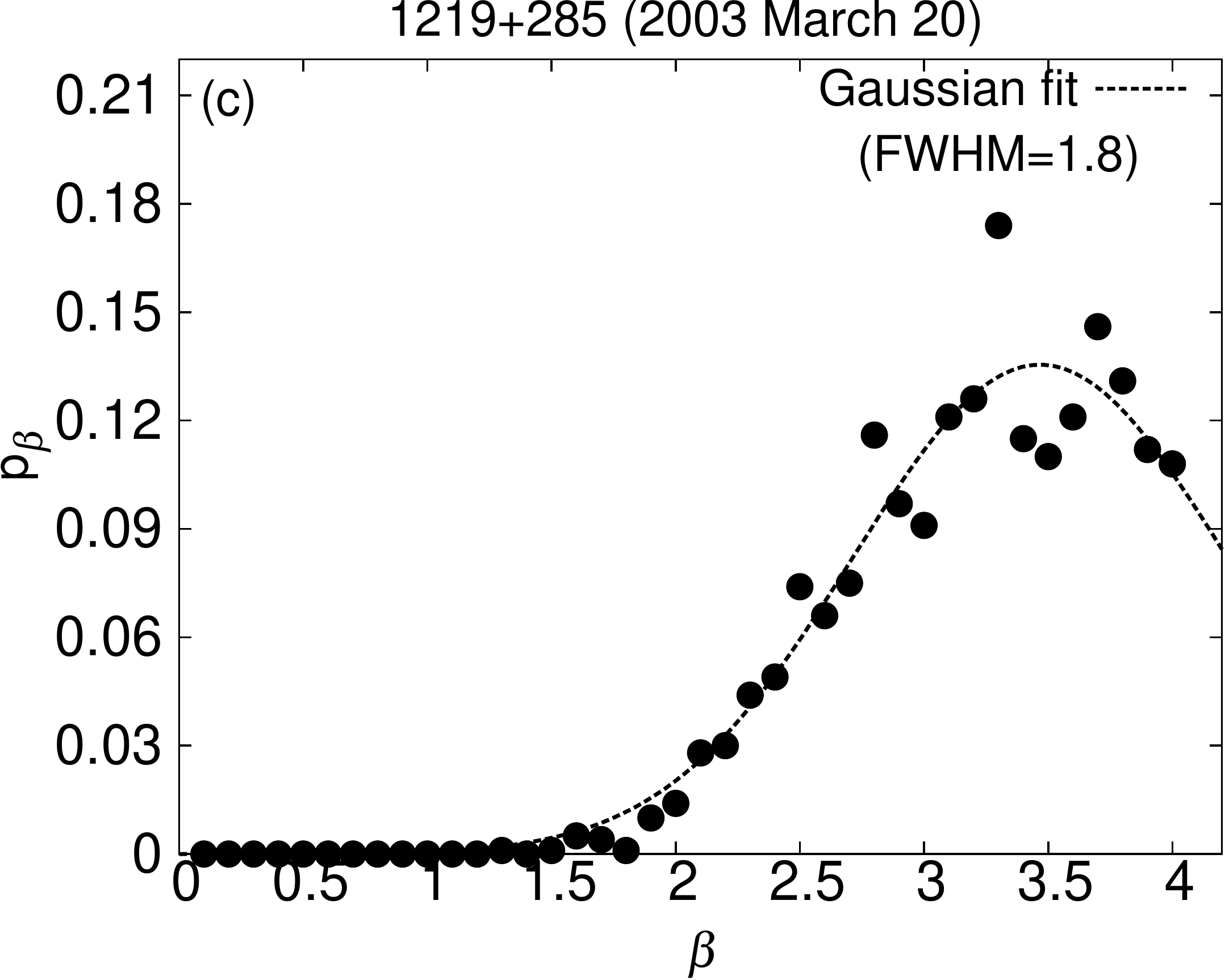}
}
\hbox{
\includegraphics[width=0.30\textwidth]{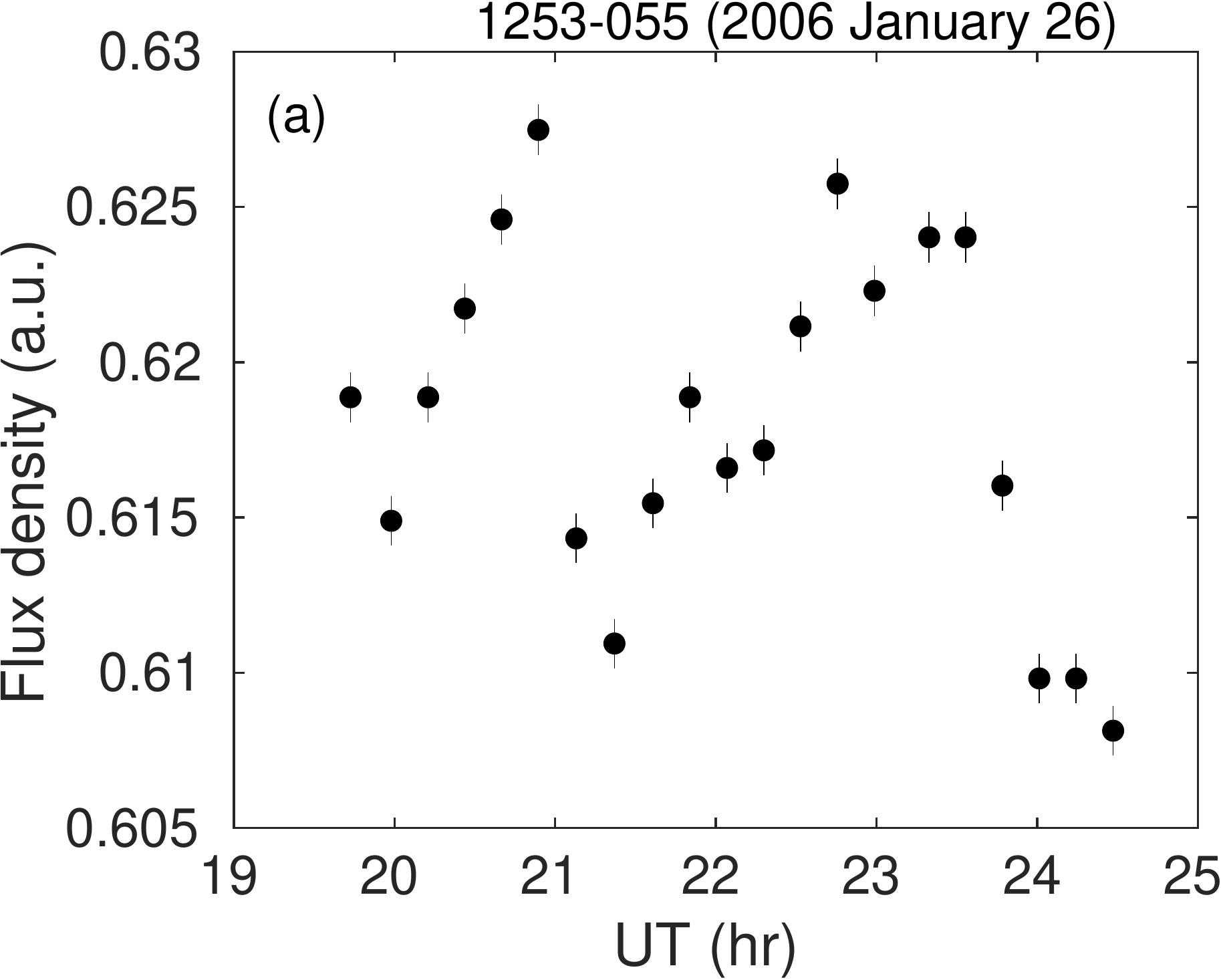}
\includegraphics[width=0.30\textwidth]{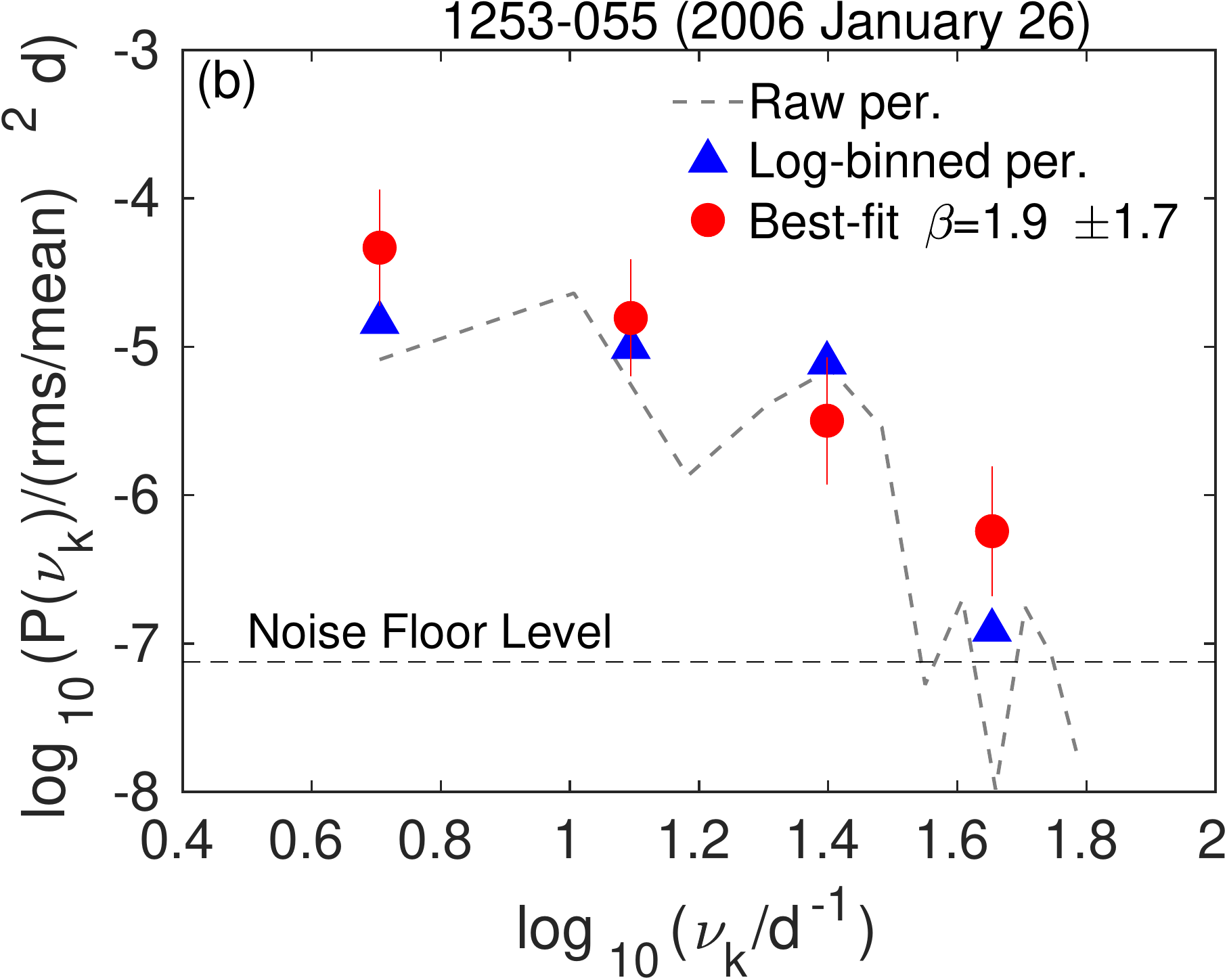}
\includegraphics[width=0.30\textwidth]{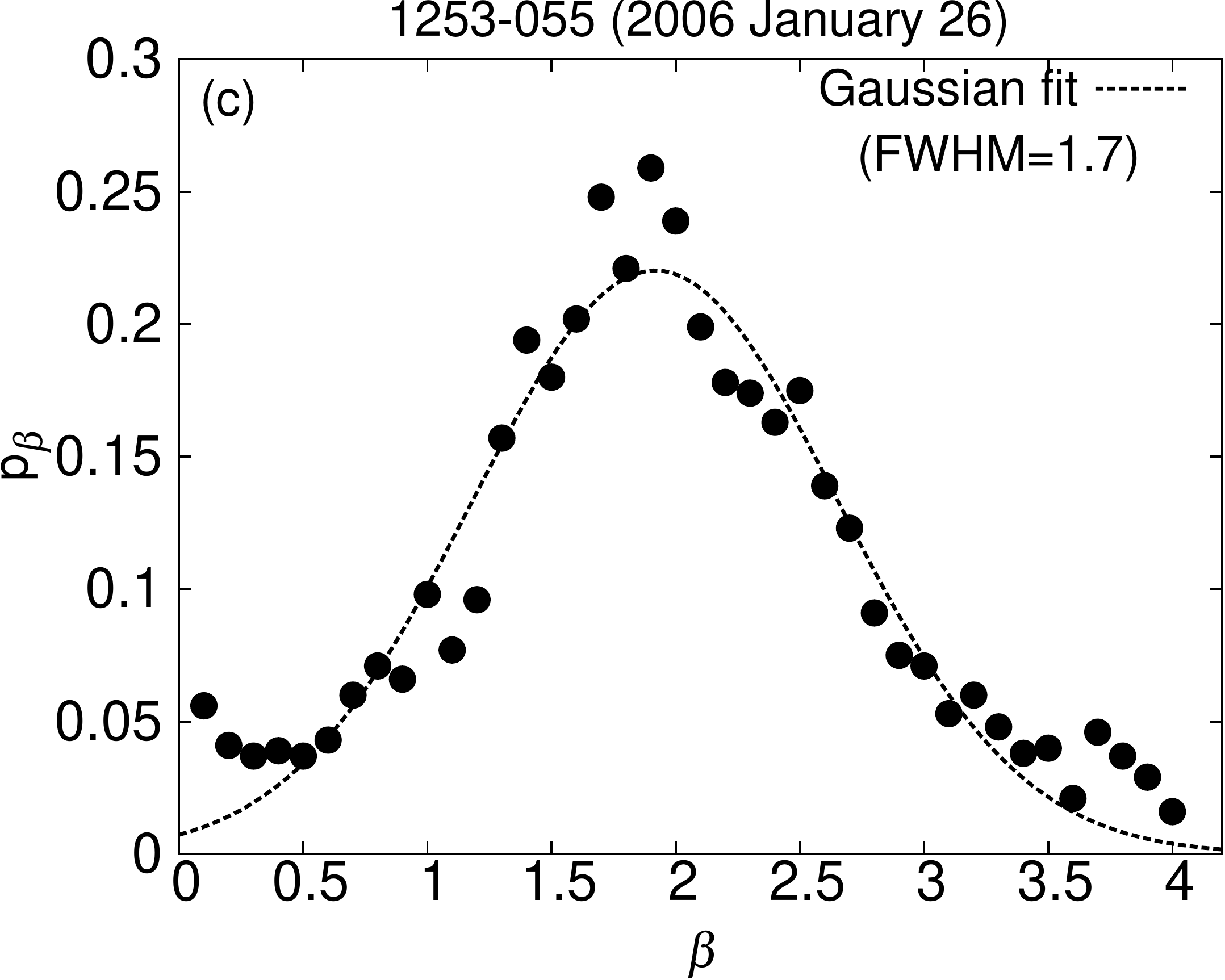}
}
\hbox{
\includegraphics[width=0.30\textwidth]{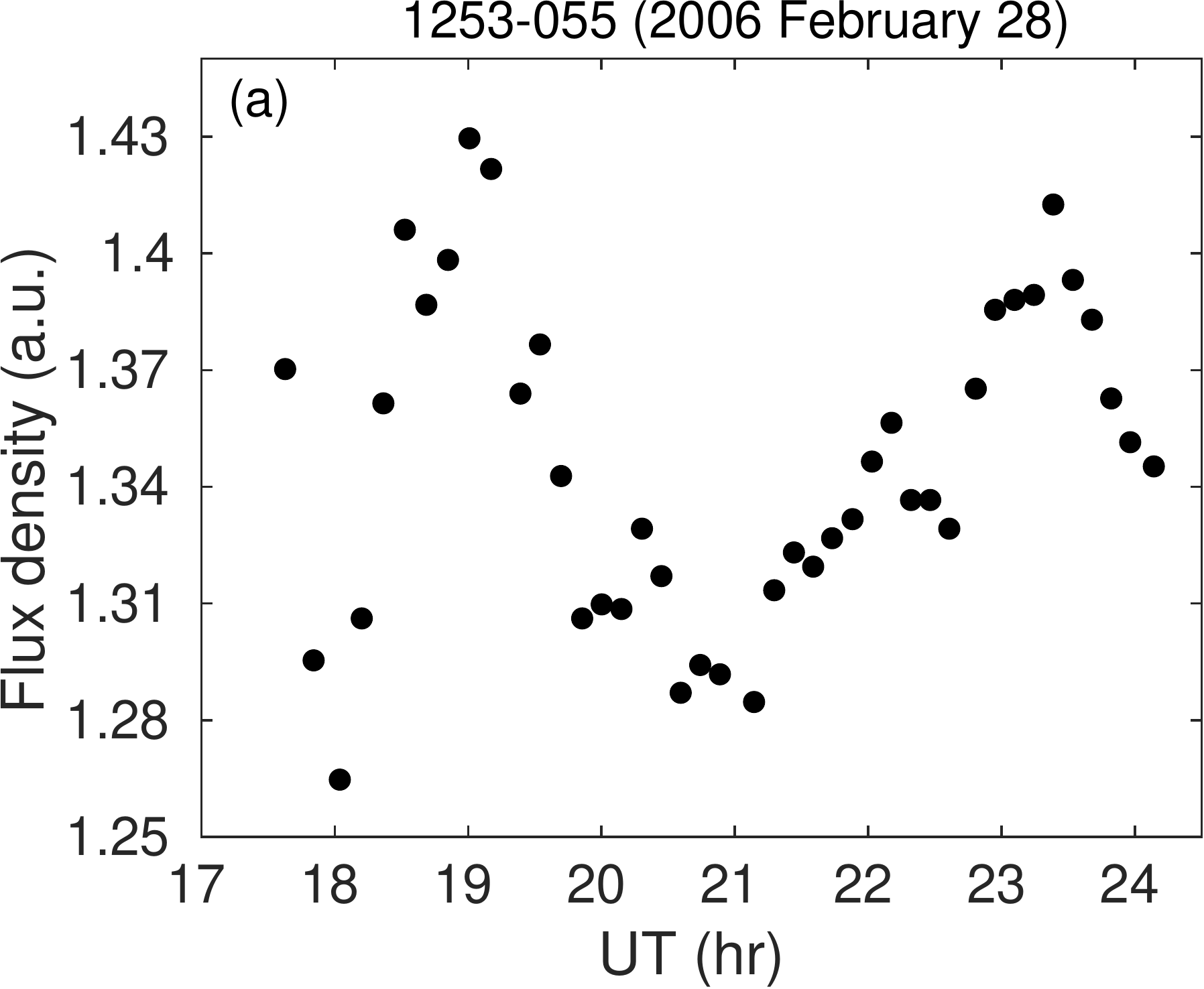}
\includegraphics[width=0.30\textwidth]{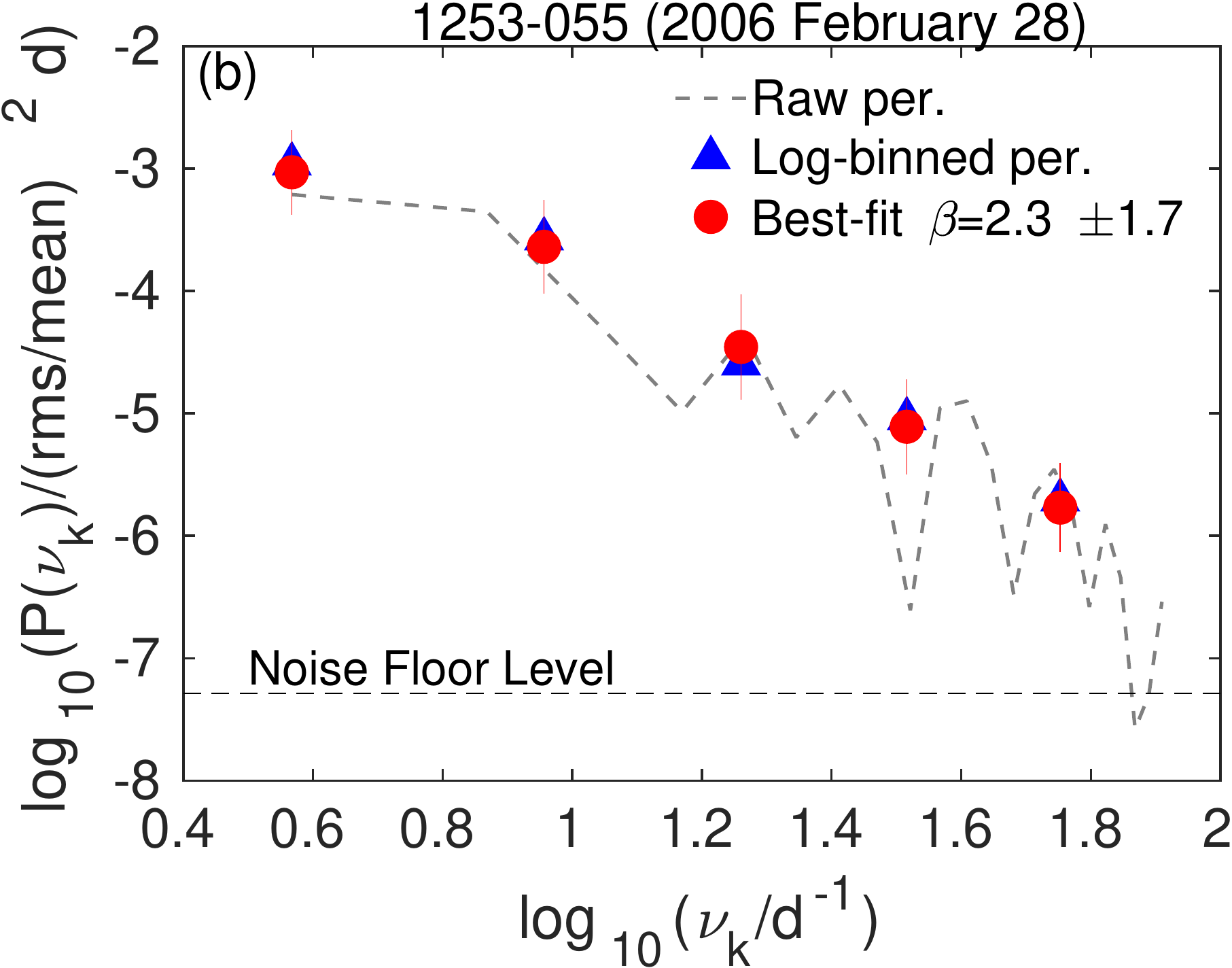}
\includegraphics[width=0.30\textwidth]{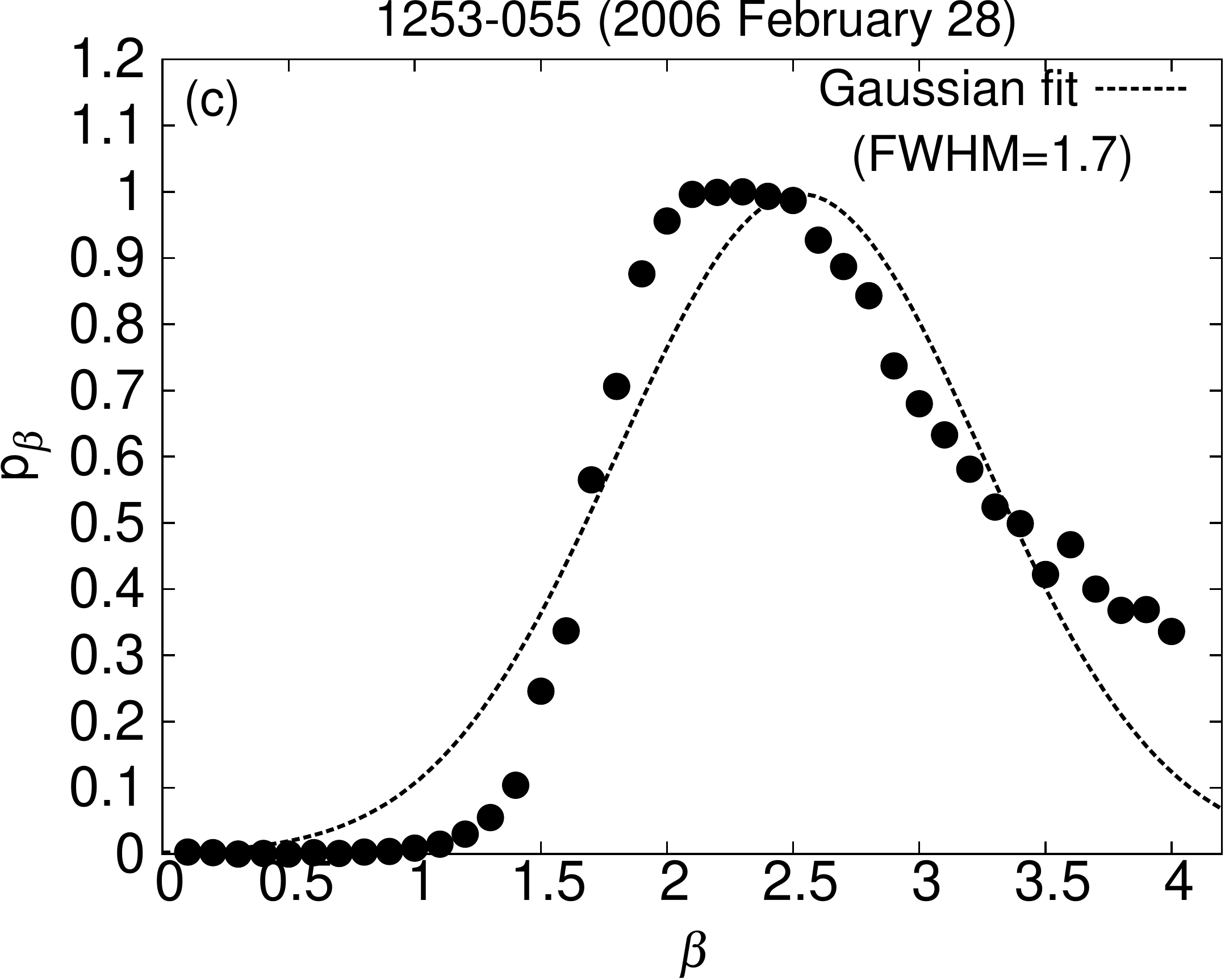}
}
\hbox{
\includegraphics[width=0.30\textwidth]{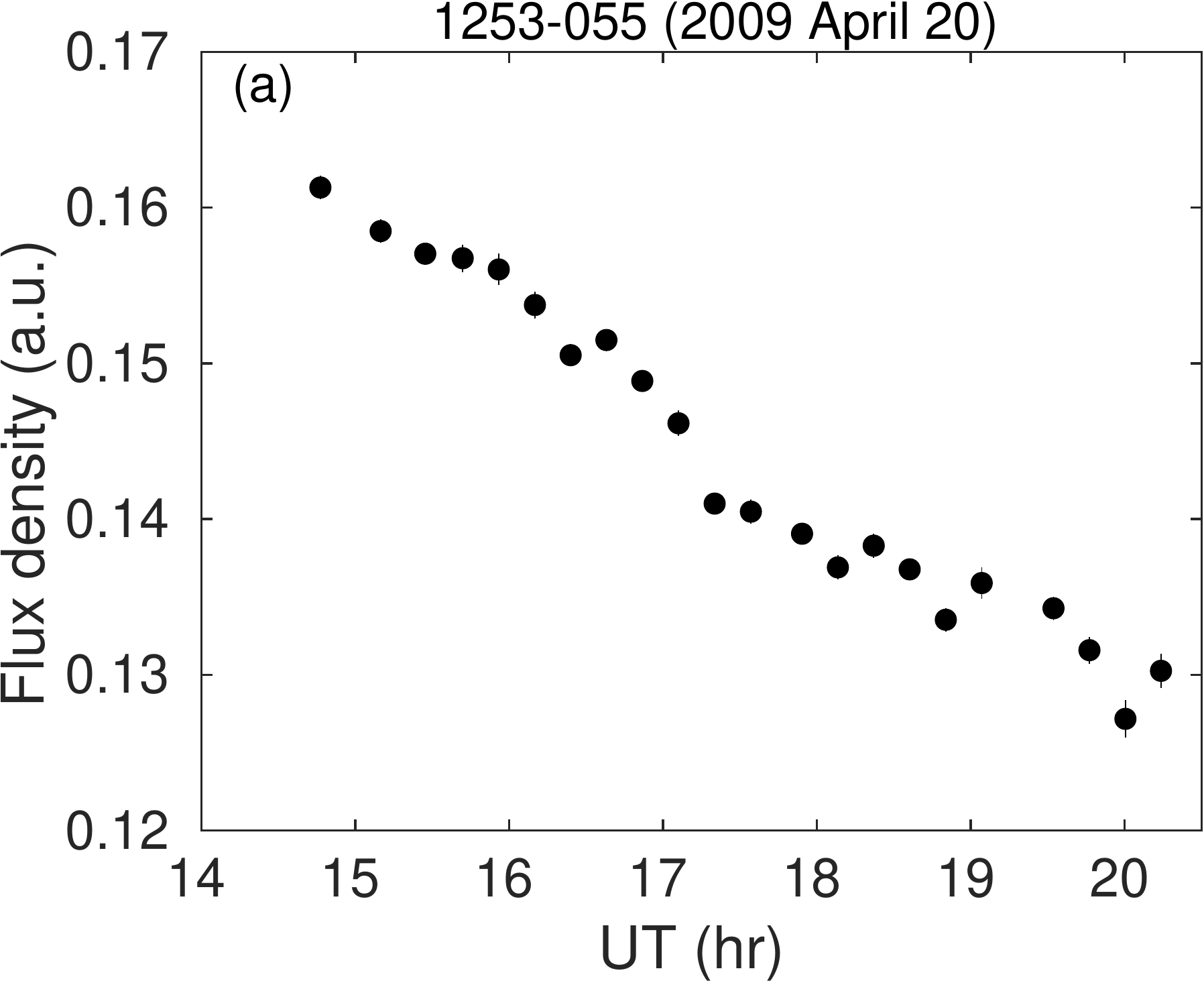}
\includegraphics[width=0.30\textwidth]{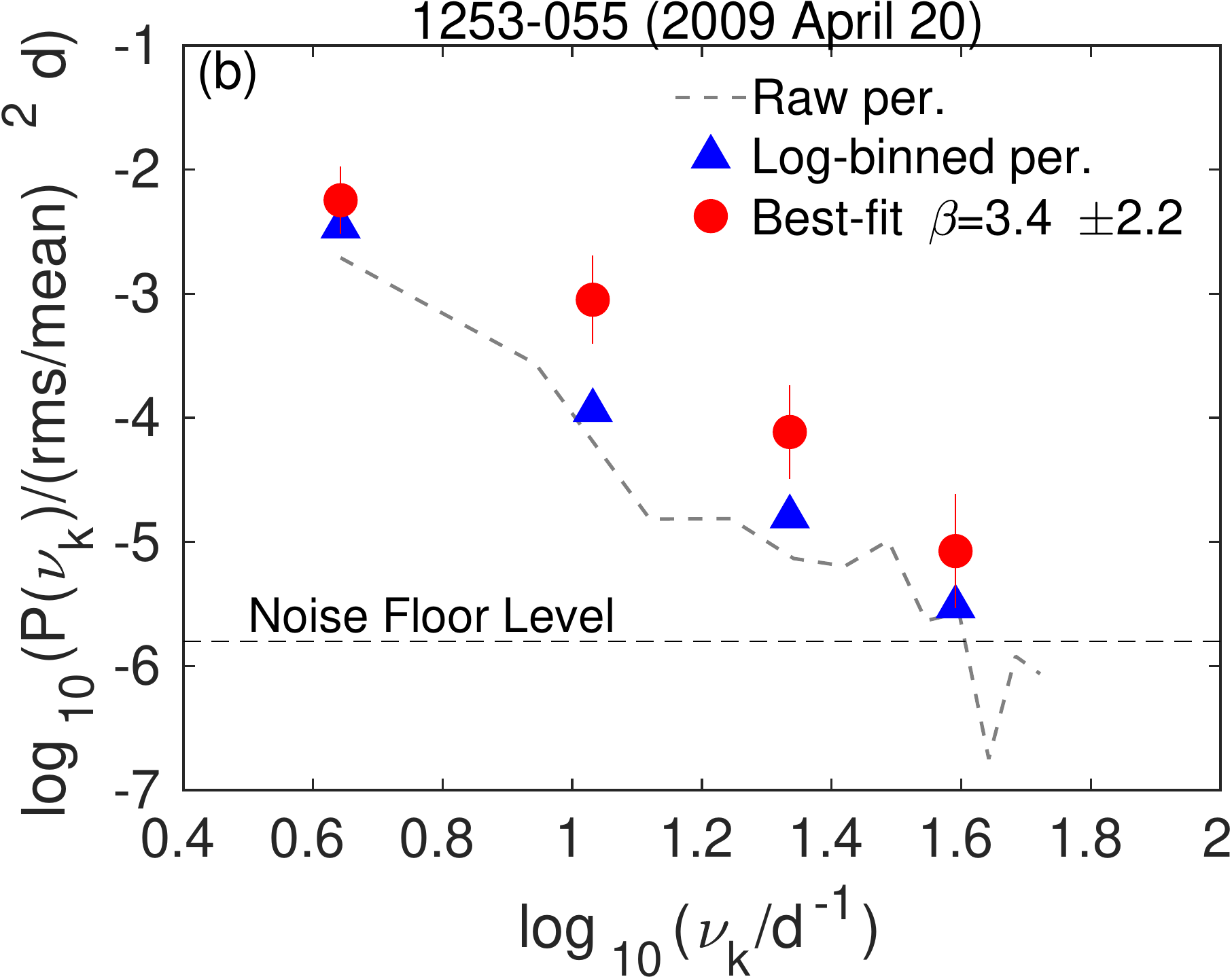}
\includegraphics[width=0.30\textwidth]{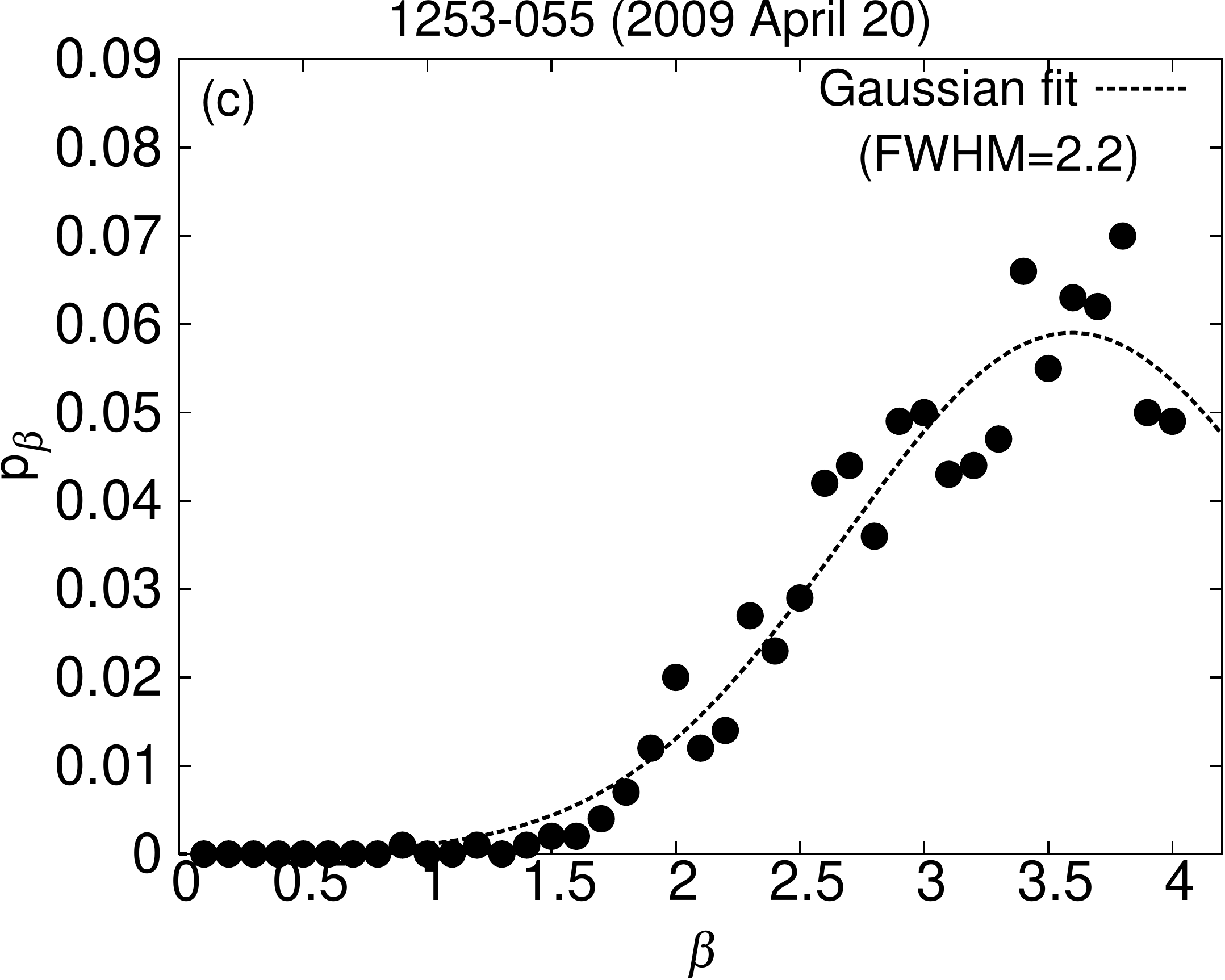}
}
\hbox{
\includegraphics[width=0.30\textwidth]{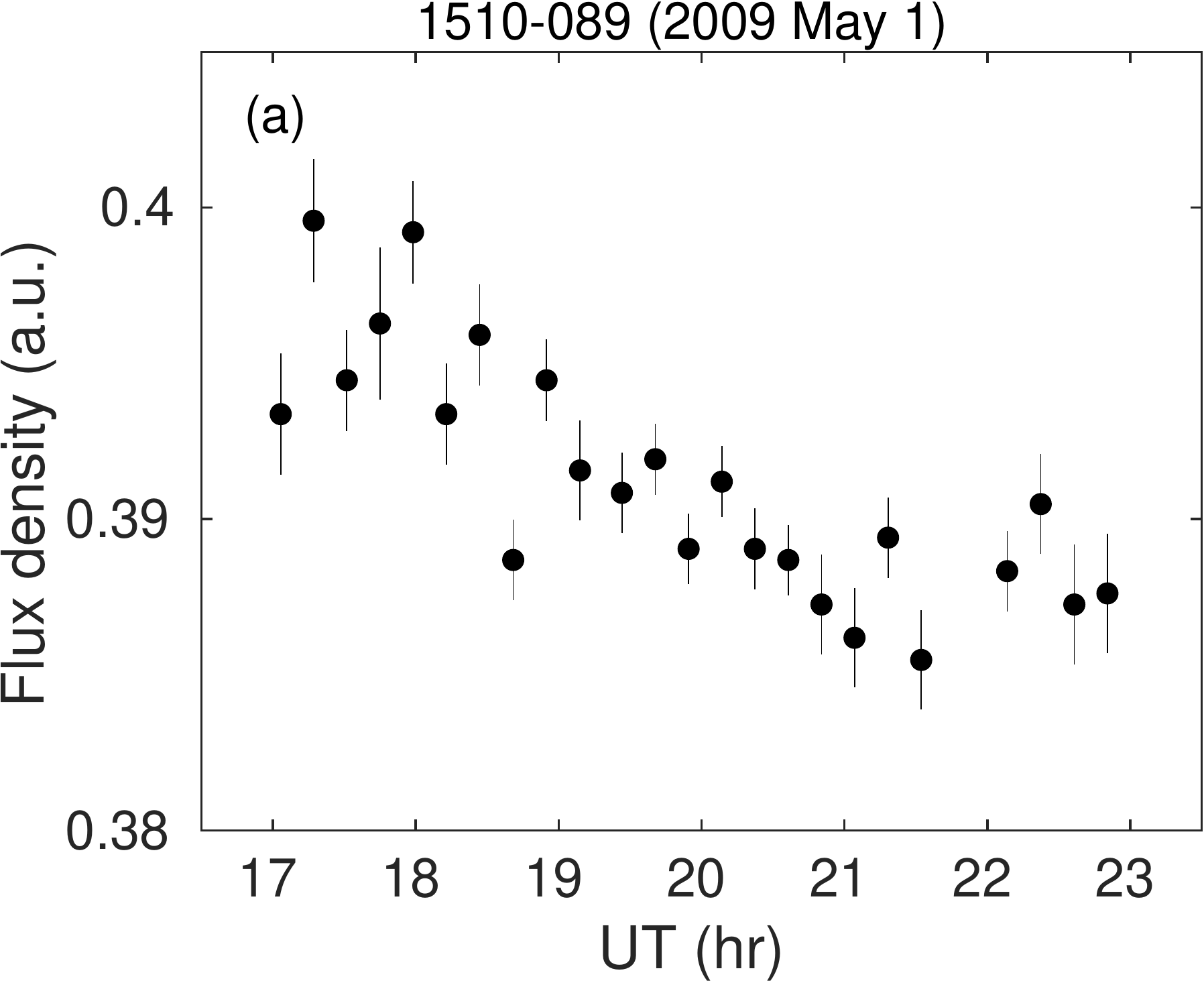}
\includegraphics[width=0.30\textwidth]{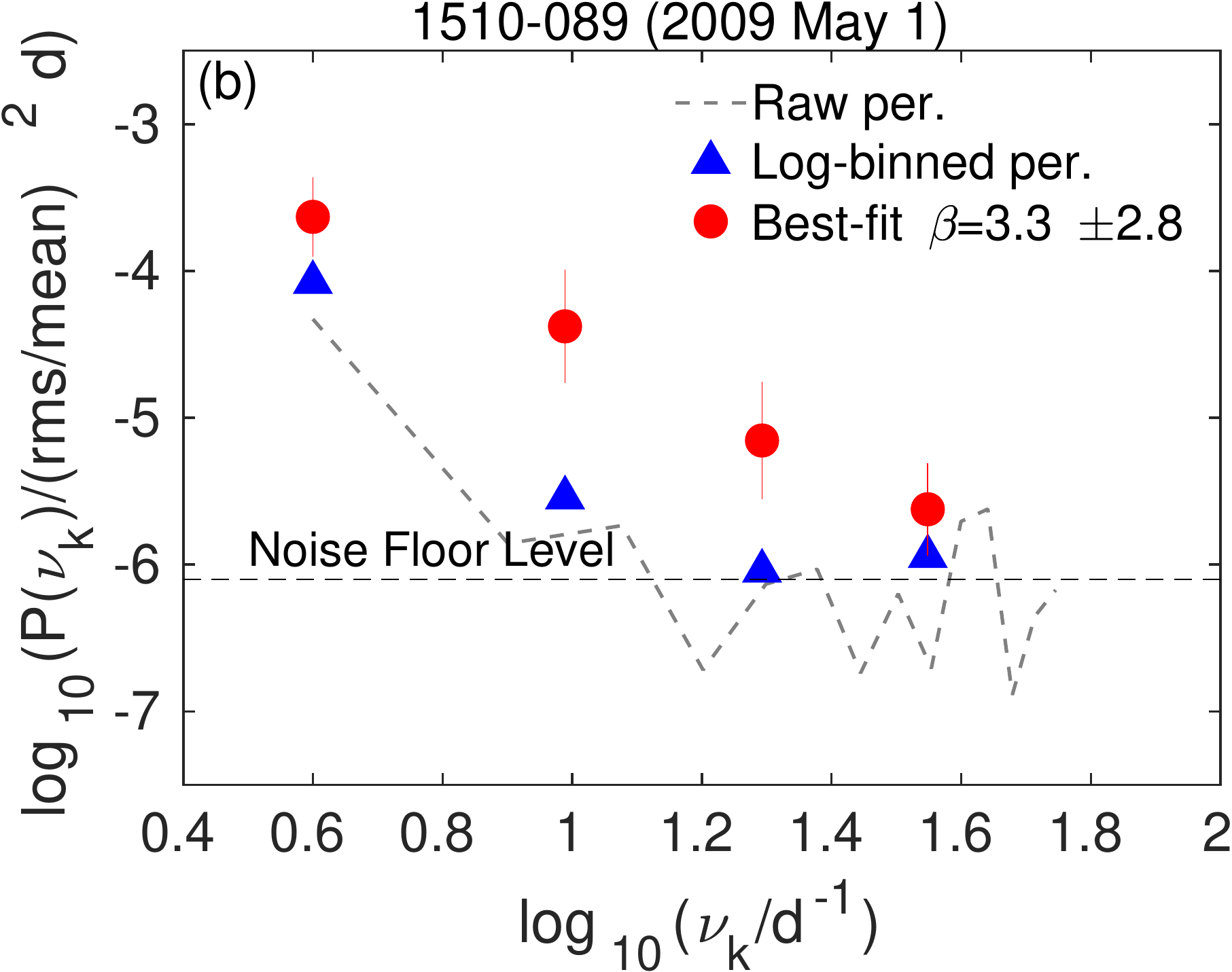}
\includegraphics[width=0.30\textwidth]{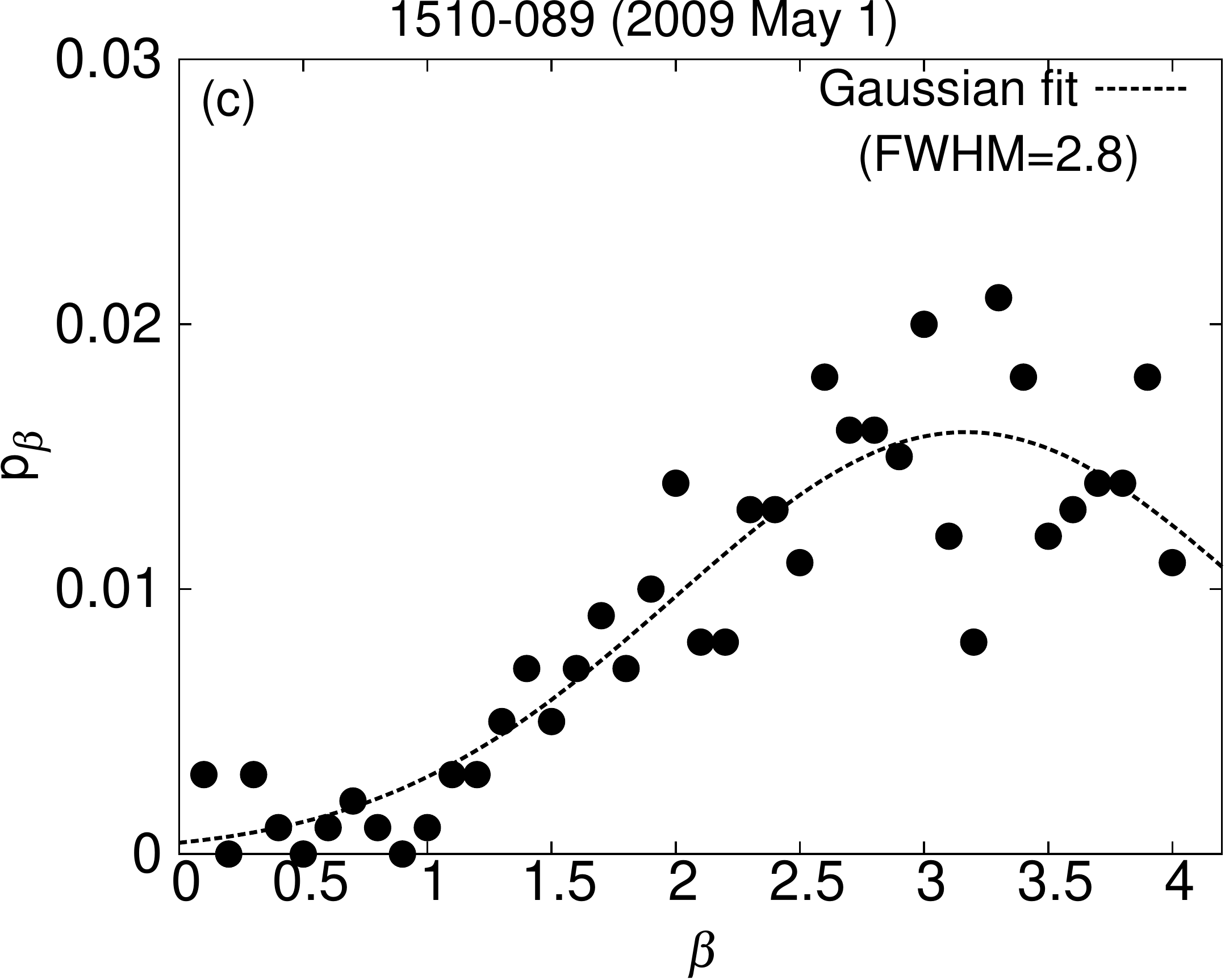}
}

\begin{minipage}{\textwidth}
\caption{(continued) }
\end{minipage}
\end{figure*}

\addtocounter{figure}{-1}
\begin{figure*}[ht!]
\hbox{
\includegraphics[width=0.30\textwidth]{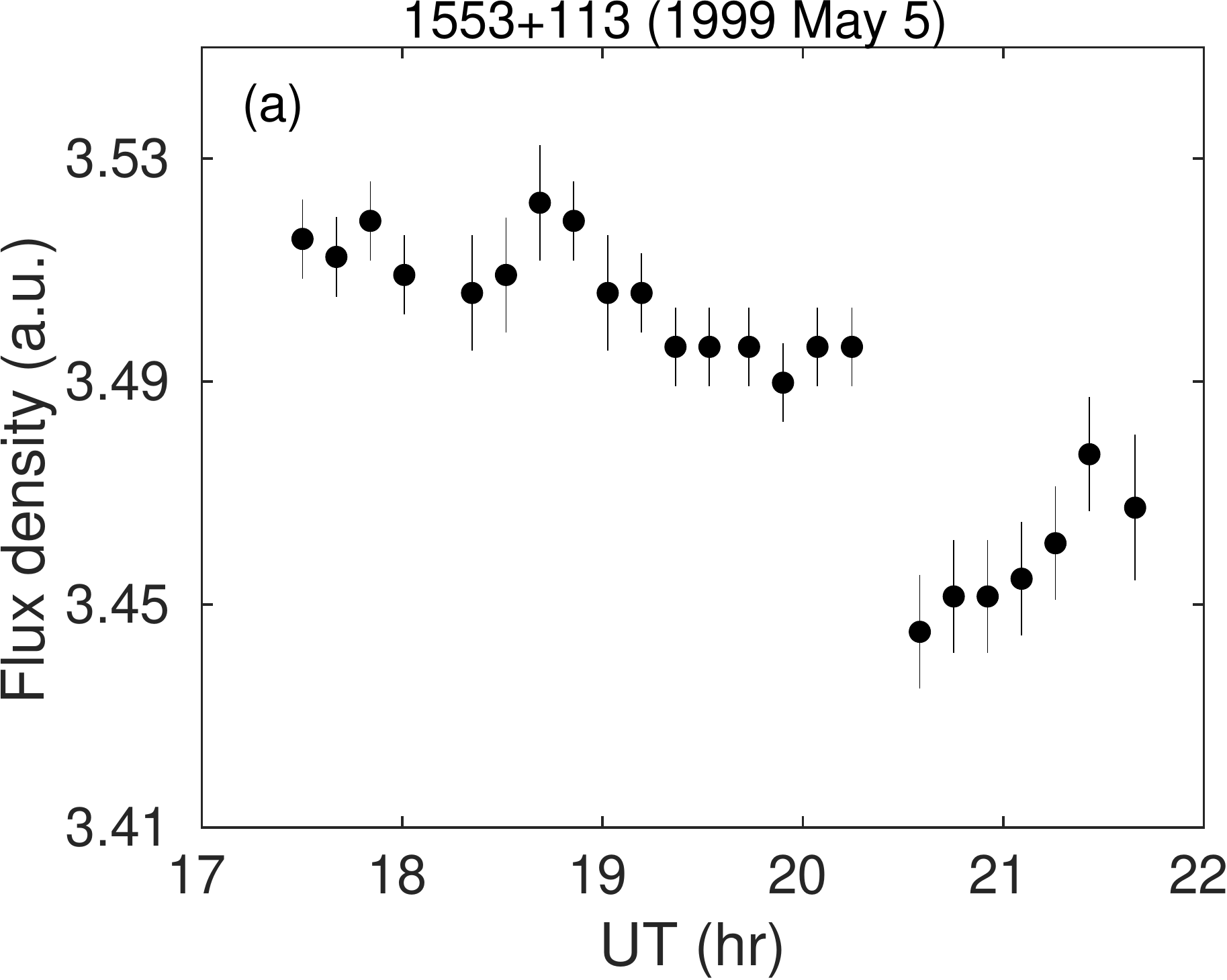}
\includegraphics[width=0.30\textwidth]{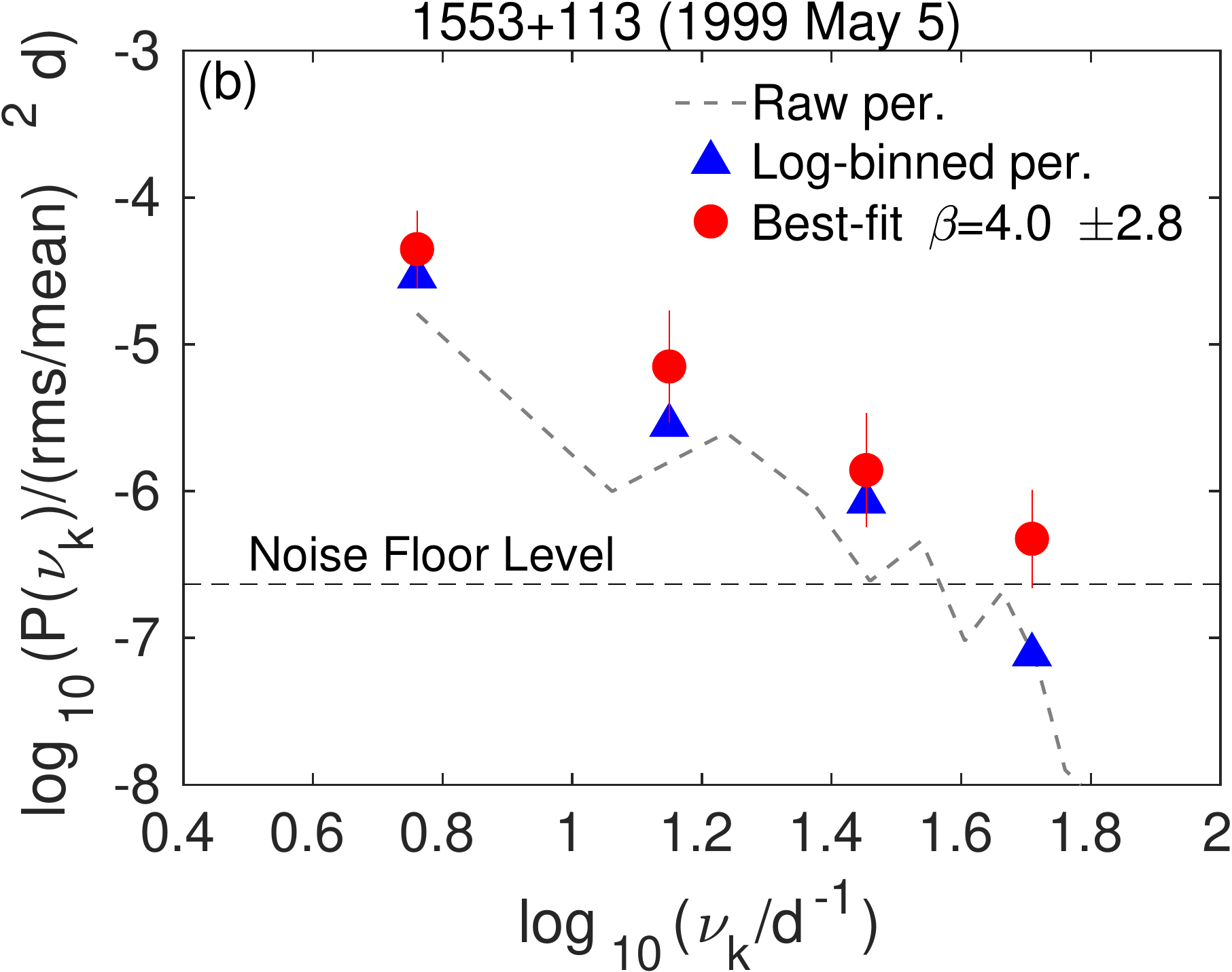}
\includegraphics[width=0.30\textwidth]{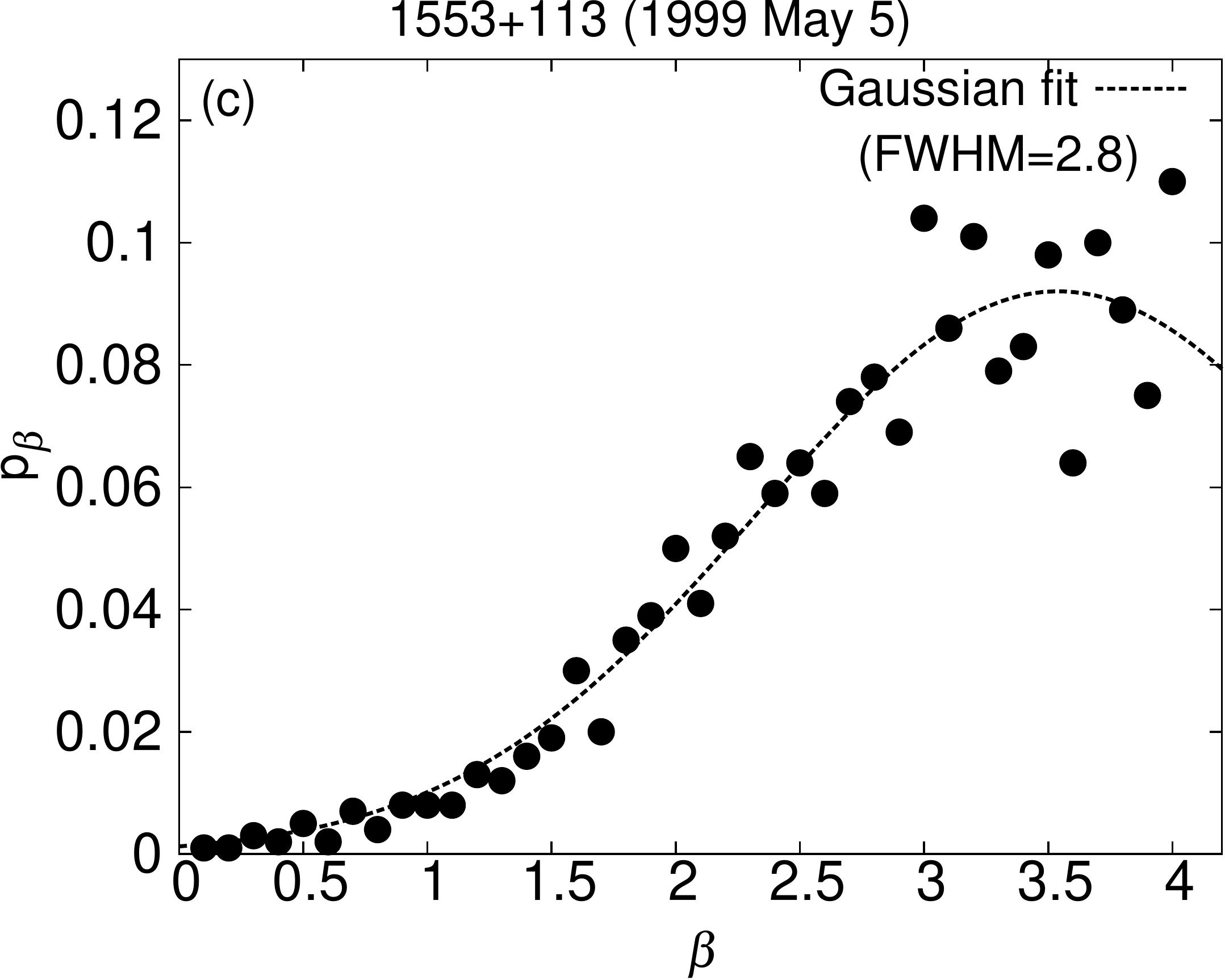}
}
\hbox{
\includegraphics[width=0.30\textwidth]{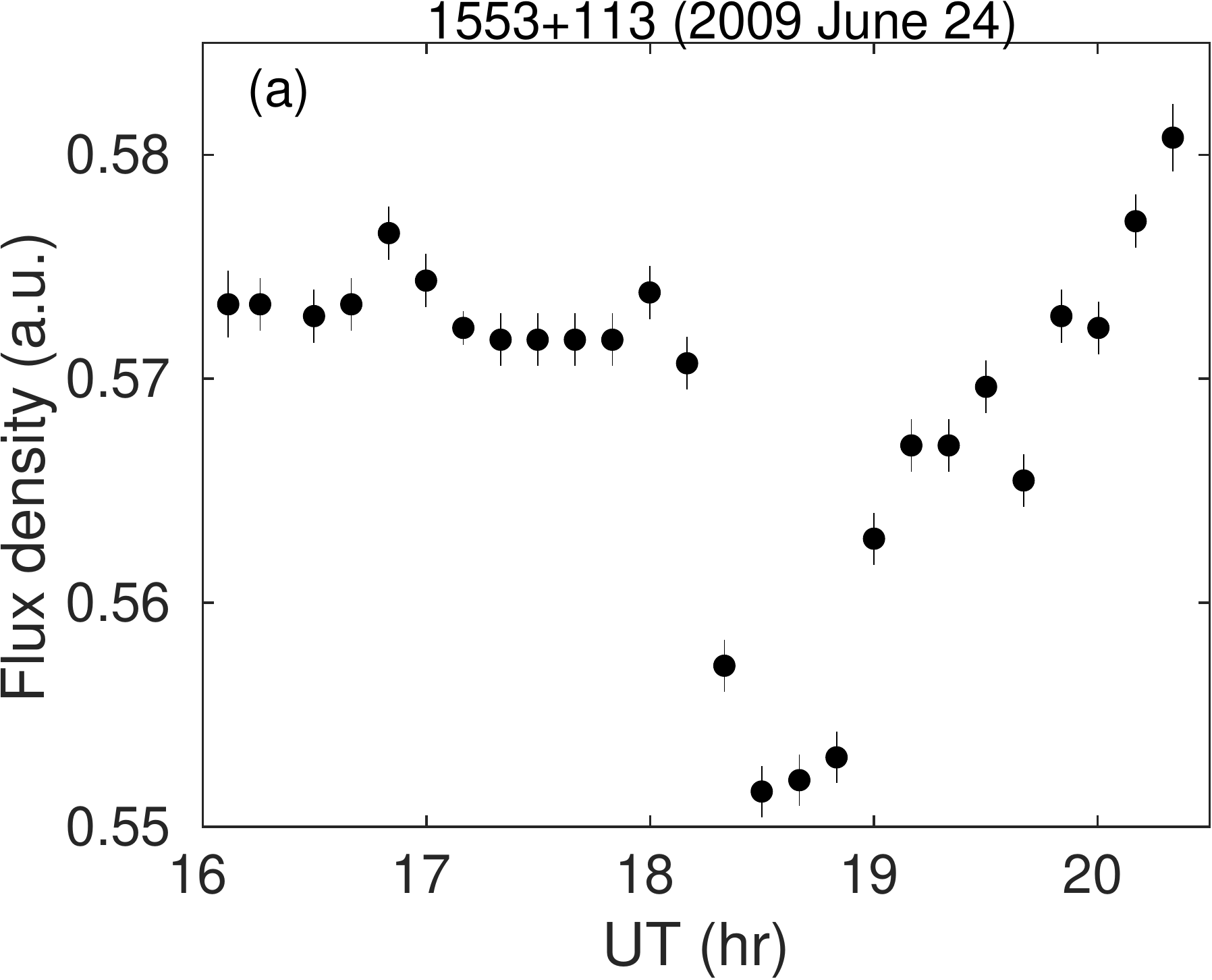}
\includegraphics[width=0.30\textwidth]{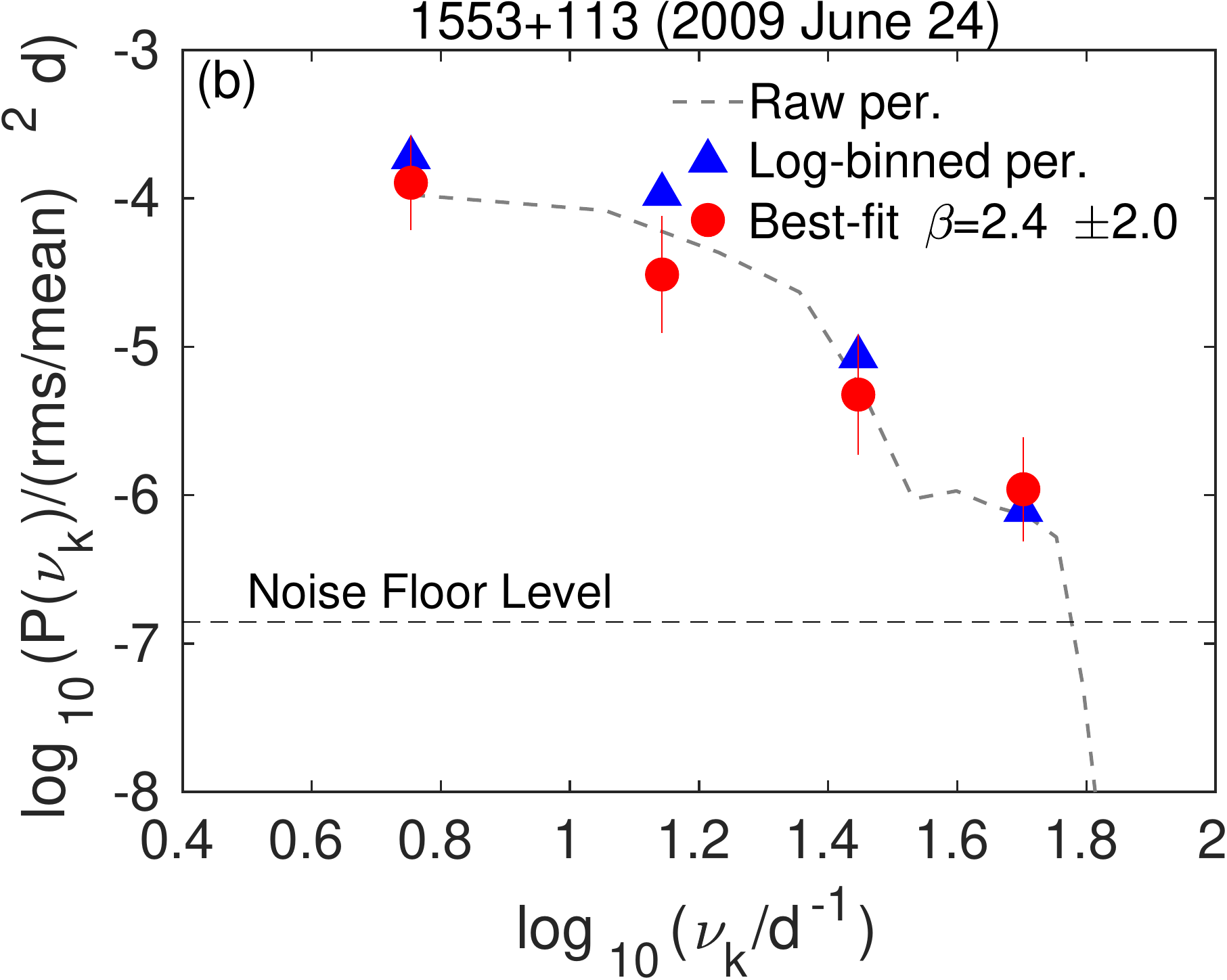}
\includegraphics[width=0.30\textwidth]{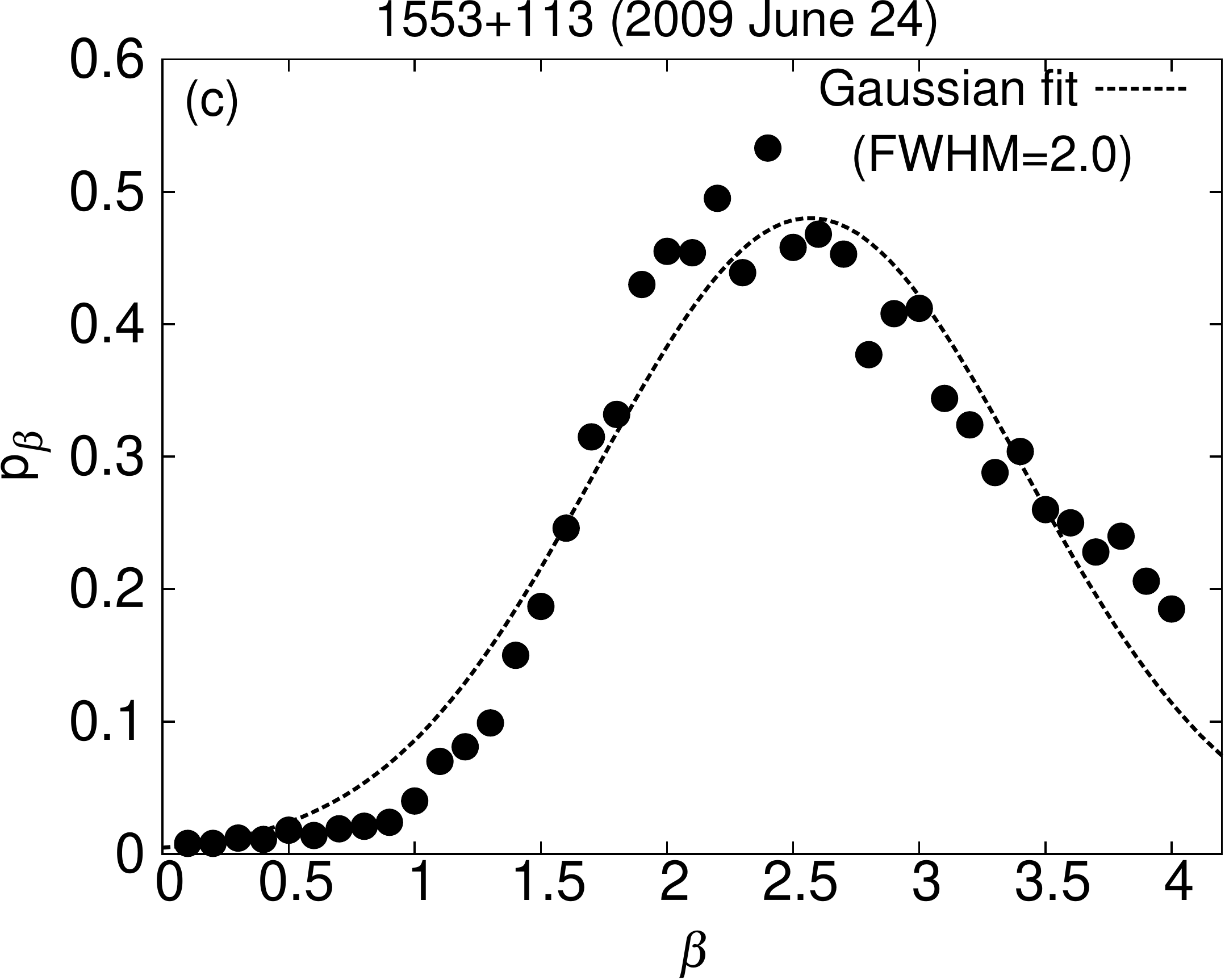}
}
\hbox{
\includegraphics[width=0.30\textwidth]{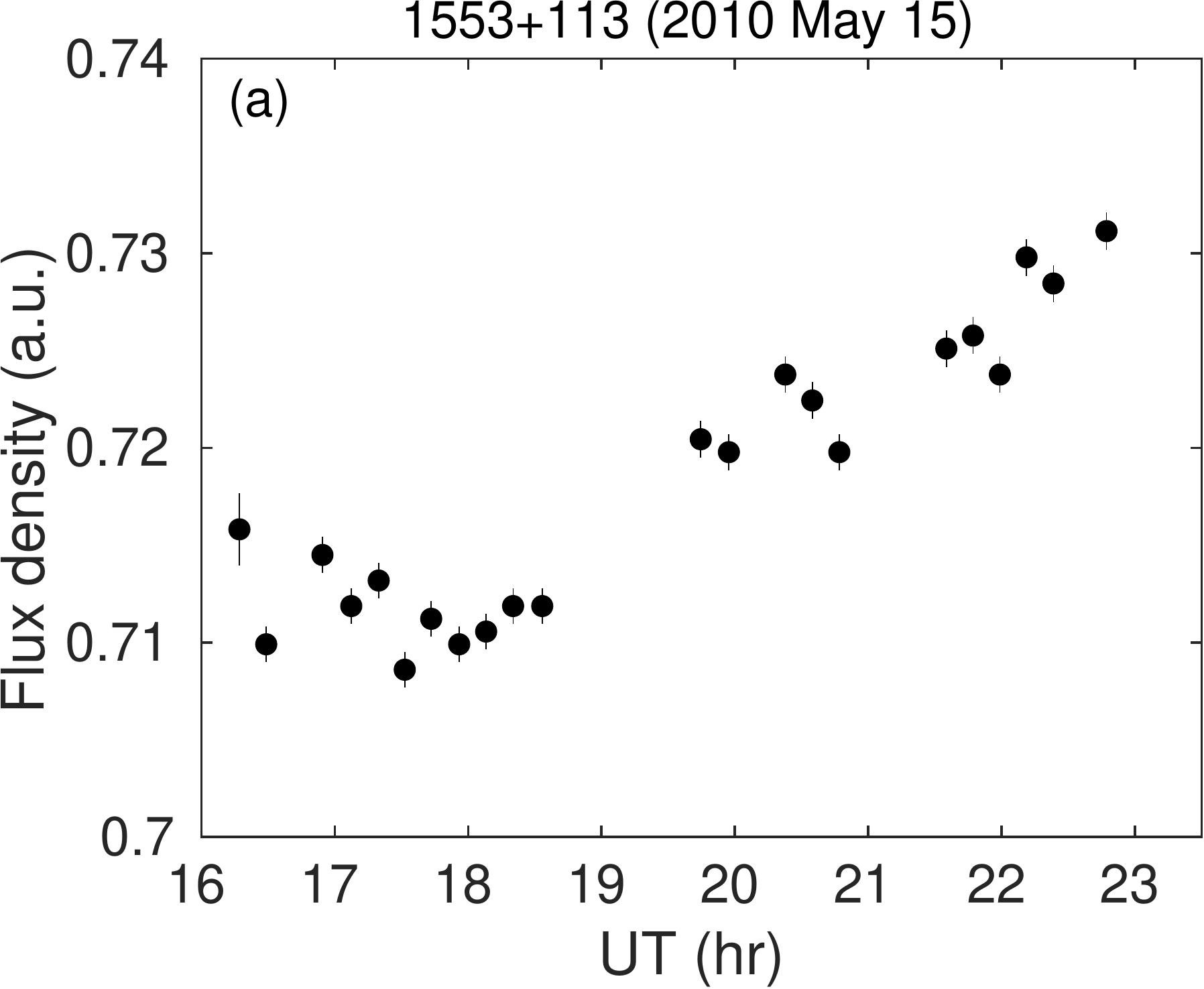}
\includegraphics[width=0.30\textwidth]{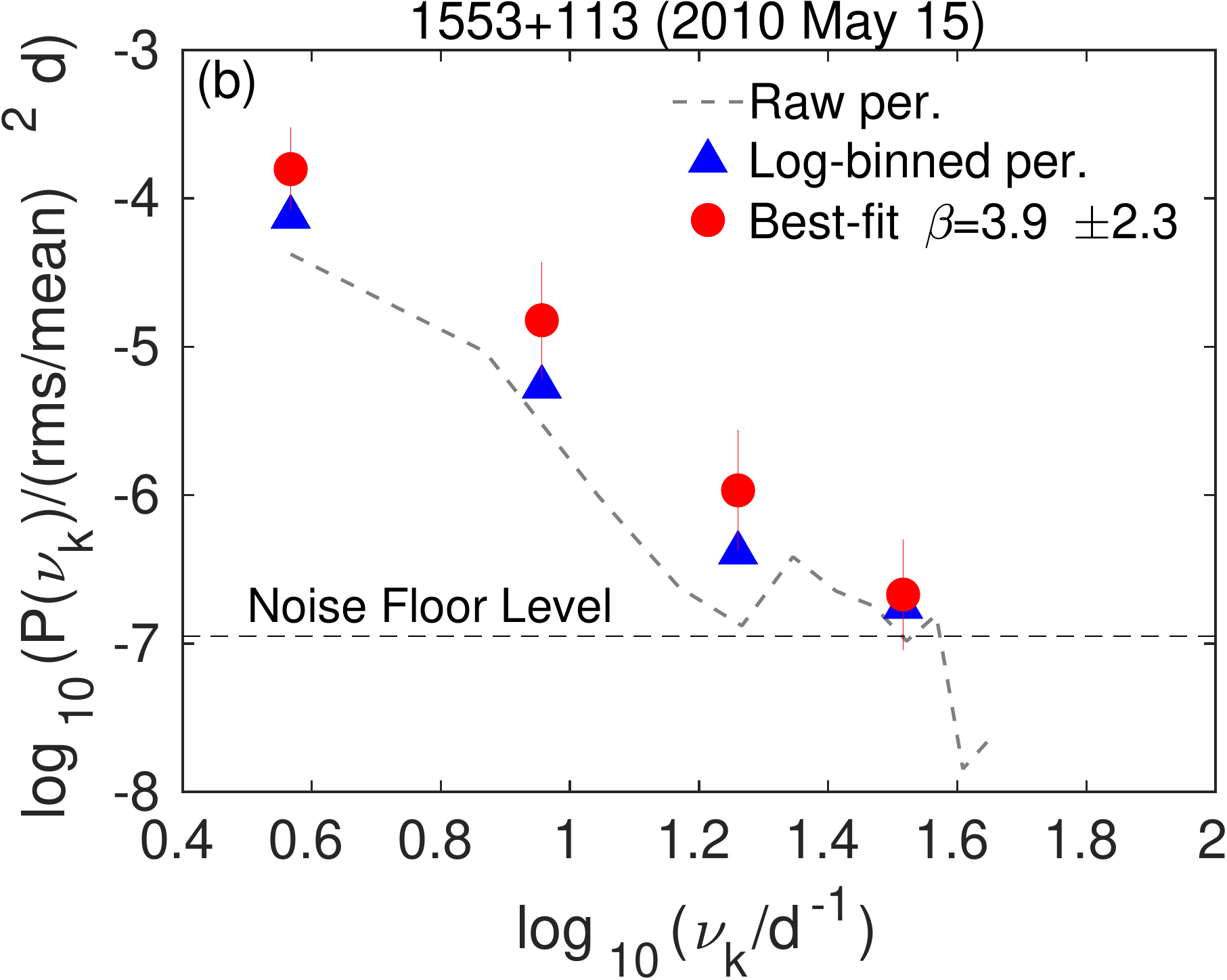}
\includegraphics[width=0.30\textwidth]{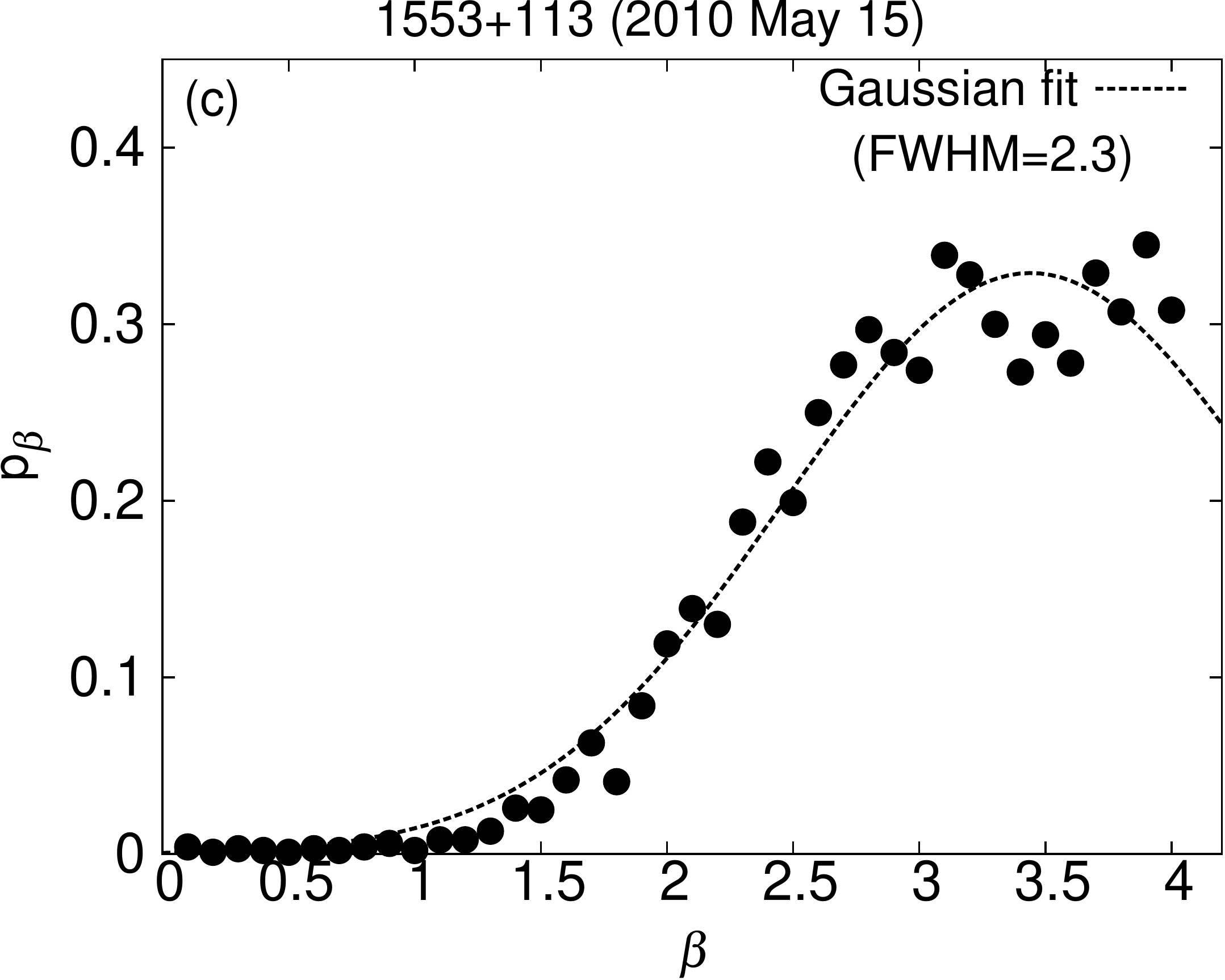}
}
\hbox{
\includegraphics[width=0.30\textwidth]{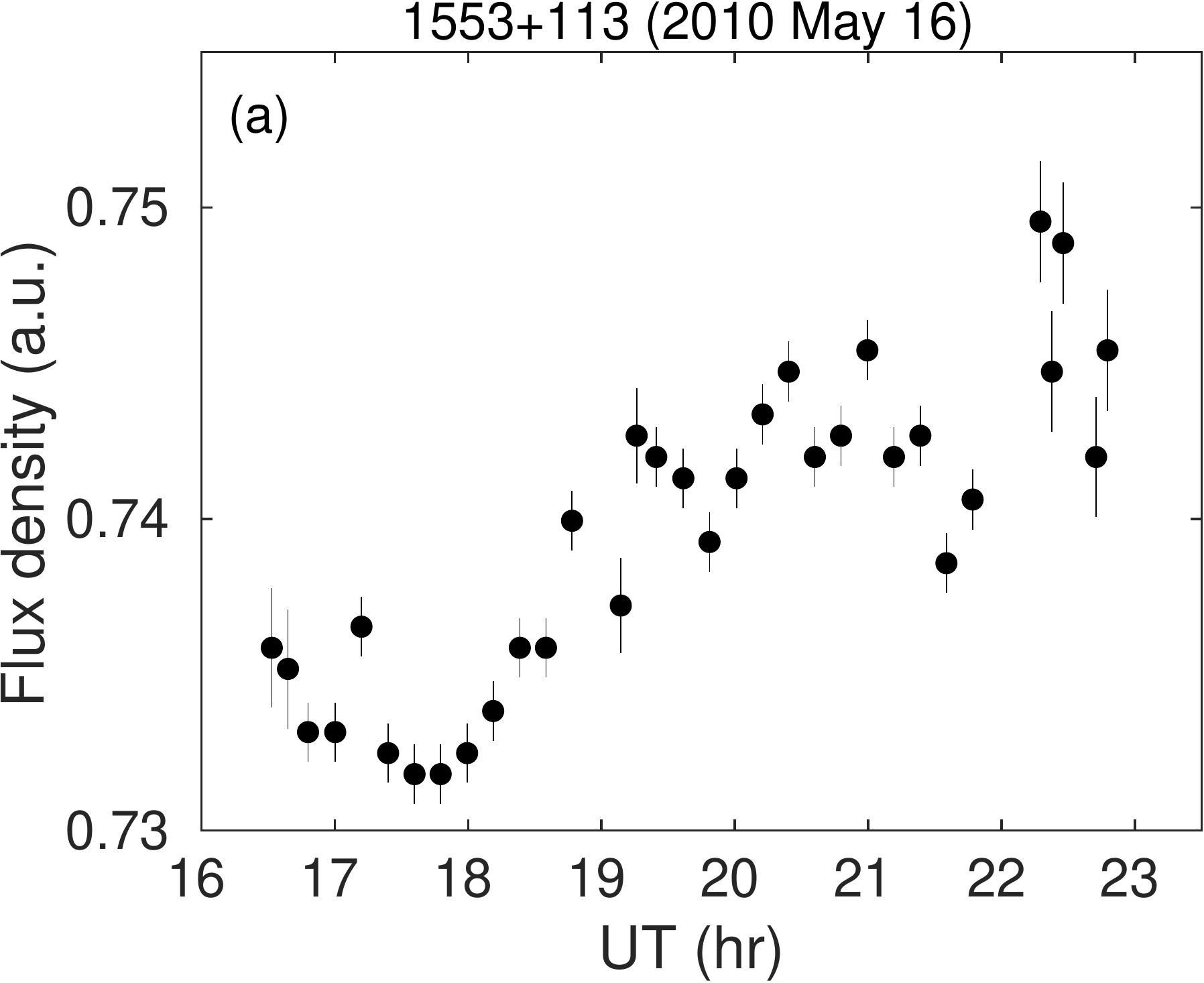}
\includegraphics[width=0.30\textwidth]{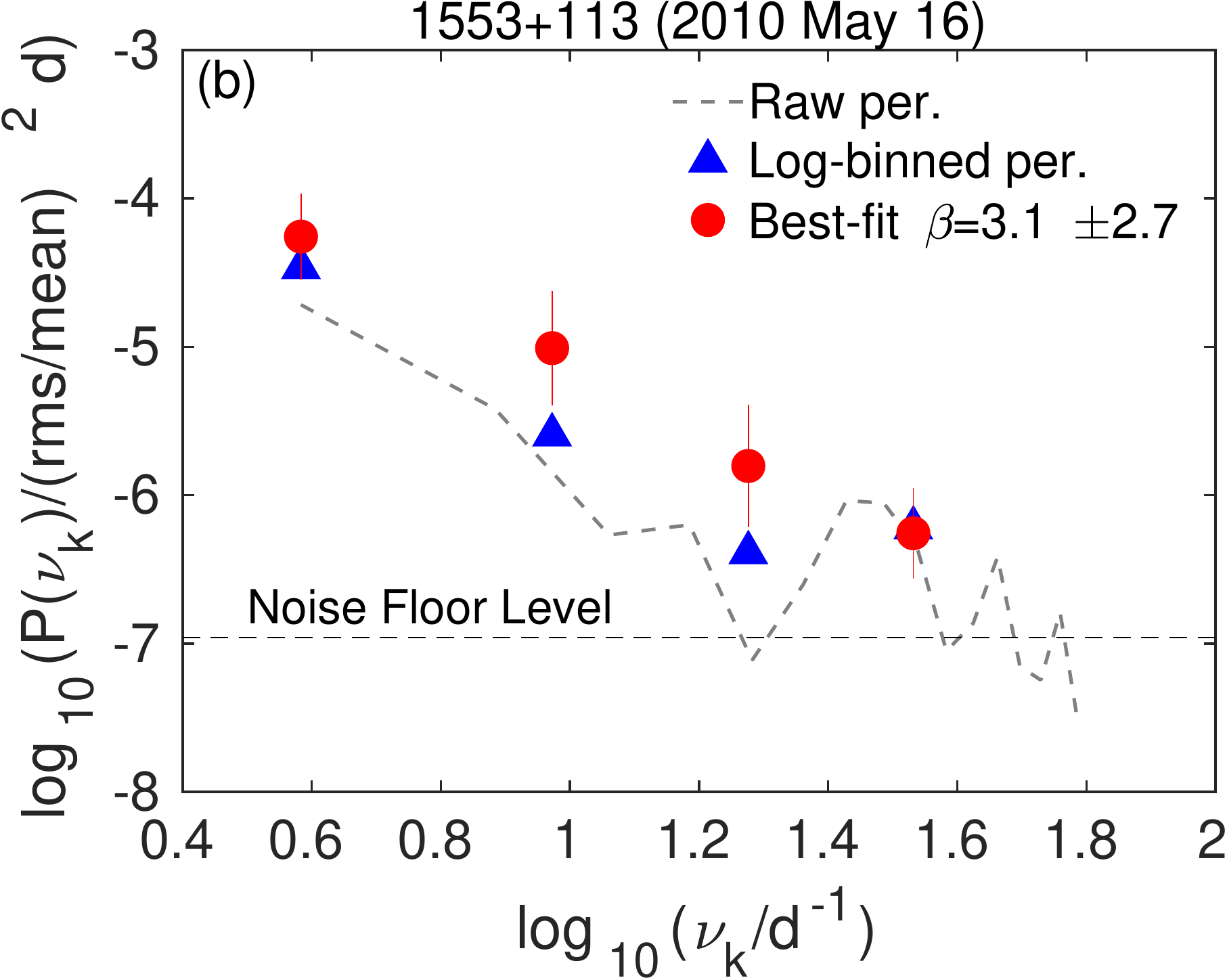}
\includegraphics[width=0.30\textwidth]{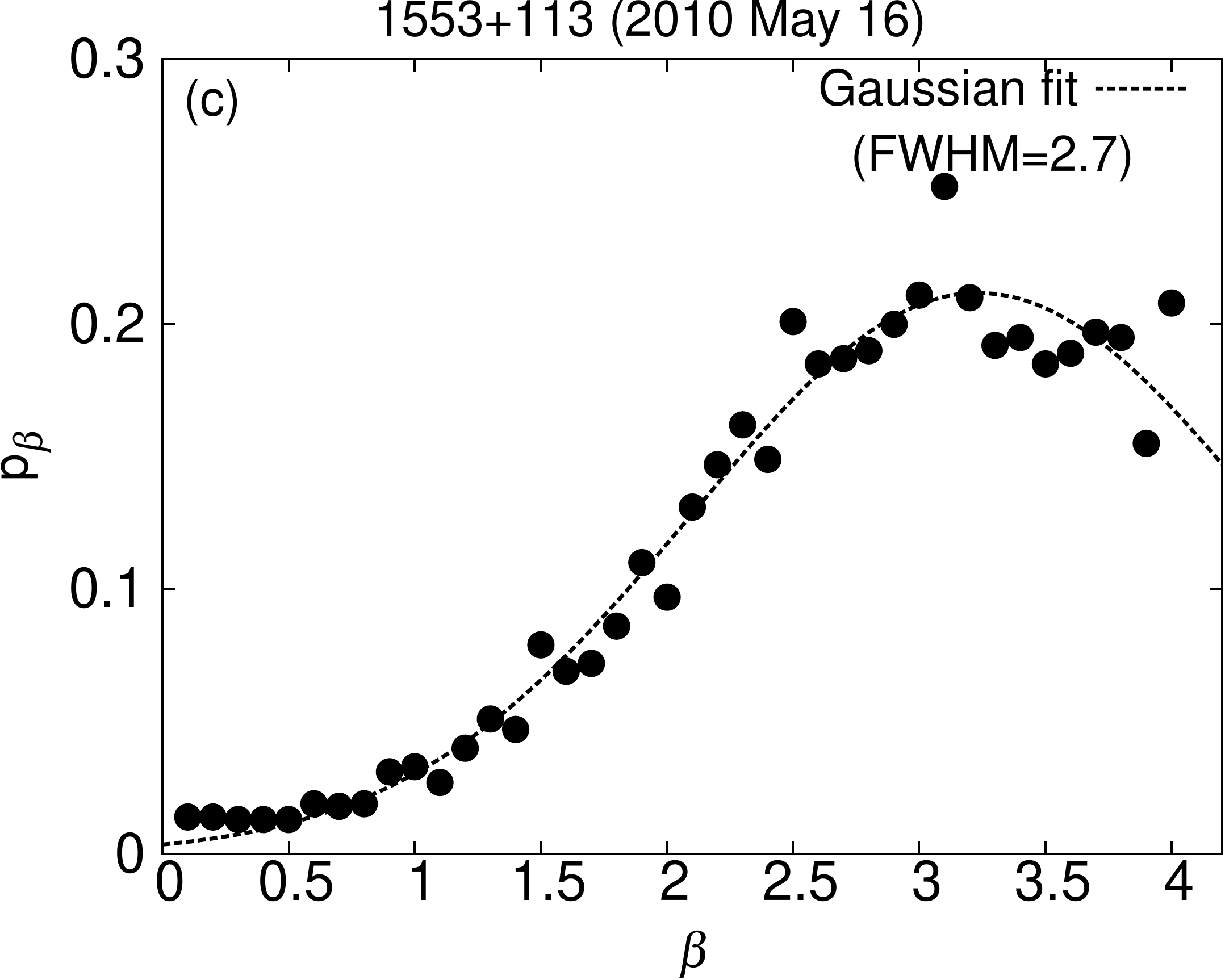}
}

\begin{minipage}{\textwidth}
\caption{(continued) }
\end{minipage}
\end{figure*}

\section{Sample}\label{sec:sample}

The blazar light curves studied here are obtained from the samples of \citet{Goyal13b} who studied the intra-night variability properties of different types of active galactic nuclei (AGN) using 262 intranight light curves. The AGNs monitored belong to radio--quiet quasar, radio--intermediate quasar, radio--loud quasar and blazar types, including BL Lacs and the FSRQs. \citeauthor{Goyal13b} blazar sample consists of 24 sources monitored on 85 sessions. The details of data gathering, reduction procedure, generation of differential light curves, and the statistical tests used to infer intranight variability are given in \citet[][]{Goyal13b} which we briefly describe here. On each monitoring session, continuous CCD observations ($>$4 hr) of the target were performed with the integration time of each frame chosen such that the flux measurement could be obtained with 0.2--0.5\% accuracies. The preprocessing (bias subtraction and flat fielding) of raw CCD frames were done using {\sc Image reduction and analysis facility (IRAF)} software and the instrumental magnitudes of the target blazar and the comparison stars on the same CCD chip were derived using aperture photometry. The relative instrumental magnitude of the blazar was computed against one steady `star--star' pair, thereby producing two differential light curves (DLCs) for a blazar on a given monitoring session. The differential photometry technique is widely used in variability studies as it counters the spurious AGN variability occurring due to varying atmospheric conditions (variable seeing or presence of thin clouds during the monitoring session) as any changes seen in blazar light should be accompanied by the same changes in comparison starlight, thereby keeping the difference unaltered. Next, we used $F-$test to infer the statistical significance of variability at significance levels, $\alpha$= 0.01 and  0.05, corresponding to $p$ value $>$0.99 and $>$0.95, respectively. If the value of $F-$test statistic for a DLC turned out to be more than the critical value at $\alpha$=0.05(0.01), the DLCs were assigned a `probable variable (confirmed variable)' status. If the test statistic turned out to be smaller than the critical value at$\alpha$=0.05 for the DLC, it was assigned a `non-variable' status. In the present study, all the monitoring sessions where both DLCs showed a `confirmed variable' status are used. We also included two monitoring sessions where one of the two DLCs showed a `probable variable' status while the other showed a `confirmed variable' status. The monitoring sessions where the blazar DLCs showed `non-variable' status are not used in this analysis as the PSDs of `non-variable' light curves will be consistent with $\beta\sim0$, resulting from fluctuations arising from measurement errors. The above criterion reduced the sample to 15 blazars and 34 intranight light curves. We further exclude the BL Lac object PKS\,0735+178 from the sample as the PSD analysis of its five intranight light curves has been reported in \citet{Goyal17}. Therefore, the current sample consists of 14 blazars which have shown statistically significant variability on 29 intranight monitoring sessions\footnote{Light curves can be obtained upon request.}. Table~\ref{tab:sample} lists the basic properties of the blazars. 

Next, we converted the differential magnitudes of blazars from logarithmic scale to linear scale using the relation F$_{obs}$$=$F$_{0}$\,$\times 10^{-0.4\times m_{obs}}$, where F$_{0}$(=1) is the arbitrary zero point magnitude flux, $m_{obs}$ is the differential blazar magnitude relative to one comparison star and F$_{obs}$ is the corresponding differential flux density as it contains the contribution from the steady comparison star flux.  The errors in were derived using standard error propagation \citep{Bevington03} and scaled by a factor 1.54 to account for the underestimation of the photometric errors by the {\sc IRAF} \citep[see, for details,][and references theirin]{Goyal13a}. We note that the differential blazar flux density can be scaled to proper fluxes using the appropriate F$_{0}$ for the R--band and the apparent magnitudes of comparison star used. Figure~\ref{fig:analysis} (panel a) shows flux density intranight light curves.

\begin{deluxetable*}{ccccccccccc}
\tablenum{2}
\tablecaption{Summary of the observations and the PSD analysis.\label{tab:psd}}
\tablewidth{0pt}
\tabletypesize{\small}
\tablehead{
\colhead{IAU name} & \colhead{Date of obs.} & \colhead{Tel.} &  \colhead{$T_{\rm obs}$} & $N_{obs}$ & $\psi$ &  \colhead{$T_{\rm mean}$} &
\colhead{$\rm \log_{10}(P_{stat})$} & \colhead{$\log_{10}(\nu_k) $ range} & \colhead{{$\beta \pm err$}} & \colhead{$p_\beta^\ast$}\\
 \colhead{} &  \colhead{} & \colhead{}  & \colhead{(hr)} &   \colhead{} &    \colhead{} & \colhead{(min)} & \colhead{($\frac{\mathrm rms}{\mathrm mean})^2$d} & \colhead{(d$^{-1}$)} &   \colhead{} &  \colhead{} 
}

\decimalcolnumbers
\startdata
0109$+$224    &  2005 Oct 29  &  ST    &  7.1  & 36  & 3.98 & 11.9    &  $-$7.08   &  0.53 to 1.47   &2.7$\pm$4.3  & 0.119   \\
0235$+$164    &  1999 Nov 12  &  ST    &  6.6  & 40  & 13.25& 9.70    &  $-$5.87   &  0.56 to 1.74   &3.9$\pm$2.8  & 0.011   \\
              &  1999 Nov 14  &  ST    &  6.2  & 34  & 10.59& 10.26   &  $-$5.35   &  0.59 to 1.53   &3.7$\pm$3.2  & 0.057   \\
              &  2003 Nov 18  &  ST    &  7.8  & 41  & 8.25 & 9.75    &  $-$6.35   &  0.48 to 1.67   &4.0$\pm$2.5  & 0.007   \\
0420$-$014    &  2003 Nov 19  &  ST    &  6.7  & 38  & 2.22 & 10.55   &  $-$7.18   &  0.55 to 1.73   &3.7$\pm$2.8  & 0.559   \\
              &  2009 Oct 25  &  ST    &  4.5  & 21  & 5.14 & 12.73   &  $-$5.94   &  0.73 to 1.67   &2.7$\pm$2.9  & 0.108   \\
0716$+$714    &  2005 Feb 1   &  ST    &  1.7  & 26  & 3.36 & 3.88    &  $-$7.28   &  1.15 to 2.10   &3.5$\pm$2.3  & 0.048   \\
0806$+$315    &  1998 Dec 28  &  ST    &  7.3  & 36  & 16.66& 12.16   &  $-$5.49   &  0.52 to 1.46   &3.0$\pm$2.4  & 0.134   \\
0806$+$524    &  2005 Feb 4   &  ST    &  7.2  & 29  & 1.31 & 14.98   &  $-$7.01   &  0.52 to 1.46   &2.8$\pm$3.8  & 0.898   \\
0851$+$202    &  1999 Dec 31  &  ST    &  5.6  & 29  & 4.81 & 11.61   &  $-$6.56   &  0.63 to 1.58   &3.8$\pm$2.8  & 0.404   \\
              &  2000 Mar 28  &  ST    &  4.2  & 22  & 5.34 & 11.55   &  $-$6.50   &  0.75 to 1.44   &2.9$\pm$2.8  & 0.979   \\
              &  2001 Feb 17  &  ST    &  6.9  & 47  & 2.78 & 8.82    &  $-$6.88   &  0.54 to 1.73   &3.3$\pm$2.4  & 0.832   \\
              &  2005 Apr 12  &  ST    &  4.8  & 56  & 9.07 & 5.10    &  $-$6.95   &  0.70 to 1.88   &3.9$\pm$2.9  & 0.006   \\
1011$+$496    &  2010 Feb 19  &  ST    &  5.6  & 43  & 2.59 & 8.43    &  $-$6.80   &  0.59 to 1.78   &2.1$\pm$2.9  & 0.248   \\
              &  2010 Mar 7   &  ST    &  5.5  & 36  & 3.95 & 9.16    &  $-$6.73   &  0.63 to 1.58   &3.3$\pm$2.8  & 0.057   \\
1156$+$295    &  2012 Mar 31  &  IGO   &  5.9  & 26  & 9.06 & 19.76   &  $-$5.51   &  0.60 to 1.29   &3.5$\pm$2.9  & 0.506   \\
              &  2012 Apr 1   &  IGO   &  8.4  & 26  & 11.50& 19.38   &  $-$5.19   &  0.45 to 1.40   &1.4$\pm$3.1  & 0.343   \\
              &  2012 Apr 2   &  IGO   &  7.2  & 20  & 22.34& 21.67   &  $-$4.85   &  0.52 to 1.21   &3.5$\pm$4.0  & 0.047   \\
1216$-$010    &  2002 Mar 16  &  ST    &  8.2  & 22  & 14.27& 22.35   &  $-$5.97   &  0.46 to 1.41   &4.0$\pm$2.8  & 0.014   \\
1219$+$285    &  2003 Mar 19  &  ST    &  6.2  & 60  & 5.76 & 6.20    &  $-$7.03   &  0.59 to 1.99   &2.6$\pm$2.0  & 0.852   \\
              &  2003 Mar 20  &  ST    &  6.3  & 67  & 10.16& 5.63    &  $-$6.82   &  0.58 to 1.98   &3.3$\pm$1.8  & 0.174   \\
1253$-$055    &  2006 Jan 26  &  ST    &  4.7  & 21  & 3.23 & 13.56   &  $-$7.12   &  0.70 to 1.65   &1.9$\pm$1.7  & 0.259   \\
              &  2006 Feb 28  &  ST    &  6.5  & 42  & 13.20& 9.30    &  $-$7.28   &  0.56 to 1.75   &2.3$\pm$1.7  & 1.000   \\
              &  2009 Apr 20  &  ST    &  5.5  & 22  & 25.86& 14.89   &  $-$5.80   &  0.64 to 1.51   &3.4$\pm$2.2  & 0.066   \\
1510$-$089    &  2009 May 1   &  ST    &  6.0  & 25  & 7.73 & 14.45   &  $-$6.10   &  0.59 to 1.54   &3.3$\pm$2.8  & 0.021   \\
1553$+$113    &  1999 May 5   &  ST    &  4.2  & 23  & 2.85 & 10.83   &  $-$6.63   &  0.76 to 1.70   &4.0$\pm$2.8  & 0.110   \\
              &  2009 Jun 24  &  ST    &  4.2  & 26  & 5.73 & 9.74    &  $-$6.85   &  0.75 to 1.70   &2.4$\pm$2.0  & 0.533   \\
              &  2010 May 15  &  ST    &  6.5  & 22  & 3.13 & 17.73   &  $-$6.95   &  0.56 to 1.51   &3.9$\pm$2.3  & 0.345   \\
              &  2010 May 16  &  ST    &  6.3  & 33  & 2.46 & 11.39   &  $-$6.95   &  0.58 to 1.53   &3.1$\pm$2.7  & 0.252   \\
\enddata
\tablecomments{
(1) name of the blazar following the IAU convention;
(2) the date of observations; 
(3) the telescope facility used. ST = 1\,m Sampurnanand Telescope of Aryabhatta Research Institute of Observational Sciences, India; IGO = 2\,m IUCAA-Girawali Observatory of Inter-University Centre of Astronomy and Astrophysics, India.
(4) the duration of the observed light curve;
(5) number of data points in the light curve; 
(6) peak-to-peak variability amplitude \citep[Eq. 9 of][]{Goyal13b};
(7) the mean sampling interval for the observed light curve (light curve duration/number of data points); 
(8) the noise level in PSD due to the measurement uncertainty;
(9) the temporal frequency range covered by the  binned logarithmic power spectra;
(10) the best-fit power-law slope of the PSD along with the corresponding errors representing 98\% confidence limit (see Section~\ref{sec:psresp});
(11) corresponding $p_\beta$. $^\ast$ power law model is considered as a bad-fit if $p_\beta$ $\leq$ 0.1 as the corresponding rejection confidence for the model is $\geq$90\% (Section~\ref{sec:psresp}).      
}
\end{deluxetable*}

\section{PSD Analysis}\label{sec:analysis}     
\subsection{Derivation of PSDs: discrete Fourier transform}\label{sec:dft}
Since the aim of the study is to obtain reliable shapes of PSDs, we subject the light curves Fourier transformation using the discrete Fourier transform (DFT) method \citep[see, for details,][and references therein]{Goyal20}. The fractional rms-squared-normalized periodogram is given as the squared modulus of its DFT for the evenly sampled light curve $f(t_i)$, observed at discrete times $t_i$ and consisting of $N$ data points and the total monitoring duration $T$,

\begin{multline}
 P(\nu_k)  =   \frac{2 \, T}{\mu^2 \, N^2} \, \Bigg\{ \Bigg[ \sum_{i=1}^{N} f(t_i) \, \cos(2\pi\nu_k t_i)  \Bigg]^2  + \\ \Bigg[ \sum_{i=1}^{N} f(t_i) \, \sin(2\pi\nu_k t_i)  \Bigg]^2 \, \Bigg \} ,
\label{eq:psd}
\end{multline}
where $\mu$ is the mean of the light curve and is subtracted from the flux values, $f(t_i)$. The DFT is computed for evenly spaced frequencies ranging from the total duration of the light curve down to the Nyquist sampling frequency ($\nu_{\rm Nyq}$) of the observed data. Specifically, the frequencies corresponding to $\nu_{k} = k/T$ with $k=1, ..., N/2$, $\nu_{\rm Nyq}= N/2T$, and $T = N (t_k-t_1)/(N-1)$ are considered. The normalized periodogram as defined in Eq.~\ref{eq:psd} corresponds to total excess variance when integrated over positive frequencies. The constant noise floor level from measurement uncertainties is given as \citep[e.g.,][]{Isobe15, Vaughan03}
\begin{equation}
\rm P_{stat} = \frac{2 \, T}{\mu^2 \, N} \, \sigma_{\rm stat}^2 \, .
\label{eq:poi_psd}
\end{equation}
where, $\sigma_{stat}^2= \sum_{j=1}^{j=N} \Delta f(t_j)^2 / N$ is the mean variance of the measurement uncertainties on the flux values $\Delta f\!(t_j)$ in the observed light curve at times $t_j$, with  $N$ denoting the  number of data points in the original light curve. The intranight light curves are roughly evenly sampled but the application of the DFT method requires strict even sampling of the time series, otherwise, the `spectral window function' corresponding to the sampling times gives a non-zero response in the Fourier-domain, resulting in false powers in the periodograms \citep[see, Appendix A of][]{Goyal20,Deeming75}. Therefore, in order to perform the DFT, we obtained the regular sampling only by linearly interpolating between the two consecutive observed data points with an interpolation interval of 1 minute which is roughly 5--15 times smaller than the original sampling interval (see column 7 of Table~\ref{tab:psd}). Even though the choice of interpolation interval is arbitrary, we note that it cannot be longer than the mean sampling interval. We tested our procedure by also using an interpolation interval about half of the original sampling interval which did not change the results. We refer the reader to  \citet{Goyal17} and \citet{Max-Moerbeck14a} for a discussion on the distortions introduced in the PSDs due to the discrete sampling and the finite duration of the light curve, known as `red-noise leakage' and `aliasing' respectively. To minimize the effects of red-noise leak, the PSDs are generated using the `Hanning' window function \citep[e.g.,][]{Press92, Max-Moerbeck14a}. Aliasing, on the other hand, contributes an equal amount of power (around the Nyquist frequency) to the periodograms \citep[][]{Uttley02}, hence will not distort the shape of PSDs.  

The periodogram obtained using equation~(\ref{eq:psd}), known as the `raw' periodogram, provides a noisy estimate of the spectral power as it consists of independently distributed $\chi^2$ variables with two degrees of freedom (DOF) \citep{TK95, Papadakis93, Vaughan03}. Therefore, a number of PSD estimates should be averaged in order to obtain a reliable estimate of the spectral power. The periodograms falling within a factor of 1.6 in frequency range are averaged with the representative frequency taken as the geometric mean of each bin \citep{Isobe15, Goyal17, Goyal20}. Except for the first bin, this choice of binning factor provides at least two periodograms in each frequency bin.

Since the observed power-spectrum is related to the `true' power spectrum by $P(\nu_k) = P_{true}({\nu_k}) \frac{\chi^2}{2}$ for a noise-like process \citep{Papadakis93, TK95, Vaughan03}. The transformation to log-log space, offsets the the observed periodograms as  

\begin{equation}
\log_{10}[ P(\nu_k) ] = \log_{10}[ (P_{true}({\nu_k}) ] + \log_{10}\Bigl[ \frac{\chi^2}{2} \Bigr] .
\label{logpsd}
\end{equation}

This offset is the expectation value of $\chi^2$ distribution with 2 DOF in log-log space and is equal to $-$0.25068 which is added to the observed periodograms \citep[][]{Vaughan05}.

\subsection{Estimation of the spectral shape: PSRESP method}\label{sec:psresp}
Since the aim of the present study is to derive shapes of intranight PSDs, we use the `power spectral response' (PSRESP) method \citep[e.g.,][]{Uttley02, Chatterjee08, Max-Moerbeck14a, Isobe15, Meyer19, Goyal20} which further mitigates the deleterious effects of red-noise leak and aliasing. In this method, an (input) PSD model is tested against the observed PSD. The estimation of best-fit model parameters and their uncertainties is performed by varying the model parameters. To achieve this, a large number of light curves are generated with a known underlying power-spectral shape using Monte Carlo (MC) simulations. Rebinning of the light curve to mimic the sampling pattern and interpolation is performed for the DFT application. The DFT of such light curve gives the distorted PSD due to effects mentioned above. Averaging large number of such PSDs gives the mean of the distorted model (input) power spectrum. The standard deviation around the mean gives errors on the modeled (input) power spectrum. The goodness of fit of the model is estimated by computing two functions, similar to $\chi^2$, defined as

\begin{equation}
\chi^2_{\rm obs} = \sum_{\nu_{k}=\nu_{min}}^{\nu_{k}=\nu_{max}} \frac{[\overline{ \log_{10}P_{\rm sim}}(\nu_k)-\log_{10}P_{\rm obs}(\nu_k)]^2}{\Delta \overline{\log_{10}P_{\rm sim}}(\nu_k)^2}
\label{chiobs}
\end{equation}
and 
\begin{equation}
\chi^2_{\rm dist, i} = \sum_{\nu_{k}=\nu_{min}}^{\nu_{k}=\nu_{max}} \frac{[\overline{ \log_{10}P_{\rm sim}}(\nu_k)-\log_{10}P_{\rm sim,i}(\nu_k)]^2}{\Delta \overline{\log_{10}P_{\rm sim}}(\nu_k)^2},
\label{chidist}
\end{equation}
where $\log P_{\rm obs}$ and  $\log P_{\rm {sim, i}}$ are the observed and the simulated log-binned periodograms, respectively. $\overline{ \log P_{\rm sim}}$ and $\Delta \overline{\log P_{\rm sim}}$ are the mean and the standard deviation obtained by averaging a large number of PSDs; $k$ represents the number of frequencies in the log-binned power spectrum (ranging from $\nu_{min}$ to $\nu_{max}$), while $i$ runs over the number of simulated light curves for a given $\beta$.  

Here $\chi^2_{\rm obs}$ determines the minimum $\chi^2$ for the model compared to the data and the $\chi^2_{\rm dist}$ values determine the goodness of the fit corresponding to the $\chi^2_{\rm obs}$. We note that $\chi^2_{\rm obs}$ and $\chi^2_{\rm dist}$ are not the same as a standard $\chi^2$ distribution because $\log_{10}P_{\rm obs}(\nu_k)$'s are not normally distributed variables since the number of power spectrum estimates averaged in each frequency bin are small \citep[][]{Papadakis93}. Therefore, a reliable goodness of fit is computed using the distribution of $\chi^2_{\rm dist}$ values. For this, the $\chi^2_{\rm dist}$ values are sorted in ascending order. The probability or $p_{\beta}$, that a given model can be rejected is then given by the percentile of $\chi^2_{\rm dist}$ distribution above which $\chi^2_{\rm dist}$ is found to be greater than  $\chi^2_{\rm obs}$ for a given $\beta$ \citep[success fraction;][]{Chatterjee08}. A large value of $p_{\beta}$ represents a good--fit in the sense that a large fraction of random realizations of the model (input) power spectrum are able to recover the shape of the intrinsic  PSD. Therefore, this analysis essentially uses the MC approach toward a frequentist estimation of the quality of the model compared to the data. This is a well-known approach to estimate the goodness of fit when the fitting statistic is not well understood \citep[see, for details,][]{Press92}.   

In this study, the light curve simulations are performed using the method of \citet{Emmanoulopoulos13} which preserves the probability density function (PDF) of the flux distribution as well as the underlying power spectral shape. In addition to assuming the power spectral shape, the method requires supplying a value of mean and standard deviation ($\sigma$) of the flux values to reproduce the flux distribution and match the variance \citep[][]{Meyer19}. We have assumed single power-law PSDs with a given $\beta$ (to reproduce the PSD shape) and supplied mean and $\sigma$ of the logarithmically transformed flux values which is found to be an adequate representation of flux distribution on shorter ($\lesssim$days) timescales for a few cases \citep[e.g.,][]{Hess10, Kushwaha20}.
 For this purpose, the mean and the $\sigma$ are computed by fitting a Gaussian function to the flux distribution. Finally, the measurement errors in the simulated flux values were incorporated by adding a Gaussian random variable with mean 0 and standard deviation equal to the mean error of the measurement uncertainties on the observed flux values \citep[][]{Meyer19, Goyal20}. In such a manner, 1,000 light curves are simulated in the $\beta$ range 0.1 to 4.0, with a step of 0.1 for each observed light curve. For the simulated light curve, the periodograms are derived in an identical manner as that of the observed light curve (Section~\ref{sec:dft}). The best-fit PSD slope for the observed PSD is given by the one with the highest $p_{\beta}$ value  and the uncertainty is given as 2.354$\sigma$ of the $p_{\beta}$ curve where $\sigma$ is the standard deviation of fitted Gaussian. This gives roughly a 98\% confidence limit on the best-fit PSD slope.                  

\begin{figure}
\hbox{
\includegraphics[width=0.4\textwidth]{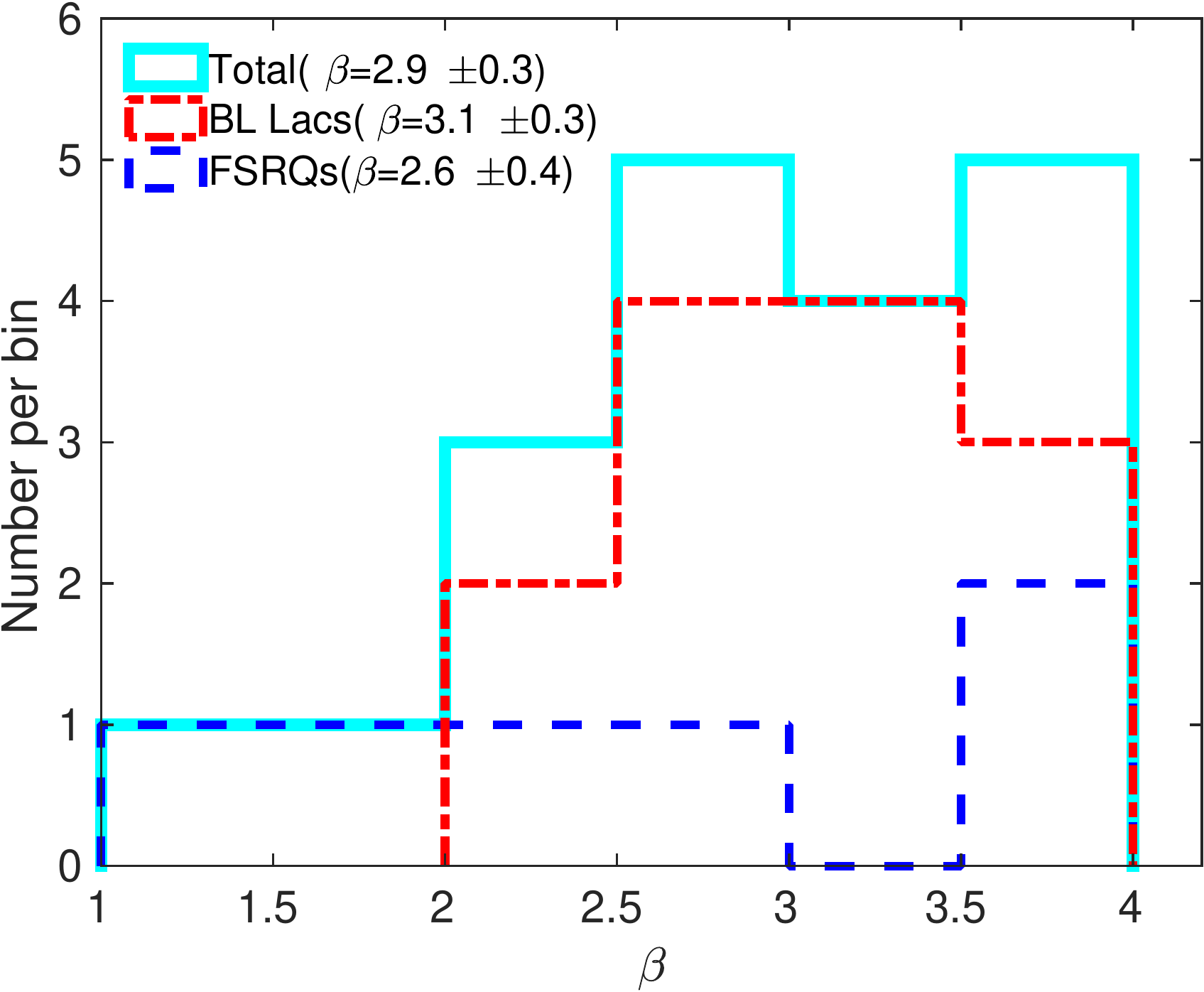}
}
\caption{Histograms of the best-fit PSD slopes derived for the entire blazar sample (cyan line; 10 sources and 19 monitoring sessions), BL Lacs (red line; seven sources and 13 monitoring sessions), and FSRQs (blue line; three sources and six monitoring sessions), respectively. The sample mean along with 1$\sigma$ uncertainty estimated using the bootstrap method for different groups is given in parentheses (see Section~\ref{sec:psresp}). }
\label{fig:hist}
\end{figure}

\begin{figure*}
\hbox{
\includegraphics[width=0.33\textwidth]{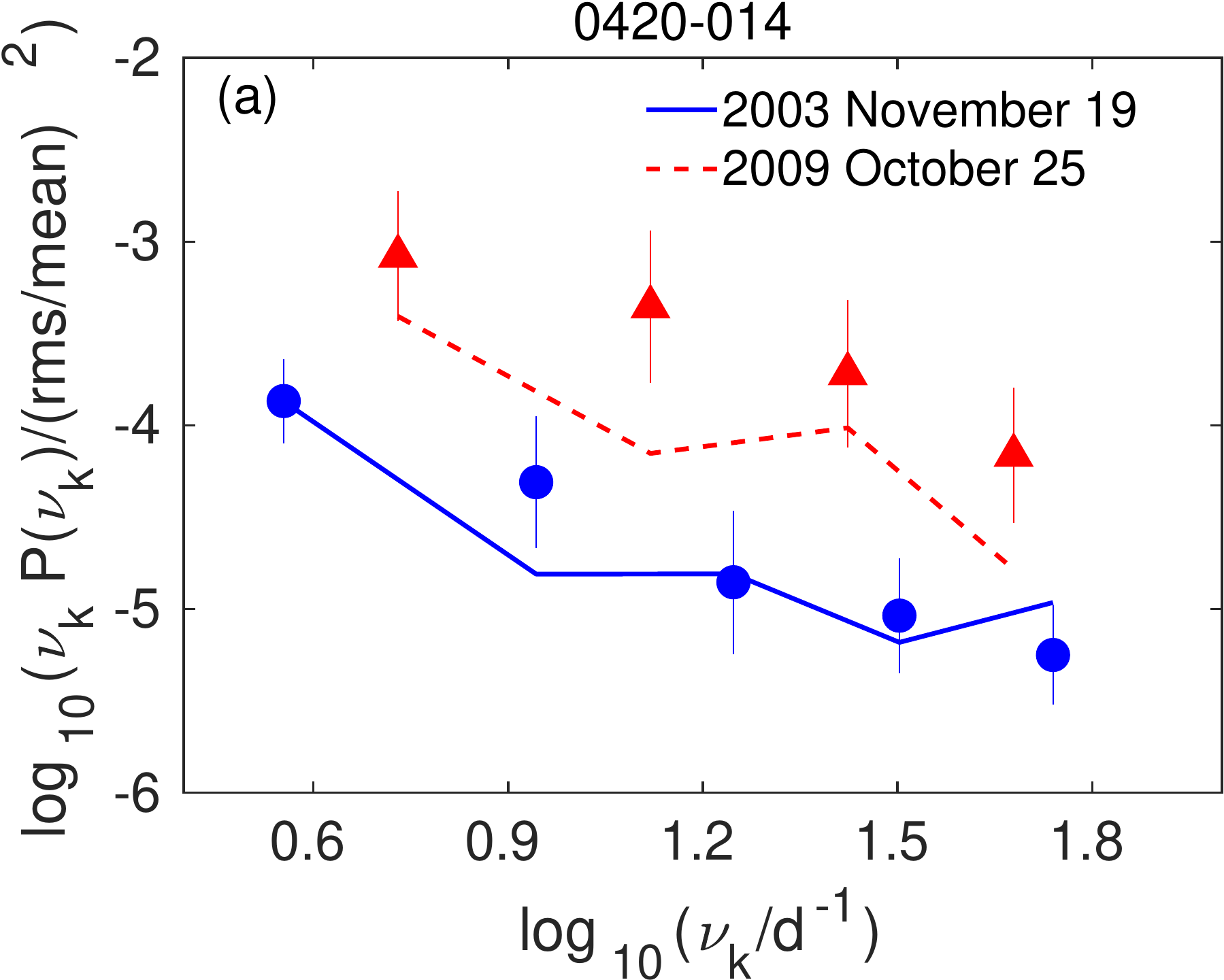}
\includegraphics[width=0.33\textwidth]{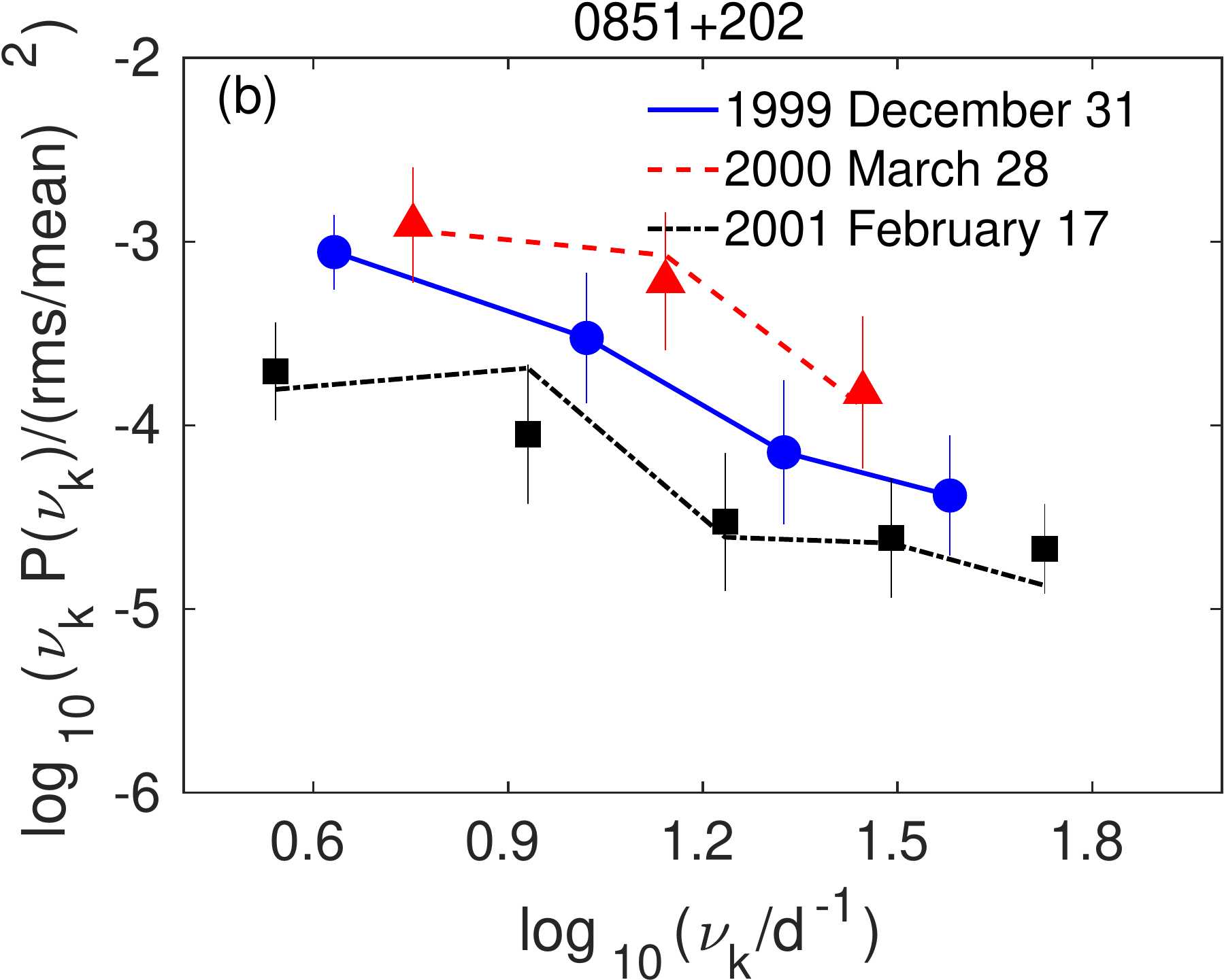}
\includegraphics[width=0.33\textwidth]{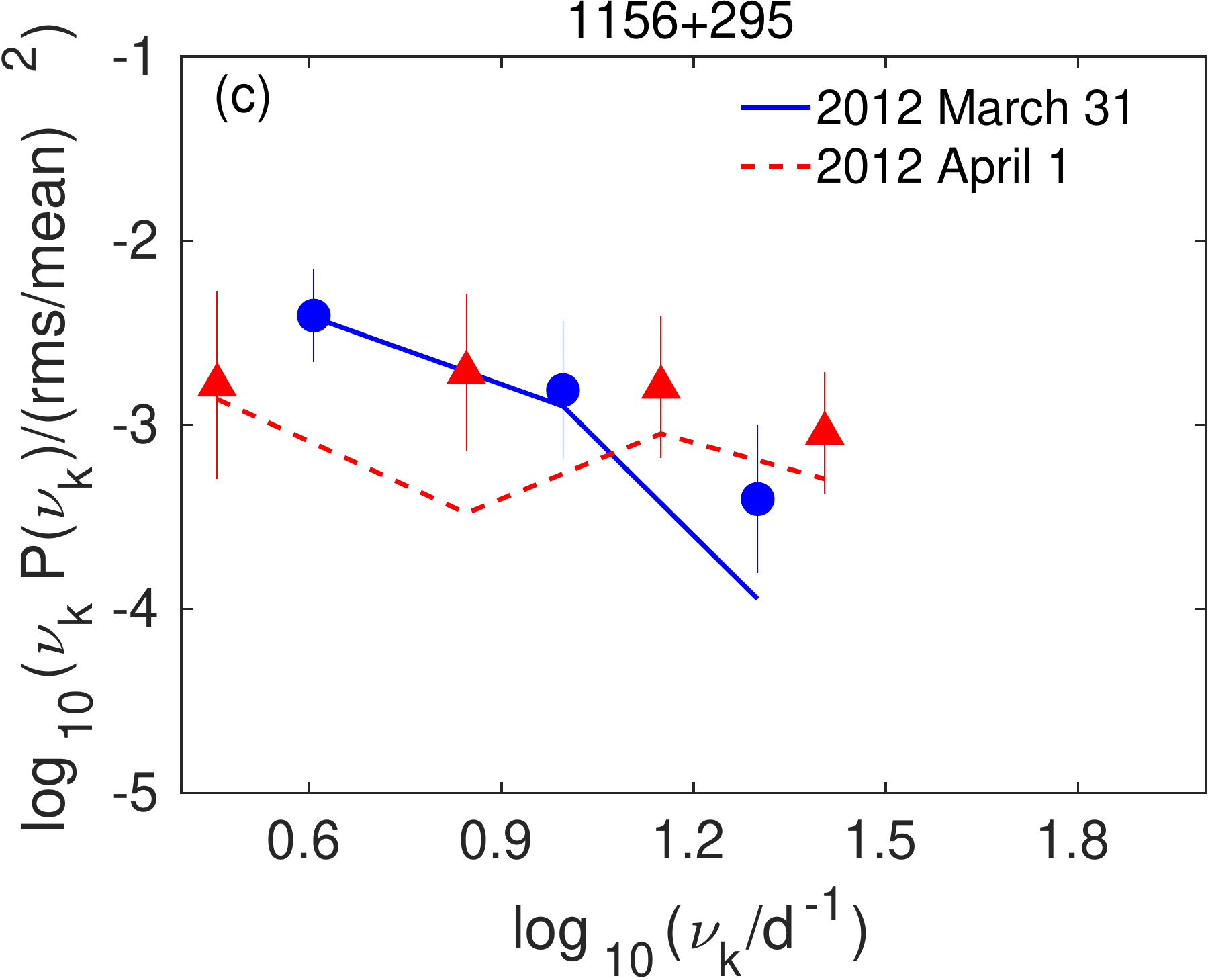}
}
\hbox{
\includegraphics[width=0.33\textwidth]{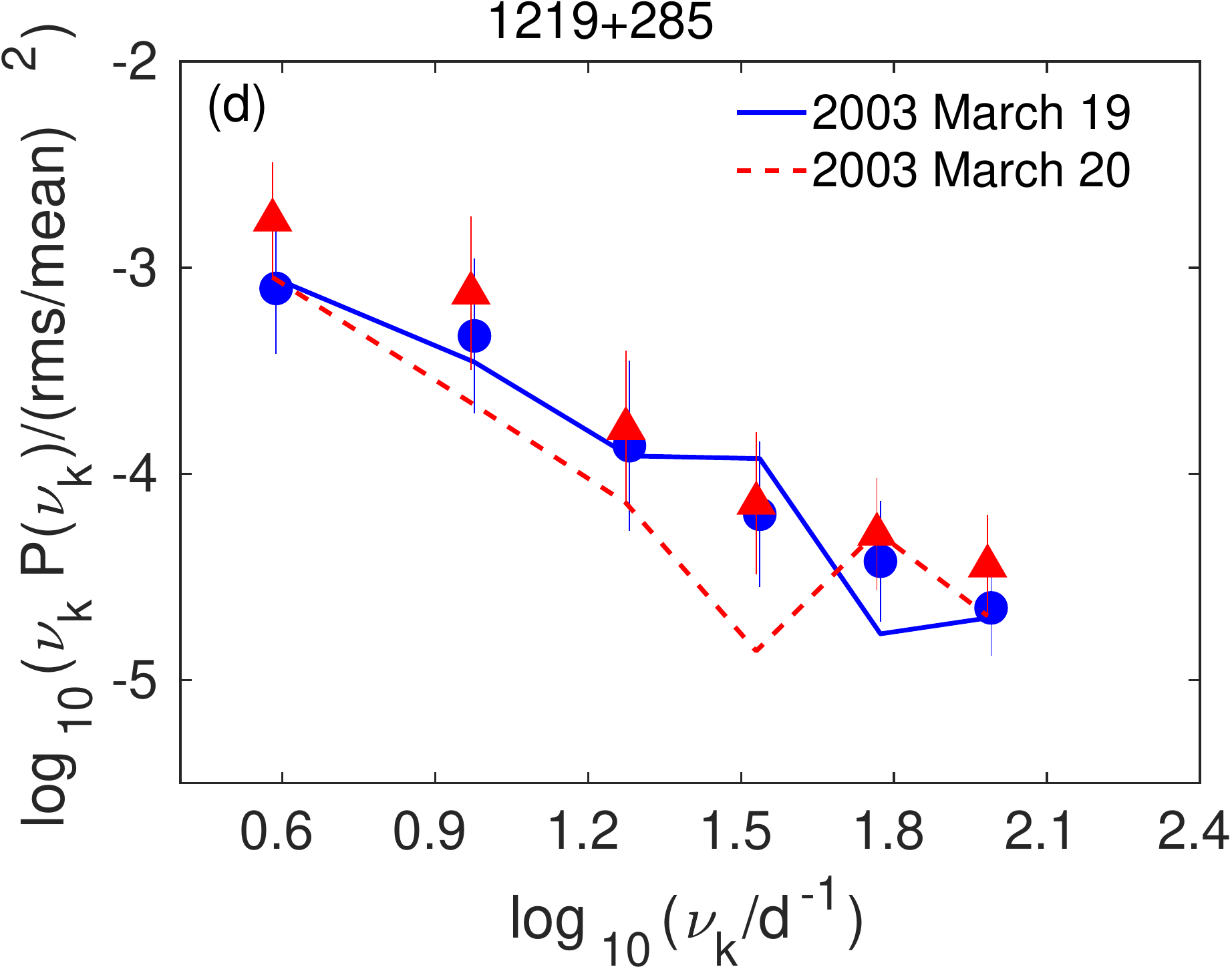}
\includegraphics[width=0.33\textwidth]{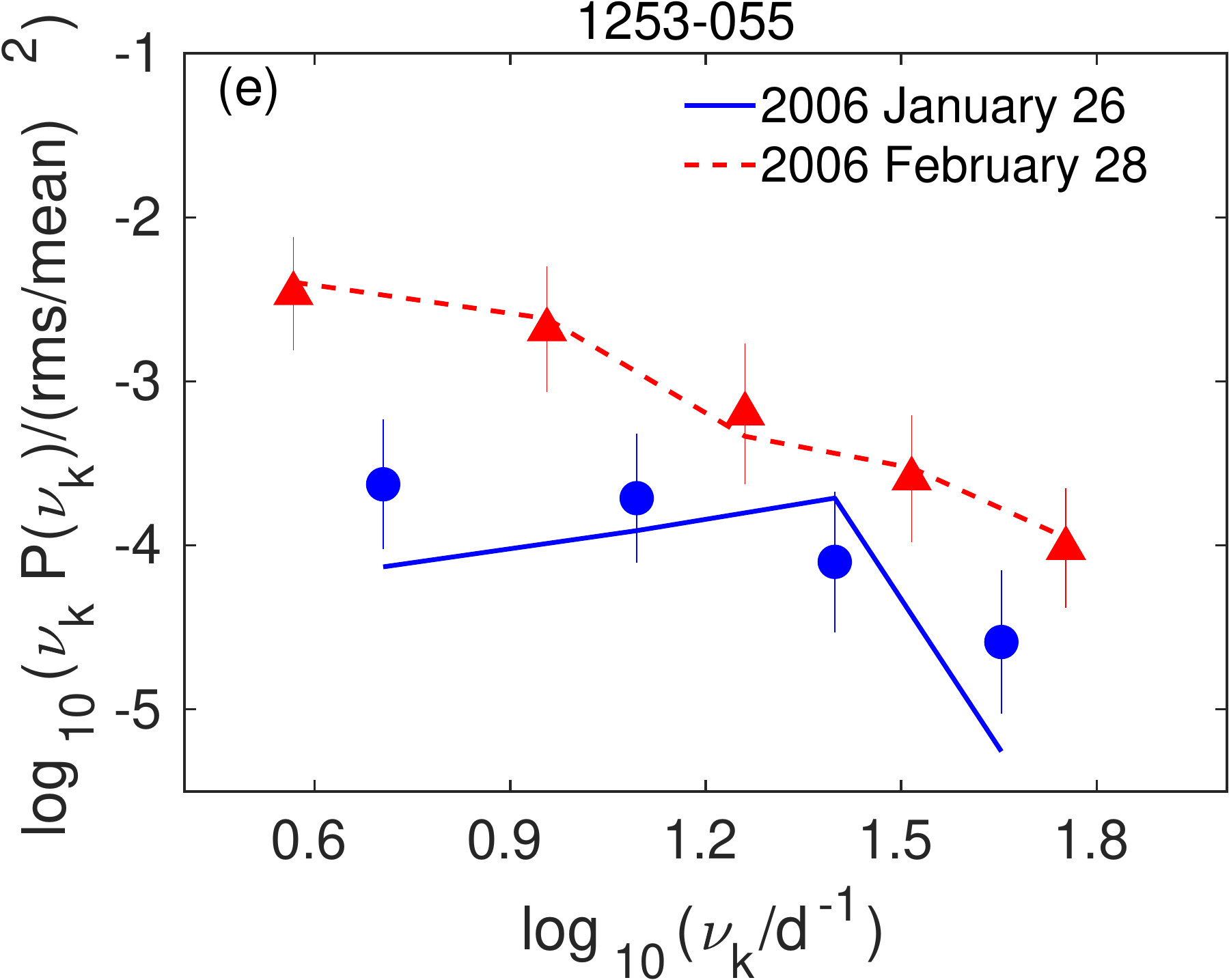}
\includegraphics[width=0.33\textwidth]{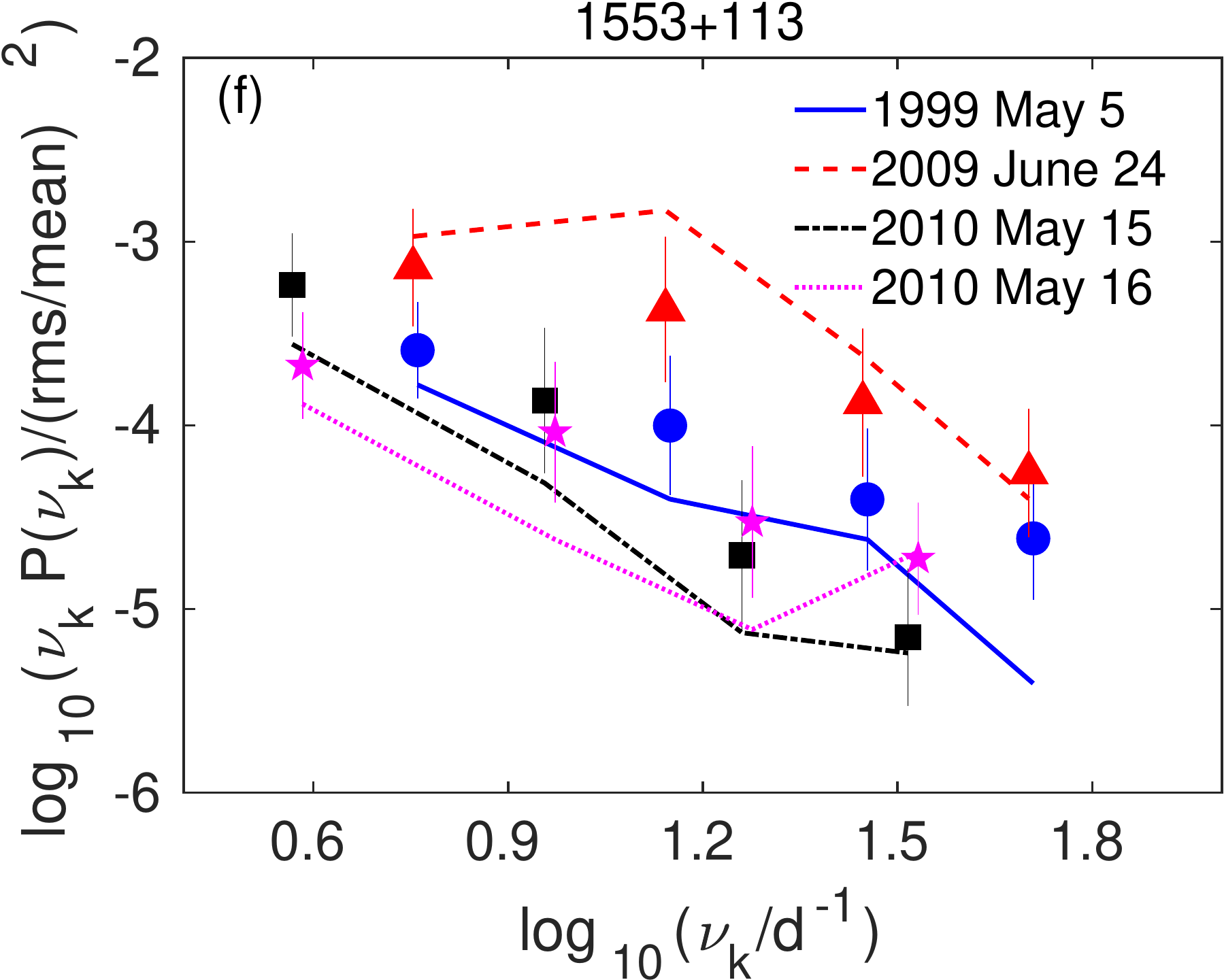}
}
\caption{$\nu_k$ P($\nu_k$) PSDs for individual blazars which showed acceptable fit in the analysis. The lines show the log-binned periodograms and the filled symbols show mean and standard deviation of best-fit PSDs given by the PSRESP method for different epochs. }
\label{fig:jointpsds}
\end{figure*}

Details on the intranight light curves used for the analysis and the derived PSDs, along with the best-fit PSDs and the maximum $p_{\beta}$, are summarized in Table~\ref{tab:psd}. Figure~\ref{fig:analysis} presents the analyzed light curves (panel a), the corresponding best-fit PSD (panel b) and the probability distribution curves ($p_\beta$ as a function of $\beta$; panel c) for the duration of the light curve down to the mean sampling intervals for the given light curve. In our analysis, we have not subtracted the constant noise floor level (shown by the dashed horizontal lines in panel b of the figure), as some of the data points are below this level. The PSRESP method also allows us to compute the rejection confidence for the input PSD shape; the maximum probability lower than 10\% means that the rejection confidence (1--$p_{\beta}$ value) is higher than 90\% for the (input) PSD model. In our analysis, we use $p_\beta$$<$0.1 as a rejection threshold for the the model, meaning that the input spectral shape does not provide a good fit to the PSD. The distribution of acceptable PSD slopes is shown in Figure~\ref{fig:hist}. The mean of a sample is computed in a straightforward manner while the error on the mean is computed using the MC bootstrap method as follows. For each sample, the $\beta$ is drawn from a Gaussian distribution of mean and standard deviation equal to $\beta$ and error/2.354 (note that Table~\ref{tab:psd} reports errors equal to  2.354$\sigma$). The mean $\beta$ is computed. These two steps are repeated 500 times. The error is given as the standard deviation of the distribution of mean $\beta$ values. In addition, we provide the $\nu_k$ $P(\nu_k)$ vs.\ $\nu_k$ curves in Figure~\ref{fig:jointpsds} for blazars observed on more than one occasion, to compare the `square' fractional variability on timescales probed by our analysis \citep[][]{Goyal20}.  

\section{Results}\label{sec:results}

In this study, we have derived the optical intranight variability PSDs of the blazar sources, covering the temporal frequency range from 10$^{0.52}$\,day$^{-1}$ to 10$^{1.99}$\,day$^{-1}$ (timescales corresponding to 7.4 hours and $\sim$15 minutes). The intranight light curves showed modest intranight variability with peak--to--peak variability amplitudes, $>$1--15\%, occasionally rising to over $15\%$, over the span of observations. Our main results are the following:
\begin{enumerate}
\item{Out of the 29 intranight light curves analyzed in the present study, the PSDs shows an acceptable fit to the single power-law spectral shapes for 19 monitoring sessions (see Section~\ref{sec:psresp}). The maximum $p_\beta$ is higher than 10\% and reaches as high as 100\% for these sessions (column 10; Table~\ref{tab:psd}; panel c of Figure~\ref{fig:analysis}). }

\item{For these 19 acceptable PSD fits, the simple power-law slopes range from 1.4 to 4.0 (albeit with a large scatter); consistent with a statistical characters of red ($\beta$$\sim$2) and black ($\beta$$\geq$3) noise stochastic processes (Table~\ref{tab:psd}, Figure~\ref{fig:hist}). The mean $\beta$ turns out to be 2.9$\pm$0.3 (1$\sigma$ uncertainty) for blazar sources.  }

\item{The computed mean value PSD slopes for the BL Lac objects (seven sources and 13 light curves) and FSRQs (three sources and six light curves) are 3.1$\pm$0.3 and 2.6$\pm$0.4, respectively; consistent with one another within 1$\sigma$ uncertainty (Figure~\ref{fig:hist}).}

\item{The PSD slopes for a few sources whose intranight PSDs show an acceptable fit to single power-law on multiple occasions are consistent with each other (column 9; Table~\ref{tab:psd}).} 

\item{The normalization of the PSDs for the sources monitored on different epochs turns out to be consistent with each another within 1$\sigma$ uncertainty for the blazars 1156+295, 1219+285. However, one order of magnitude change is noted in the normalization of PSDs between 2003 November 19 and 2009 October 25 for 0420$-$014, between 1999 Dec 31 and 2001 February 17 for 0851+2020, and 2006 January 26 and 2006 February 28 for 1253$-$055 (Figure~\ref{fig:jointpsds}).}

\end{enumerate}

Our PSD analysis using the PSRESP method returns rejection confidence higher than 90\% for 10 out of the 29 analyzed lightcurves. These are: 0235+164 on 1999 November 12 and 14, and 2003 November 18, 0716+714 on 2005 February 1, OJ\,287 on 2005 April 12, 1011+496 on 2010 March 7, 1156+295 on 2012 April 2, 1216$-$010 on 2002 March 16, 3C\,279 on 2009 April 20, 1510$-$089 on 2009 May 1 (Table~\ref{tab:psd}). This is due to the fact that for the majority of these light curves, the intensity variations are essentially monotonic, i.e., a steady rise or fall without any other feature over the span of observations. This means that PSD model with $\beta$ $>$4 could fit the observed PSDs better over the temporal frequency range probed by the observations.

Next, we note that the reported 98\% confidence limits on the acceptable best-fit PSD slopes are large, in general (Table~\ref{tab:psd}), and in some cases, larger than the value itself. These are: 0109+224 on 2005 October 29, 0420$-$014 on 2009 October 25, 0806+524 on 2005 February 4, 1011+496 on 2010 February 19, and 1156+295 on 2012 April 1. A possible cause of this could be the limited number of data points in the studied light curves (20--67; column 4 of Table~\ref{tab:psd}). \citet[][]{Aleksic15a} studied the effects of changing the number of data points in the light curve and the estimation of the best-fit PSD slope (and uncertainty) using the long-term multiwavelength light curves having $\geq$30 data points for the blazar Mrk\,421. They note that the location of the maximum in the probability distribution curve does not change noticeably for different binning factors but the width, shape, and amplitude change significantly; however, it is unclear if the broadening of the probability distribution curve and hence the estimation of the uncertainty in the best-fit PSD slope is related to the gradual increase of binning factors (i.e., a decrease of a number of data points) for different light curves \citep[see, Figure 4 of][]{Aleksic15a}. Also, we note that the reported uncertainties on the best-fit X-ray intra-night PSD slopes using the PSRESP method also show large scatter, despite having $>$300 data points in the examined light curves for the AGNs Mrk\,421, PKS\,2155$-$304, and 3C\,273 \citep[Table 4 of][]{Bhattacharyya20}. Therefore, we conclude that a limited number of data points do not play a significant role in assessing the uncertainties of the spectral shape parameters using the PSRESP method.      

\section{Discussion and conclusions}\label{sec:discussions}

We report the first systematic study to characterize the intranight variability PSD properties comprising of 14 blazar sources and 29 densely sampled light curves, covering timescales from several hours to $\sim$15 minutes.  All the analyzed light curves were of duration $\geq$ 4 hours (except for the BL Lac 0716+714 for which duration was $\sim$1.5\,hr) and could be obtained with measurement accuracies $\lesssim$0.2--0.5\% in 5--15 minutes of integration time using the 1--2\,m class telescopes irrespective of blazar flux state, sky brightness or atmospheric conditions. The intranight monitoring sessions were scheduled solely based on the target's availability in the night sky for a duration longer than 4\,hours from a given telescope site. Therefore, the intranight PSDs are derived irrespective of the flux state of a blazar. The slopes show a range from 1.4 to 4.0, indicating that the variability at synchrotron emission frequencies has a statistical character of red to black--noise stochastic process on intranight timescales with no signs of cutoff at high frequencies due to measurement noise-floor levels arising due to measurement uncertainties. The mean $\beta$ for the entire sample is $\sim$2.9, indicating a steeper than red-noise character preference of the variability. Our crude estimates of mean $\beta$'s for the BL Lacs and the FSRQs subclasses (due to the small number of sources and the intranight light curves analyzed) give $\sim$3.1 and 2.6, respectively. These two estimates are comparable with each other, indicating that processes driving the variability occur in non-thermal jets for these sources and not in the accretion disk which could be dominant in FSRQs at optical frequencies \citep[e.g.,][see, however, \citealt{Mangalam93}, for models relating variability due to hot spots or instabilities in accretion disk resulting in $\beta$ =1.4 to 2.1]{Ghisellini17}. The majority of obtained slopes could be reconciled if the intranight fluctuations are driven by changes in the bulk Lorentz factors of the jet, provided that the turbulence is dominant on smaller than few minutes timescales \citep[$\beta$$\sim$2.1--2.9;][]{Pollack16}.

The PSD slopes obtained in this analysis can be directly compared with \citet{Wehrle19} who derived blazar/AGN PSDs using the {\it Kepler-}satellite data with long duration ($>$75 days), nearly uniformly sampled light curves with sampling intervals 30 minutes (long--cadence) and 1 minute (short cadence data for the blazar OJ\,287), respectively. First, the PSD slopes obtained for their sample using the long--cadence data range between $\beta\sim$1.8 and 3.8 and covers temporal frequencies between log $\nu_k$$\sim$-6.5\,Hz and $\sim$-5.0\,Hz with the slope tending to white noise at higher frequencies (timescales $\leq$18 hours). Second, the PSDs slopes obtained for the BL Lacs and FSRQ types are indistinguishable from one another within this frequency range. Their results are comparable to ours (see above), even though our analysis cover variability frequencies higher than {\it Kepler}'s long--cadence data ($\sim$1.5 decades in frequency range down to sub--hour timescales). Moreover, using short--cadence data for the blazar OJ\,287, they obtain $\beta$ $\sim$2.8 in the frequency range log $\nu_k$ = $-$2.7\,Hz to $-$5.7\,Hz with the PSD flattening to white-noise at $\nu_k$ $\geq$ 10$^{-2.7}$\,Hz. Our intranight PSD slopes for the OJ\,287, obtained on three separate occasions, are 3.8, 2.9, and 3.1, respectively, consistent with their result on overlapping variability frequencies (Table~\ref{tab:psd}).

In \citet[][]{Goyal17}, \citet[][]{Goyal18}, and \citet[][]{Goyal20}, based on PSD analyses of long-term variability using decade--long GHz--band radio--to--TeV$\gamma-$ray light curves of a few selected blazar sources, we hypothesized that the broadband emission is generated in an extended yet highly turbulent jet. The variability appeared to be driven by a single stochastic process at synchrotron frequencies but seemed to require the linear superposition of two stochastic processes at IC frequencies with relaxation timescales $\geq$1,000 days and $\sim$ days, respectively. Stochastic fluctuations in the local jet conditions (e.g., velocity fluctuations in the jet plasma or magnetic field variations) lead to energy dissipation over all spatial scales. The radiative response of the accelerated particles is delayed with respect to the input perturbations and this forms the red--noise segment of the PSD at synchrotron frequencies. At IC frequencies, however, due to inhomogeneities in the local photon population available for upscattering, the additional relaxation timescale of about $\sim$ one day, i.e., the light crossing time of the emission region, can result in a jet with Doppler boosting factor, 30, forming the pink-noise segment of the PSD. The steeper than red--noise PSD slopes on intranight timescales obtained in this analysis against the strict red--noise character of long--term variability at optical frequencies \citep[$\beta\sim$2;][]{Chatterjee08, Goyal17, Nilsson18, Goyal20}, indicate a cutoff of variability power on timescales around $\sim$days. We note that such a cutoff of variability power on timescales $\sim$days have been noted in the X--ray PSD of the blazar Mrk\,421 for which the X-ray emission, although {\it it} originates in the non-thermal jet, it is believed that the variability process is driven by accretion disk processes \citep[][]{Chatterjee18}. Moreover, our conclusion is only tentative, as joint analysis of full variability spectrum using long-term and intranight data, covering many orders of frequencies without gaps, is needed to reach robust conclusions. The normalization of PSDs for a few sources which were monitored on multiple occasions turns out to be consistent with one another within 1$\sigma$ uncertainty with a few exceptions. For the blazars 0420$-$014, OJ\,287 and 3C\,279, the normalization changes by one order of magnitude between different epochs (Figure~\ref{fig:jointpsds}). This indicates a hint of non-stationarity of the variability process on intranight timescales \citep[similar conclusions are obtained for the intranight X--ray variability of the blazar Mrk\,421 for which the intranight light curves are modeled as a non-stationary stochastic process;][]{Bhattacharyya20}.    

 At this point, we note that the duty cycle of intranight blazar variability at optical frequencies is found to be $\sim$40\% when monitored for a duration $>$4\,hours and measurement accuracies 0.2--5\% in few minutes of integration time \citep[][]{Goyal13b}. Almost always, these blazars are variable on longer timescales \citep[$>$days to years;][]{Stalin04, Sagar04, Gopal-Krishna11, Goyal12} which exhibits a red--noise character down to a few days timescales (see above). However, $\sim$60\% of the monitoring sessions, these sources turned out to be non-variable at short timescales meaning that small-scale flux variability, if present, is below the measurement uncertainties. This would imply, occurring intermittently, a cutoff of variability power on timescales longer than $\sim$0.5\,day. Neglecting these cases clearly introduces a bias in the understanding of these results as the PSDs are derived only when the statistically significant variability is found. The implication would be that the energy dissipation processes within the jet generating the flux variations on these timescales are transitory in nature and as such should be taken into consideration when modeling the jet emission and its variability down to intranight timescales \citep[see, in this context,][who derives the PSDs over 5 decades of temporal frequency range down to days timescale]{Pollack16}. 

Finally, we note that variability timescales smaller than the light--crossing time of the event horizon of the SMBH provide natural scales for the cutoff of variability if the dominant particle acceleration mechanism arise from disturbances or instabilities near the jet base \citep[$\sim$16 minutes for a 10$^8$ solar mass SMBH;][]{Begelman08}. Using the black hole masses of blazars studied here (column 7; Table~\ref{tab:sample}), the light crossing time of the event horizon ranges from $\sim$4\,minutes for the smallest SMBH mass of the blazar 1219+285 to $\sim$2\,hours for the largest SMBH mass for blazars 0420$-$014, 1156+295, 0806+524, and 3C\,279, respectively. These timescales translate to $\sim$24\,seconds--12\,minutes in the observer's frame assuming typical bulk Lorentz factor, 10, for the jet plasma \citep[][]{Lister16}. Such timescales are not covered by us, given the typical sampling intervals $\sim$5--15\,minutes. This will be explored in future studies with dedicated blazar monitoring programs on $>$2\,m aperture telescopes enabling flux measurements down to sub-percent accuracies in a few seconds of integration time to reach the smallest energy dissipation sites in jets.   

\acknowledgments

I thank the referee for careful reading of the manuscript and providing many insightful comments which have improved both the content and the presentation. AG acknowledges the financial support from the Polish National Science Centre (NCN) through the grant 2018/29/B/ST9/02298. The light curve simulations have been performed at the Prometheus cluster of the Cyfronet PL grid under the computing grant `lcsims2'. I thank Micha{\l} Ostroswski, Paul J. Wiita, and Marian Soida for discussions.  

\vspace{5mm}
\facilities{NED, ST:1.0m, IGO:2.0m}

\clearpage
\newpage



\end{document}